\documentclass[aps,onecolumn,notitlepage,footinbib,superscriptaddress,showpacs,longbibliography]{revtex4-1}%
\let\revappendix\appendix
\usepackage[english]{babel}
\usepackage[T1]{fontenc}
\usepackage[dvipsnames]{xcolor}
\usepackage{endnotes}
\usepackage{amssymb,amsmath,amsfonts}
\usepackage{textcomp}
\usepackage{graphicx}
\usepackage[utf8x]{inputenc}
\usepackage{float}
\usepackage[normalem]{ulem}
\usepackage{siunitx}
\usepackage{tikz}
\usepackage{epstopdf}
\usepackage{booktabs}
\usepackage{hyperref}
\usepackage{graphicx}
\usepackage[export]{adjustbox}
\draft 
\graphicspath{{./figures/}
              {./figures/valid/}}

\begin{document}

\newcommand{\lsolid}  [1]{\raisebox{2pt}{\tikz{\draw[#1,solid     ,line width=1pt](0,0)--(07mm,0);}}}
\newcommand{\ldash}   [1]{\raisebox{2pt}{\tikz{\draw[#1,dashed    ,line width=1pt](0,0)--(07mm,0);}}}
\newcommand{\ldott}   [1]{\raisebox{2pt}{\tikz{\draw[#1,dotted    ,line width=1pt](0,0)--(07mm,0);}}}
\newcommand{\ldashdot}[1]{\raisebox{2pt}{\tikz{\draw[#1,dashdotted,line width=1pt](0,0)--(07mm,0);}}}
\newcommand{\ldashlon}[1]{\raisebox{2pt}{\tikz{\draw[#1,longdash  ,line width=1pt](0,0)--(07mm,0);}}}
\definecolor{brick}{rgb}{0.75,0,0}
\definecolor{blight}{rgb}{0.7,0.65,1}
\definecolor{yellow1}{rgb}{1,0.85,0.1}
\definecolor{yellow2}{rgb}{1,0.95,0.5}
\definecolor{lila}{rgb}{0.7,0.3,0.5}
\definecolor{brick}{rgb}{0.75,0,0}
\definecolor{rltgreen}{rgb}{0,0.5,0}
\definecolor{oneblue}{rgb}{0,0,0.75}
\definecolor{marron}{rgb}{0.64,0.16,0.16}
\definecolor{forestgreen}{rgb}{0.13,0.54,0.13}
\definecolor{purple}{rgb}{0.62,0.12,0.94}
\definecolor{dockerblue}{rgb}{0.11,0.56,0.98}
\definecolor{freeblue}{rgb}{0.25,0.41,0.88}
\definecolor{myblue}{rgb}{0,0.2,0.4}
\definecolor{rosy}{rgb}{.971,.951,.92} 
\definecolor{gris}{rgb}{0.6,0.6,0.6}
\definecolor{grisclair}{rgb}{0.9,0.9,0.9}
\definecolor{rouge}{rgb}{0.6,0.,0.}
\definecolor{darkgreen}{rgb}{0.12, 0.3, 0.17}

\title{Shock impingement on a transitional hypersonic high-enthalpy boundary layer}

\author{D. Passiatore}
\affiliation{DMMM, Politecnico di Bari, via Re David 200, 70125 Bari, Italy}
\email[Corresponding author (currently at Stanford University): ]{donatella.passiatore@poliba.it.}

\author{L. Sciacovelli}
\affiliation{Laboratoire DynFluid, Arts et M\'{e}tiers ParisTech, 151 Bd. de l'H\^{o}pital, 75013 Paris, France}
\author{P. Cinnella}
\affiliation{Sorbonne Universit\'{e}, Institut Jean Le Rond d'Alembert , 4 Place Jussieu, 75005 Paris, France}
\author{G. Pascazio}
\affiliation{DMMM, Politecnico di Bari, via Re David 200, 70125 Bari, Italy}

\date{\today}

\begin{abstract}
The dynamics of a shock wave impinging on a transitional high-enthalpy boundary layer out of thermochemical equilibrium is investigated for the first time by means of a direct numerical simulation. The freestream Mach number is equal to 9 and the oblique shock impinges with a cooled flat-plate boundary layer with an angle of $10^\circ$, generating a reversal flow region. In conjunction with freestream disturbances, the shock impingement triggers a transition to a fully turbulent regime shortly downstream of the interaction region. Accordingly, wall properties emphasize the presence of a laminar region, a recirculation bubble, a transitional zone and fully turbulent region. In the entire transitional process, the recognized mechanisms are representative of the second mode instability combined with stationary streaky structures, their destabilization being eventually promoted by shock impinging. The breakdown to turbulence is characterized by an anomalous increase of skin friction and wall heat flux, due to the particular shock pattern. 
At the considered thermodynamic conditions the flow is found to be in a state of thermal non-equilibrium throughout, with non-equilibrium effects enhanced and sustained by the shock-induced laminar/turbulent transition, while chemical activity is almost negligible due to wall cooling. In the interaction region, relaxation towards thermal equilibrium is delayed and the fluctuating values of the rototranslational and the vibrational temperatures strongly differ, despite the strong wall-cooling.
The fully turbulent portion exhibits evolutions of streamwise velocity, Reynolds stresses and turbulent Mach number in good accordance with previous results for highly-compressible cooled-wall boundary layers in thermal nonequilibrium, with turbulent motions sustaining thermal nonequilibrium.
Nevertheless, the vibrational energy is found to contribute little to the total wall heat flux.
\end{abstract}

\maketitle

\section{Introduction}
Shock-wave boundary layer interactions (SWBLI) have been extensively investigated over the last decades due to their importance for both aeronautical and aerospace applications. The impingement of a shock wave on a fully developed boundary layer may indeed occur for several reasons and both in internal and external flow configurations. For instance, a shock wave interacting with a boundary layer can be driven by the complex geometry of the body itself (e.g. ramps, wedges) or it can impinge on the boundary layer after generating from an external body. The latter can be the case of supersonic intakes or multi-body launch vehicles, and it is the focus of the present work. The physics underlying such configuration is complex and strongly multiscale. The dynamics of high-speed compressible turbulent boundary layers becomes tightly coupled with strong gradients of the thermodynamic properties, leading to an increase of thermomechanical loads (see e.g., \cite{delery2009some} for an overview of relevant physical processes). In hypersonic and high-enthalpy regimes, thermochemical non-equilibrium effects (i.e., chemical reactions and vibrational excitation) must be taken into account as well \cite{candler2019rate}, further complicating the picture. Given the coexistence of several critical features, computational approaches based on averaged Navier--Stokes equations are unable to faithfully predict the flow field behavior, hence the necessity of performing high-fidelity spatial- and time-resolving simulations.\\
When an incident shock impinges on a fully developed boundary layer, the latter experiences a strong adverse pressure gradient. If the shock is strong enough, a recirculation bubble occurs and the flow separates. Several additional flow features are generated by this interaction, depending on the nature of the incoming boundary layer.
Laminar boundary layers have the advantage of a lower drag but are more sensitive to separation in adverse pressure gradients, resulting in wider recirculation bubbles with respect to turbulent flows. In such configurations, investigations have been carried out concerning shock-induced instabilities \cite{robinet2007bifurcations,hildebrand2018simulation} as well as shock-induced transition to turbulence \cite{fu2021shock}, that can be obtained when the shock angle is sufficiently high.\\
The interaction between an oblique shock and a fully turbulent boundary layer has been the object of intensive research efforts \cite{andreopoulos2000shock,dupont2005space,dupont2006space,pirozzoli2006direct,adler2018dynamic,yu2022spectral}. One of the most remarkable results of the interaction is the amplification of turbulence downstream of the incident shock and the emergence of oscillatory motions. For strong interactions, the ensemble of the separation bubble and the shock system is subjected to unsteady motions that spread on a wide range of characteristic frequencies. For instance, an oscillatory behavior of the reflected shock has been observed for high-frequency ranges \cite{babinsky2011shock}, representative of the most energetic turbulent scales of the incoming boundary layers. Kelvin--Helmholtz waves are also depicted, destabilizing the occurring shear layer and leading to vortex shedding. Another oscillatory motion is the so-called bubble ``breathing'', a low-frequency instability corresponding to enlargement and shrinkage of the bubble, observed both numerically and experimentally \cite{delery2009some}. Studies on the topic are numerous \cite{sandham2011shock,aubard2013large,clemens2014low}, although no clear consensus on the specific source of this flow unsteadiness (e.g., upstream boundary layer fluctuations, shear layer entertainment mechanism, intermittency) has been found yet. Several parameters can affect the overall SWBLI dynamics, among which it is noteworthy mentioning the effect of non-adiabatic walls. It has been indeed found that wall cooling tends to reduce the interaction scales and the bubble size, while increasing pressure fluctuations \cite{volpiani2018effects,volpiani2020effects}.\\
More recently, attention has been paid to shock-wave/transitional boundary layer interactions \cite{vanstone2013shock,sandham2014transitional,schulein2014effects,willems2015experiments,currao2020hypersonic}. These studies are meant to mimic more realistic configurations in which the boundary layer is not completely unperturbed, but may be subjected to random disturbances deriving from the external flow or naturally generated in wind tunnel facilities.\\
Of particular interest is the high-enthalpy hypersonic regime (encountered in re-entry and low-altitude hypersonic flight problems), in which the assumption of calorically perfect gas is no longer valid and out-of-equilibrium processes can be triggered at the high temperatures induced by the intense wall friction and the strong shock waves. 
In such configurations, chemical dissociation and vibrational relaxation phenomena interplay with the SWBLI physics. High-enthalpy effects on turbulent flows have gained renewed attention for smooth boundary layer configurations \cite{duan2011direct4,passiatore2021finite,direnzo2021direct,passiatore2022thermochemical,li2022wall}, where their influence was found to be often important depending on the thermodynamic operating regime. On the other hand, most of the research on shock-wave boundary layer interactions is limited to calorically perfect gas assumptions and low-enthalpy conditions. To the authors' knowledge, the presence of high-enthalpy effects in SWBLI has only been considered by Volpiani \cite{volpiani2021numerical} and Passiatore \textit{et al.} \cite{passiatore2022high}, with the purpose in both cases of assessing the capabilities of numerical methods in robustly handling such severe configurations.\\
The main objective of this study is therefore to extend the knowledge about high-enthalpy wall-bounded turbulent flows to configurations involving the impingement of shock waves. For that purpose we perform for the first time a high-fidelity numerical simulation of a shock-wave/hypersonic boundary layer interaction, in the presence fo both chemical and thermal non-equilibrium effects. The boundary layer considered in this work is excited by means of superposed freestream disturbances and is found to be in a transitional state at the location of shock impingement.\\
The paper is organized as follows. Section~\ref{sec:equations} describes the governing equations and the thermochemical models used for the computation. The numerical strategy adopted and the problem setup are reported in sections~\ref{sec:numerics} and \ref{sec:setup}, respectively. Section~\ref{sec:results} presents the main results, providing a general overview of the flow dynamics, particular insights on the interaction region, inspections on turbulent statistics and a detailed analysis of the thermochemical flow field. Concluding remarks are then provided in section~\ref{sec:results}.
%
\section{Governing equations}\label{sec:equations}
The fluid under investigation is air at high-temperature in thermochemical non-equilibrium, modeled as a five-species mixture of N$_2$, O$_2$, NO, O and N. Such flows are therefore governed by the compressible Navier--Stokes equations for multicomponent chemically-reacting and thermally-relaxing gases, which read:
\begin{align}
\frac{\partial \rho}{\partial t} + \frac{\partial \rho u_j }{\partial x_j} & =0 \\
\label{eq:momentum}
\frac{\partial \rho u_i}{\partial t} + \frac{\partial \left( \rho u_i u_j + p \delta_{ij} \right)}{\partial x_j} &=\frac{\partial \tau_{ij}}{\partial x_j} \\
\label{eq:energy}
\frac{\partial \rho E}{\partial t} + \frac{\partial \left[\left(\rho E + p \right)u_j\right]}{\partial x_j} & =\frac{\partial (u_i \tau_{ij})}{\partial x_j} - \frac{\partial (q^{\text{TR}}_j + q^{\text{V}}_j)}{\partial x_j} -\frac{\partial}{\partial x_j}\left(\sum_{n=1}^\text{NS} \rho_n u_{nj}^D h_n \right) \\
\label{eq:species}
\frac{\partial \rho_n}{\partial t} + \frac{\partial \left( \rho_n u_j \right)}{\partial x_j} &= -\frac{\partial \rho_n u_{nj}^D }{\partial x_j} + \dot{\omega}_n \qquad (n = 1,..,\text{NS}-1)\\
\label{eq:vibr}
\frac{\partial \rho e_\text{V}}{\partial t} + \frac{\partial \rho e_\text{V} u_j}{\partial x_j} &= \frac{\partial}{\partial x_j} \left( -q^\text{V}_j - \sum_{m=1}^\text{NM} \rho_m u_{mj}^D e_{\text{V}m} \right) + \sum_{m=1}^{\text{NM}} \left( Q_{\text{TV}m} + \dot{\omega}_m e_{\text{V}m} \right).
\end{align}
In the preceding formulation, $\rho$ is the mixture density, $t$ the time coordinate, $x_j$ the space coordinate in the $j$-th direction of a Cartesian coordinate system, with $u_j$ the velocity vector component in the same direction, $p$ is the pressure, $\delta_{ij}$ the Kronecker symbol and $\tau_{ij}$ the viscous stress tensor, modeled as
\begin{equation}
 \tau_{ij} = \mu\left(\frac{\partial u_i}{\partial x_j} + \frac{\partial u_j}{\partial x_i} \right) -\frac{2}{3}\mu\frac{\partial u_k}{\partial x_k}\delta_{ij},
 \end{equation}
with $\mu$ the mixture dynamic viscosity.
In equation~\eqref{eq:energy}, $E = e + \frac{1}{2}u_i u_i$ is the specific total energy (with $e$ the mixture internal energy), $q^\text{TR}_j$ and $q^\text{V}_j$ the roto-translational and vibrational contributions to the heat flux, respectively; $u_{nj}^D$ denotes  the diffusion velocity and $h_n$  the specific enthalpy for the $n$-th species. In the species conservation equations \eqref{eq:species}, $\rho_n = \rho Y_n$ represents the $n$-th species partial density ($Y_n$ being the mass fraction) and $\dot{\omega}_n$ the rate of production of the $n$-th species. The sum of the partial densities is equal to the mixture density $\rho{=}\sum_{n=1}^\text{NS} \rho_n$, NS being the total number of species. To ensure total mass conservation, the mixture density and NS$-1$ species conservation equations are solved, while the density of the NS-th species is computed as $\rho_\text{NS} = \rho - \sum_{n=1}^{\text{NS}-1} \rho_n$. We set such species as molecular nitrogen being the most abundant one throughout the computational domain. As for equation~\eqref{eq:vibr}, $e_\text{V}=\sum_{m=1}^\text{NM} Y_m e_{\text{V}m}$ represents the mixture vibrational energy, with $e_{\text{V}m}$ the vibrational energy of the $m$-th molecule and NM their total number. In the same equation, $Q_\text{TV}=\sum_{m=1}^\text{NM} Q_{\text{TV}m}$ represents the energy exchange between vibrational and translational modes (due to molecular collisions and linked to energy relaxation phenomena) and $\sum_{m=1}^\text{NM} \dot{\omega}_m e_{\text{V}m}$ the vibrational energy lost or gained due to molecular depletion or production. Each species is assumed to behave as a thermally-perfect gas; Dalton's pressure mixing law leads then to the thermal equation of state:
\begin{equation}
   p =  \rho T \sum_{n=1}^\text{NS} \frac{\mathcal{R} Y_n}{\mathcal{M}_n} = T \sum_{n=1}^\text{NS} \rho_n R_n,
\end{equation}
$R_n$ and ${\cal M}_n$ being the gas constant and molecular weight of the $n$-th species, respectively, and $\mathcal{R} = 8.314$ J/mol~K the universal gas constant. The thermodynamic properties of high-$T$ air species are computed considering the contributions of translational, rotational and vibrational (TRV) modes; specifically, the internal energy reads:
\begin{equation}
e = \sum_{n=1}^\text{NS} Y_n h_n - \frac{p}{\rho},
\quad \text{with} \quad
h_n = h^0_{f,n} + \int_{T_\text{ref}}^T (c^\text{T}_{p,n}+c^\text{R}_{p,n}) \text{ d}T' + e_{Vn} (T_V).
\end{equation}
Here, $h^0_{f,n}$ is the $n$-th species enthalpy of formation at the reference temperature ($T_\text{ref} = \SI{298.15}{K}$), $c^\text{T}_{p,n}$ and $c^\text{R}_{p,n}$ the translational and rotational contributions to the isobaric heat capacity of the $n$-th species, computed as
\begin{equation}
   c^\text{T}_{p,n} = \frac{5}{2} R_n \qquad \text{and} \qquad
   c^\text{R}_{p,n} = \begin{cases}
    R_n & \text{for diatomic species} \\
    0 & \text{for monoatomic species}
   \end{cases}
\end{equation}
and $e_{Vn}$ the vibrational energy of species $n$, given by
\begin{equation}\label{eq:evib}
 e_{Vn} =  \frac{T^\star_{V,n} R_n}{\exp{(T^\star_{V,n}/T_\text{V})} - 1},
\end{equation}
with $T^\star_{V,n}$ the characteristic vibrational temperature of each molecule ($\SI{3393}{K}$, $\SI{2273}{K}$ and $\SI{2739}{K}$ for N$_2$, O$_2$ and NO, respectively). After the numerical integration of the conservation equations, the roto-translational temperature $T$ is computed from the specific internal energy (devoid of the vibrational contribution) directly, whereas an iterative Newton--Raphson method is used to compute $T_\text{V}$ from $e_\text{V}=\sum_{m=1}^\text{NM} Y_m e_{\text{V}m}$.\\
Both the heat fluxes are modeled by means of Fourier's law, $q^\text{TR}_j = -\lambda_\text{TR} \frac{\partial T}{\partial x_j}$ and $q^\text{V}_j = -\lambda_\text{V} \frac{\partial T_\text{V}}{\partial x_j}$, $\lambda_\text{TR}$ and $\lambda_\text{V}$ being the roto-translational and vibrational thermal conductivities, respectively. 
To close the system, we use the two-temperatures (2T) model of Park \cite{park1988two} to take into account the simultaneous presence of thermal and chemical non-equilibrium for the computation of $\dot{\omega}_n$ and $Q_\text{TV}$. Specifically, the five species interact with each other through a reaction mechanism consisting of five reversible chemical steps \cite{park1989nonequilibrium}:
\begin{alignat}{3}
 \notag \text{R1}: & \qquad  \text{N}_2 + \text{M} && \Longleftrightarrow 2\text{N} + \text{M} \\
 \notag \text{R2}: & \qquad  \text{O}_2 + \text{M} && \Longleftrightarrow 2\text{O} + \text{M} \\ \label{eq:reactions}
        \text{R3}: & \qquad  \text{NO}  + \text{M} && \Longleftrightarrow  \text{N} + \text{O} + \text{M} \\
 \notag \text{R4}: & \qquad \text{N}_2 + \text{O}  && \Longleftrightarrow  \text{NO}+ \text{N} \\
 \notag \text{R5}: & \qquad \text{NO}  + \text{O}  && \Longleftrightarrow  \text{N} + \text{O}_2
\end{alignat}
where M is the third body (any of the five species considered). Dissociation and recombination processes are described by reactions R1, R2 and R3, whereas the shuffle reactions R4 and R5 represent rearrangement processes. The mass rate of production of the \textit{n}-th species is governed by the law of mass action:
\begin{equation}
 \dot{\omega }_n =  \mathcal{M}_n \sum_{r=1}^\text{NR} \left( \nu_{nr}'' - \nu_{nr}' \right) \times \left[ k_{f,r} \prod_{n=1}^\text{NS} \left(\frac{\rho Y_n}{\mathcal{M}_n}\right)^{\nu_{nr}'} - k_{b,r} \prod_{n=1}^\text{NS} \left(\frac{\rho Y_n}{\mathcal{M}_n}\right)^{\nu_{nr}''} \right],
 \label{eq:mass_action}
\end{equation}
where $\nu_{nr}'$ and  $\nu_{nr}''$ are the stoichiometric coefficients for reactants and products in the $r$-th reaction for the $n$-th species, respectively, and NR is the total number of reactions. Furthermore, $k_{f,r}$ and $k_{b,r}$ denote the forward and backward rates of the $r$-th reaction, modeled by means of Arrhenius' law. The coupling between chemical and thermal nonequilibrium is taken into account by means of a modification of the temperature values used for computing the reaction rates. Indeed, a geometric-averaged temperature is considered for the dissociation reactions R1, R2 and R3 in~\eqref{eq:reactions}, computed as $T_\text{avg}=T^{q}T_\text{V}^{1-q}$ with $q=0.7$ \cite{park1988two}.\\
Lastly, the vibrational-translational energy exchange is computed as:
\begin{equation}\label{eq:qtv}
 Q_\text{TV} =  \sum_{m=1}^\text{NM} Q_{\text{TV},m} = \sum_{m=1}^\text{NM} \rho_m \frac{e_{\text{V}m}(T) - e_{\text{V}m}(T_\text{V})}{t_m},
\end{equation}
where $t_m$ is the corresponding relaxation time evaluated by means of the expression \cite{millikan1963systematics}:
\begin{equation}
 t_m = \sum_{n=1}^\text{NS}\frac{\rho_n}{\mathcal{M}_n} \sum_{n=1}^\text{NS}\frac{t_{mn}}{\rho_n/\mathcal{M}_n}.
\end{equation}
Here, $t_{mn}$ is the relaxation time of the $m$-th molecule with respect to the $n$-th species, computed as the sum of two contributions
\begin{equation}
 t_{mn} = t_{mn}^{MW} + t_{mn}^c.
\end{equation}
The first term writes:
\begin{equation}
\label{eq:relax_time}
 t_{mn}^\text{MW} = \frac{p}{p_\text{atm}}\exp\left[{ a _{mn}(T^{-\frac{1}{3}} -b_{mn})-18.42}\right],
\end{equation}
where $p_\text{atm}=\SI{101325}{Pa}$ and $a_{mn}$ and $b_{mn}$ are coefficients reported in \cite{park1993review}. Since this expression tends to underestimate the experimental data at temperatures above \SI{5000}{K}, a high-temperature correction was proposed by Park \cite{park1989assessment}:
\begin{equation}
t_{mn}^c = \sqrt{\frac{\phi_{mn}}{\mathcal{M}_m \sigma}},
\end{equation}
where $\phi_{mn} = \frac{\mathcal{M}_m \mathcal{M}_n}{\mathcal{M}_m + \mathcal{M}_n}$ and $\sigma = \sqrt{\frac{8\mathcal{R}}{T \pi}}\frac{7.5 \times 10^{-12} \text{NA}}{T}$, NA being Avogadro's number. Note that this correction is taken into account in our numerical simulations, albeit such high temperatures are hardly reached in the computational domain under the thermodynamic conditions currently investigated.\\
As for the computation of the transport properties, pure species' viscosity and thermal conductivities are computed using curve-fits by Blottner \cite{blottner1971chemically} and Eucken's relations \cite{hirschfelder}, respectively. The corresponding mixture properties are evaluated by means of Wilke's mixing rules \cite{wilke1950viscosity}. Mass diffusion is modeled by means of Fick's law:
\begin{equation}
 \rho_n u^D_{nj}= -\rho D_{n} \left(\frac{\partial Y_n}{\partial x_j} +  Y_n \frac{\partial \text{ln}\mathcal{M}}{\partial x_j}  \right) + \rho_n \sum_{n=1}^\text{NS} D_n \left(  \frac{\partial Y_n}{\partial x_j} + Y_n \frac{\partial \text{ln}\mathcal{M}}{\partial x_j}  \right),
\end{equation}
where $\mathcal{M}$ is the mixture molecular weight. Here, the first term on the r.h.s. represents the effective diffusion velocity and the second one is a mass corrector term that should be taken into account in order to satisfy the continuity equation when dealing with non-constant species diffusion coefficients \cite{poinsot2005theoretical}. Specifically, $D_n$ is an equivalent diffusion coefficient of species $n$ into the mixture, computed following Hirschfelder's approximation \cite{hirschfelder}, starting from the binary diffusion coefficients which are curve-fitted in \cite{gupta1990review}. 
\section{Numerical methodology}\label{sec:numerics}
The numerical solver described in Sciacovelli \textit{et al.} \cite{sciacovelli2021assessment} is used for the present computation. The Navier--Stokes equations are integrated numerically by using a high-order finite-difference scheme. The convective fluxes are discretized by means of central tenth-order differences, supplemented with a high-order adaptive nonlinear artificial dissipation. The latter consists in a blend of a ninth-order-accurate dissipation term based on tenth-order derivatives of the conservative variables, used to damp grid-to-grid oscillations, along with a low-order shock-capturing term. This term is equipped with a highly-selective pressure-based sensor. For the vibrational energy equation, a sensor based on second-order derivatives of the vibrational temperature is used. Time integration is carried out using a low-storage 3$^\text{rd}$-order Runge--Kutta scheme. The numerical strategy has been validated for thermochemical nonequilibrium flows, including SWBLI laminar configurations \cite{passiatore2022high}.

\section{Problem setup}\label{sec:setup}
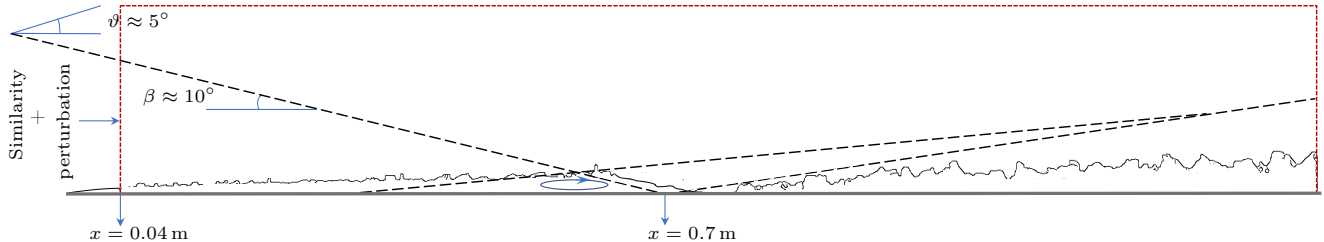
\begin{figure}
 \centering
   \begin{tikzpicture}
   \node[anchor=south west,inner sep=0] (image) at (0,0) {
   \includegraphics[width=\columnwidth,trim={10 150 20 150},clip]{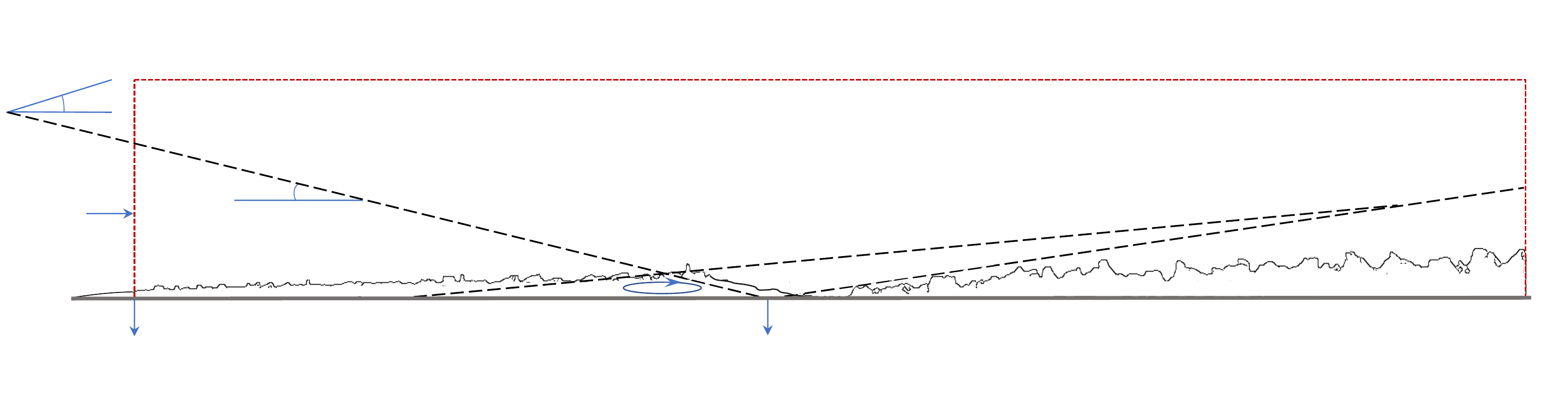}};
   \begin{scope}[x={(image.south east)},y={(image.north west)}]
   \node [align=center] at (0.51,0.15){\scriptsize $x=\SI{0.7}{m}$};
   \node [align=center] at (0.1,0.15){\scriptsize $x=\SI{0.04}{m}$};
   \node [align=center] at (0.1,0.76) {\scriptsize $\vartheta \approx 5^\circ$};
   \node [align=center] at (0.13,0.54) {\scriptsize $\beta \approx 10^\circ$};
   \node [align=center,rotate=90] at (0.01,0.49) {\scriptsize Similarity};
   \node [align=center] at (0.025,0.48) {\scriptsize +};
   \node [align=center,rotate=90] at (0.045,0.47) {\scriptsize perturbation};
  \end{scope}
  \end{tikzpicture}
 \caption{Sketch of the configuration under investigation.}\label{fig:config}
\end{figure}
The configuration under investigation, displayed in figure~\ref{fig:config}, consists in a shock wave that impinges on a thermally and chemically out-of-equilibrium flat-plate boundary layer. We note in the following with $\vartheta$ and $\beta$ the deflection angle and the shock angle, respectively. The current setup stems from the one of Sandham \textit{et al.} \cite{sandham2014transitional}, in which inflow freestream perturbations were applied on a $M_\infty=6$ laminar perfect-gas boundary layer; the original case is presented in Appendix~\ref{app:sandham} for validation purposes.
Here, we consider the post-shock conditions of a scramjet, approximated by a $\SI{6}{\degree}$ planar wedge, flying at Mach 12 at an altitude of \SI{36}{km}. The freestream conditions are then $T_\infty=\SI{405}{K}$, $p_\infty=\SI{2258.6}{Pa}$ and $M_\infty = 9$, corresponding to a stagnation enthalpy of $H^0_\infty=\SI{6.86}{MJ/kg}$. To deal with calorically imperfect gas mixtures, the generalized Rankine-Hugoniot relations for an arbitrary equation of state are iteratively solved to obtain post-shock conditions.  For the selected freestream Mach number and a shock angle of $\beta=\SI{10}{\degree}$, the relations give $\vartheta \approx \SI{5}{\degree}$ and the post-shock (PS) variables $T_\text{PS}=\SI{554}{K}$, $p_\text{PS}=\SI{6235}{Pa}$, $u_\text{PS}=\SI{3620}{m/s}$. These conditions are imposed at the left boundary of the rectangular domain as a jump on the inlet profiles. The latter are obtained by solving the locally self-similar equations reported in \cite{sciacovelli2021assessment}, extended to thermochemical non-equilibrium. The inflow plane of the computational domain and the ideal impingement station are located at $x_0 = \SI{0.04}{m}$ and $x_\text{sh}=\SI{0.7}{m}$ from the leading edge, respectively.
The wall temperature is fixed equal to \SI{2500}{K} for both the translational and vibrational temperature, and non-catalytic conditions are applied. Characteristic outflow boundary conditions are imposed at the top and right boundaries, whereas periodicity is enforced in the spanwise direction.
The extent of the computational domain is $(L_x \times L_y \times L_z)/\delta^\star_\text{in} = 800  \times 80  \times 60 $, $\delta^\star_\text{in}=\SI{1.77e-3}{m}$ being the displacement thickness at the inlet, defined in equation~\eqref{eq:delta_star}. As for the discretization, a total number of $N_x \times N_y \times N_z =  6528 \times 402 \times 720$ grid points is used, with constant grid size in the streamwise and spanwise directions and a constant grid stretching of 1\% in the wall-normal direction, the height of the first cell away from the wall being $\Delta y_w = \SI{2.5e-5}{m}$. Unless otherwise stated, in the following we will make use of a dimensionless streamwise coordinate computed as $\hat{x}= (x-x_\text{sh})/\delta^\star_\text{in}$.

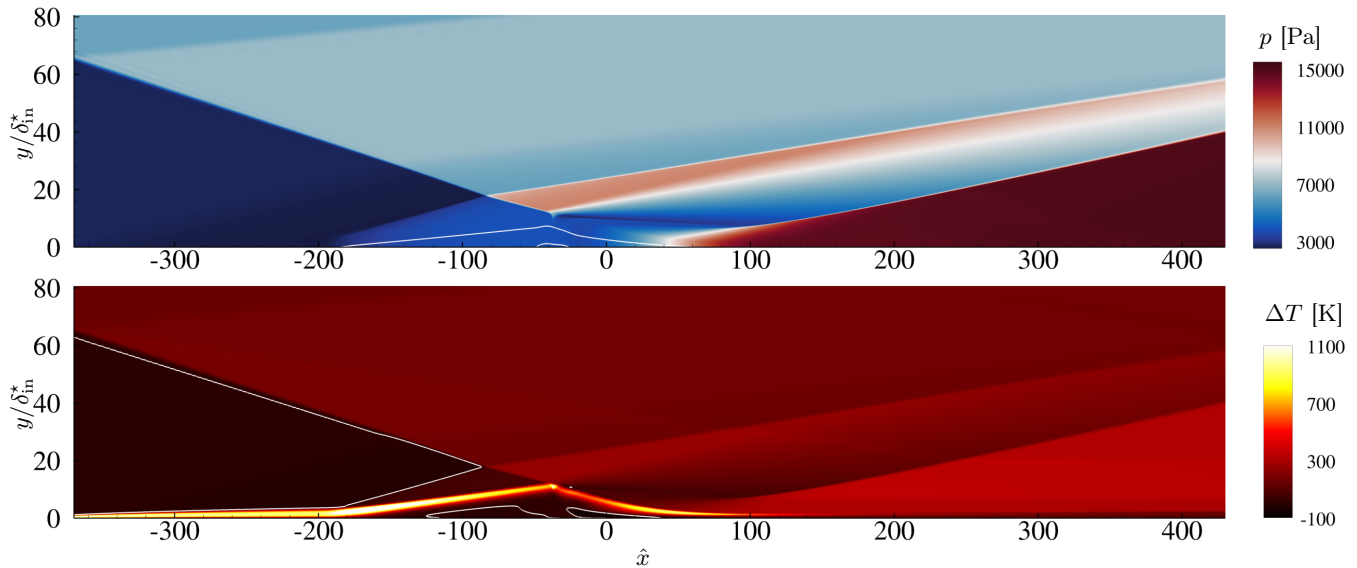
\begin{figure}
 \centering
   \begin{tikzpicture}
   \node[anchor=south west,inner sep=0] (image) at (0,0) {
   \includegraphics[width=\columnwidth,trim={5 5 5 5},clip]{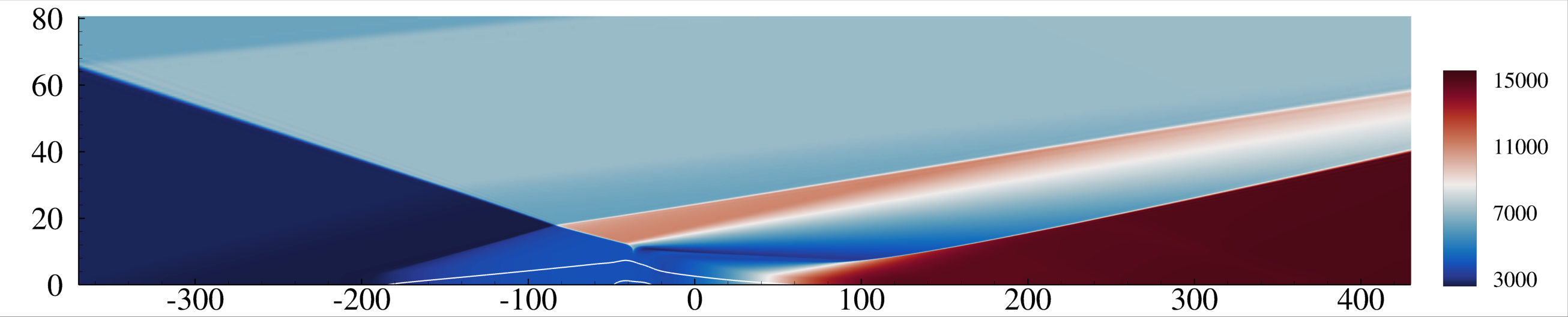}};
   \begin{scope}[x={(image.south east)},y={(image.north west)}]
   \node [align=center] at (0.95,0.87) {\small $p$ [Pa]};
   \node [align=center,rotate=90] at (0.01,0.52) {\small $y/\delta^\star_\text{in}$};
   \end{scope}
  \end{tikzpicture}\\
  \begin{tikzpicture}
   \node[anchor=south west,inner sep=0] (image) at (0,0) {
   \includegraphics[width=\columnwidth,trim={5 5 5 5},clip]{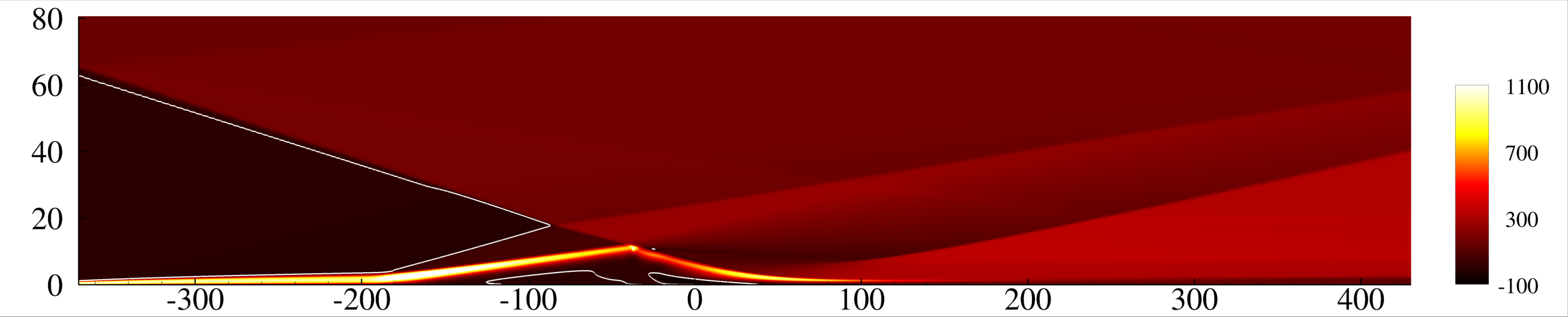}};
   \begin{scope}[x={(image.south east)},y={(image.north west)}]
   \node [align=center] at (0.96,0.85) {\small $\Delta T$ [K]};
   \node [align=center] at (0.47,-0.05) {\small $\hat{x}$};
   \node [align=center,rotate=90] at (0.01,0.52) {\small $y/\delta^\star_\text{in}$};
   \end{scope}
  \end{tikzpicture}
 \caption{Visualization of the base flow without perturbation. Top: isocontours of pressure, $p$; the white line denotes the extent of the recirculation bubble ($u=0$ isoline). Bottom: isocontours of the temperature difference, $\Delta T = T-T_V$; the white line denotes the $\Delta T = 0$ contour. The $y$ axis has been stretched for better visualization.}\label{fig:base_flow}
\end{figure}

Figure~\ref{fig:base_flow} reports the pressure and temperature difference ($\Delta T = T-T_V$) isocontours of the base flow used to initialize the three-dimensional computation. The adverse pressure gradient induced by the incident shock generates a recirculation bubble, marked with a white line in the top figure. Upstream of the separation bubble, a series of compression waves occur, which then coalesce into the separation shock; the latter interacts with the incident shock that penetrates the separated flow. Downstream of the separation bubble, a reattachment shock is generated which readjusts the previously deflected flow.
Globally, the characteristic features of SWBLI are not altered by high-enthalpy effects; on the other hand, such a complex dynamics strongly influences the thermochemical activity. Coherently with the inlet temperature profiles, the amount of thermal non-equilibrium before the bubble is extremely high while chemical activity is essentially negligible. The rise of the temperatures and pressure in the separation zone enhances chemical dissociation whereas the gap between the two temperatures is reduced, moving towards a quasi thermally-equilibrated state right after the recirculation bubble. The $\Delta T = 0$ isoline, dividing the thermally under- and over-excited regions, shows that the flow is under-excited everywhere except for the freestream pre-shock region and in the recirculation bubble, where a slight vibrational over-excitation is observed. A comparison of the configurations with and without shock impingement reveals that, in the latter case, the flow remains in a state of stronger thermal non-equilibrium and quasi-frozen chemical activity throughout the entire boundary layer. Therefore, the pressure rise caused by the incident shock is responsible for a reduction of the amount of thermal non-equilibrium and an increase of the chemical activity.
Additional details about the present base flow can be found in Passiatore \textit{et al.}\cite{passiatore2022high}.

Laminar-to-turbulent transition for the three dimensional simulation is favoured by superimposing disturbances on the described base flow. Following \cite{sandham2014transitional}, the density self-similar profile is perturbed as follows:
\begin{equation}\label{eq:forcing}
\frac{\rho'}{\rho_\infty} = a W \left(\frac{y}{\delta^\star_\text{in}}\right) \sum_{j=1}^{N_J} \text{cos} \left(  \frac{2 \pi j z}{L_z} + \Phi_j\right)\sum_{k=1}^{N_K} \text{sin}\left( 2 \pi \hat{f}_k \frac{u_\infty}{\delta^\star_\text{in}} t + \Psi_k\right)
\end{equation}
Here, $a=\num{5e-4}$ is the amplitude of the perturbation, $W(y/\delta^\star_\text{in})= 1- \text{exp}(-(y/\delta^\star_\text{in})^3)$ is a function that damps the disturbances near the wall, $\Phi_j$ and $\Psi_k$ are the phases that correspond to random numbers between 0 and $2\pi$, with $N_J=16$ and $N_K=20$. The dimensionless frequency $\hat{f}_k$ is set equal to 0.02$k$.

\section{Results}\label{sec:results}
\begin{figure}
 \centering
    \begin{tikzpicture}
   \node[anchor=south west,inner sep=0] (a) at (0,0) {\includegraphics[width=1\textwidth, trim={5 8 8 35}, clip]{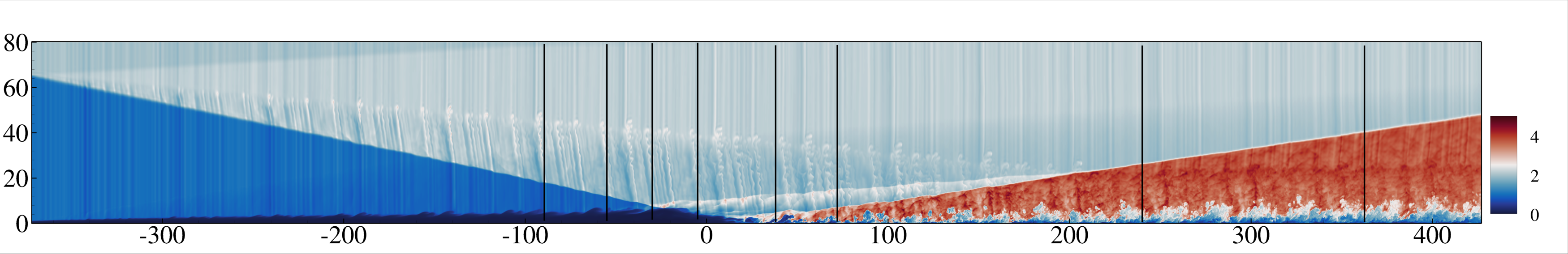}};
   \begin{scope}[x={(a.south east)},y={(a.north west)}]
     \node [align=center] at (0.5,0.0) {\small $\hat{x}$};
     \node [align=center] at (0.97,0.75) {\scriptsize $\frac{\rho u}{(\rho u)_\infty}$};
     \node [align=center,rotate=90] at (-0.015,0.54) {\scriptsize $y/\delta^\star_\text{in}$};
     \node [align=center,rotate=90] at (0.337,0.68) {\tiny $\hat{x}=-90$};
     \node [align=center,rotate=90] at (0.38,0.68) {\tiny $\hat{x}=-55$};
     \node [align=center,rotate=90] at (0.41 ,0.68) {\tiny $\hat{x}=-30$};
     \node [align=center,rotate=90] at (0.437,0.7) {\tiny $\hat{x}=-5$};
     \node [align=center,rotate=90] at (0.487 ,0.7) {\tiny $\hat{x}=38$};
     \node [align=center,rotate=90] at (0.525,0.7) {\tiny $\hat{x}=72$};
     \node [align=center,rotate=90] at (0.72 ,0.68) {\tiny $\hat{x}=240$};
     \node [align=center,rotate=90] at (0.86,0.68) {\tiny $\hat{x}=363$};
       \end{scope}
 \end{tikzpicture}
  \begin{tikzpicture}
   \node[anchor=south west,inner sep=0] (a) at (0,0) {\includegraphics[width=1\textwidth, trim={5 8 8 170}, clip]{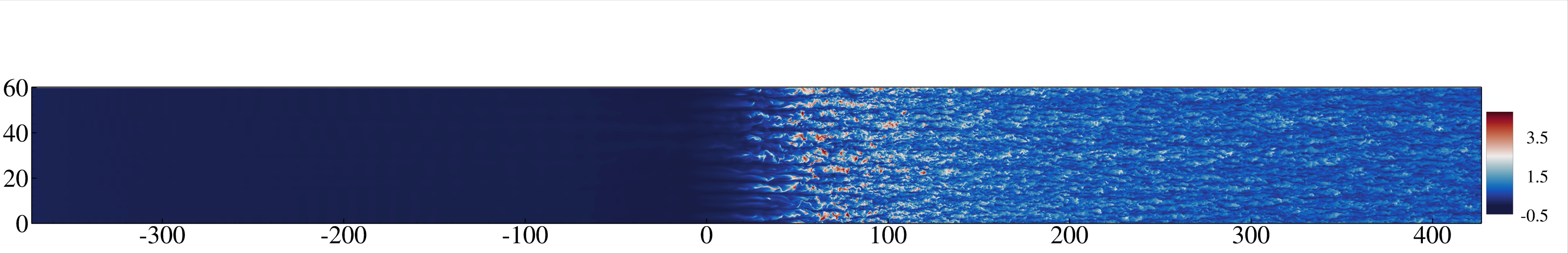}};
   \begin{scope}[x={(a.south east)},y={(a.north west)}]
     \node [align=center] at (0.97,0.92) {\scriptsize $\frac{\rho u}{(\rho u)_\infty}$};
     \node [align=center] at (0.5,0.0) {\small $\hat{x}$};
     \node [align=center,rotate=90] at (-0.015,0.58) {\scriptsize $z/\delta^\star_\text{in}$};
     \end{scope}
 \end{tikzpicture}
 \caption{Instantaneous visualization of streamwise momentum in a $xy$-plane (top) and in a $xz$-plane at $y/\delta^\star_\text{in}=0.5$ (bottom). The $y$ axis has been stretched for better visualization.}\label{fig:global_inst_view}
 \end{figure}
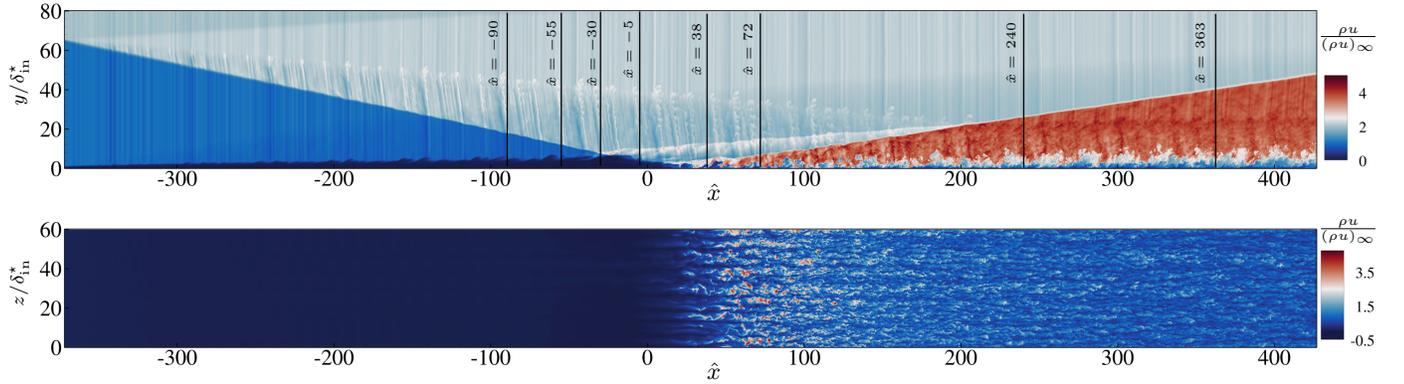

In the results presented below, flow statistics are computed by averaging in time and in the spanwise homogeneous direction, after that the initial transient has been evacuated. For a given variable $f$, we denote with $\overline{f} = f - f'$ the standard time- and spanwise average, being $f'$ the corresponding fluctuation, whereas $\widetilde{f}= f - f''= \overline{\rho f} / \overline{\rho}$ denotes the density-weighted Favre averaging, with $f''$ the Favre fluctuation. Statistics are collected for a period corresponding to $T_\text{stats} \approx  2050 \delta^\star_\text{in}/u_\infty$, for a total of approximately 40000 temporal snapshots. The sampling frequency is approximately 1000 and 60 times larger than the lowest and the highest harmonic in the perturbation function, respectively (i.e. $\Delta t_\text{stats} \approx 4.38 \times 10^{-2} \delta^\star_\text{in}/u_\infty$).\\
Profiles of various dynamic, thermodynamic and thermochemical flow quantities are extracted at eight stations in the laminar region, before the shock impingement, in the interaction zone and in the fully turbulent state. The location of such stations are marked in figure~\ref{fig:global_inst_view}, displaying the isocontours of the streamwise momentum in a $xy$-slice (top) and a near wall $xz$-slice (bottom). In the top panel, a shock pattern similar to the one in the base flow of figure~\ref{fig:base_flow} can be appreciated, as well as the effect of the superposed density disturbances. Due to the thickening of the perturbed boundary layer upstream of the impingement, it is possible to detect a distinguishable separation shock only after the interaction with the incident one. The angulation of the separation shock is such that it impacts the reattachment shock, differently from the base flow. Table~\ref{tab:stations} reports some boundary layer properties at the selected positions. 
Throughout the paper, the superscript ``$\bullet^+$'' denotes normalization with respect to the viscous length scale $\ell_v = \overline{\mu}_w/(\overline{\rho}_w u_\tau)$, $u_\tau = \sqrt{ \overline{\tau}_w/\overline{\rho}_w }$ being the wall friction velocity. The boundary layer displacement thickness, the momentum thickness and the shape factor are defined as:
\begin{align}
\label{eq:delta_star}
\delta^\star  =\int_0^\delta \left(1 - \frac{\rho u}{\rho_\delta u_\delta}\right) \text{ d}y, \qquad
\theta =\int_0^\delta \frac{\rho u}{\rho_\delta u_\delta} \left(1-\frac{\rho u}{\rho_\delta u_\delta}\right)\text{ d}y, \qquad
H =\frac{\delta^\star}{\theta},
\end{align}
where the subscript $\delta$ denotes the variables computed at the edge of the boundary layer. Mesh spacing in wall units, reported in the table, shows that DNS-like resolution is achieved throughout the domain. The wall-normal profiles of statistics are mostly displayed in inner semi-local units $y^\star=\overline{\rho} u_\tau^\star y/\overline{\mu}$, with $u_\tau^\star=\sqrt{\overline{\tau}_w/\overline{\rho}}$, or in outer scaling $y/\delta$.
\begin{table}
\begin{center}
 \def~{\hphantom{0}}
 \begin{tabular}{lcccccccccc}
  \midrule
   & Legend & $\hat{x}$ & $Re_x \times 10^5$ & $Re_\theta$ &  $Re_\tau$ & $Re_{\delta^\star}$& $H$ & $\Delta x^+$ & $\Delta z^+$ & $\Delta y_w^+$ \\
  \midrule
   Laminar      &   \ldashdot{blue}      &  -90 & 16.43 & 864.12 & -     & 20635  & 23.88   & -    & -     & -     \\
   Separation   &   \ldashdot{rltgreen}  &  -55 & 18.57 & 984.61 & -     & 23550  & 23.92   & -    & -     & -     \\
   Inside bubble&   \ldashdot{green}     &  -30 & 19.74 & 1003.0 & -     & 29318  & 29.23   & -    & -     & -     \\
   Reattachment &   \ldashdot{yellow}    &  -5  & 21.17 & 962.57 & -     & 19244  & 19.99   & -    & -     & -     \\
   Transition   &   \ldashdot{orange}    &  38  & 23.41 & 1169.4 & 245.6 & 8942   & 7.64     & 8.87 & 6.02  & 1.02   \\
   Transition   &   \ldashdot{red}       &  72  & 25.21 & 1108.2 & 392.5 & 7755   & 7.00     & 7.43 & 5.05  & 0.86   \\
   Turbulent   &   \ldashdot{marron}     &  240 & 34.29 & 1519.8 & 517.9 & 13688  & 9.00    & 8.89 & 6.04  & 1.02  \\
   Turbulent   &   \ldashdot{black}      &  363 & 40.90 & 1957.0 & 562.4 & 16858  & 8.61    & 8.76 & 5.96  & 1.00  \\
   \midrule
  \end{tabular}
 \end{center}
 \caption{Boundary layer properties at eight selected streamwise stations. In the table, Re$_x= \rho_\infty u_\infty x / \mu_\infty$ and Re$_\theta=\rho_\infty u_\infty \theta / \mu_\infty$ are the Reynolds numbers based on the distance from the leading edge and on the local momentum thickness, respectively; Re$_\tau= \rho_w u_\tau \delta / \overline{\mu}_w$  is the friction Reynolds number, $Re_{\delta^\star} = \rho_\infty u_\infty \delta^\star /\mu_\infty$ is the Reynolds number based on the displacement thickness and $H$ is the shape factor. Lastly, $\Delta x^+$,  $\Delta y^+_\text{w}$ and  $\Delta z^+$ denote the grid sizes in inner variables in the $x$-direction, $y$-direction at the wall and in the $z$-direction, respectively.}\label{tab:stations}
\end{table}

\subsection{Overview of flow dynamics}

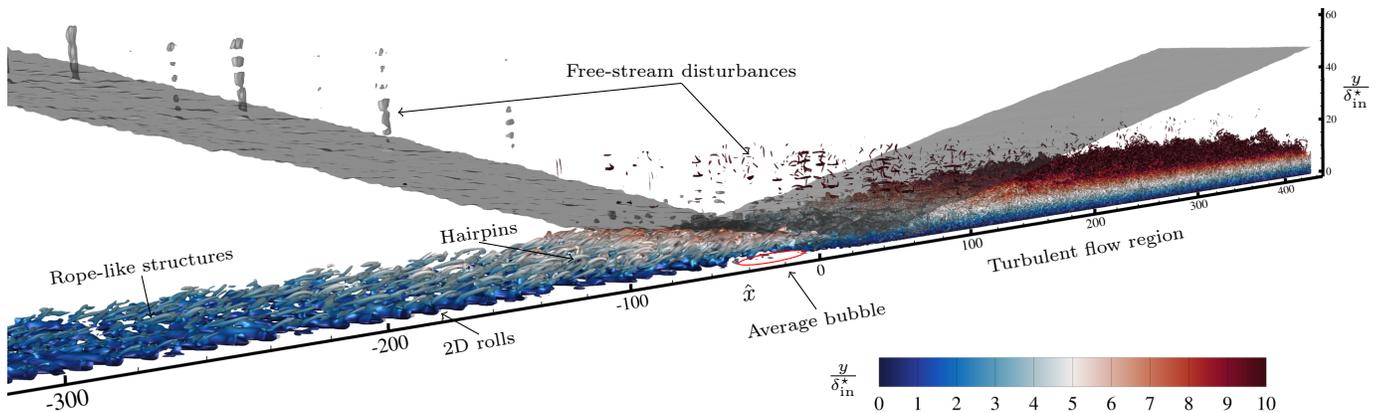
\begin{figure}
 \centering
   \begin{tikzpicture}
   \node[anchor=south west,inner sep=0] (image) at (0,0) {
   \includegraphics[width=\textwidth,trim={0 320 5 400},clip]{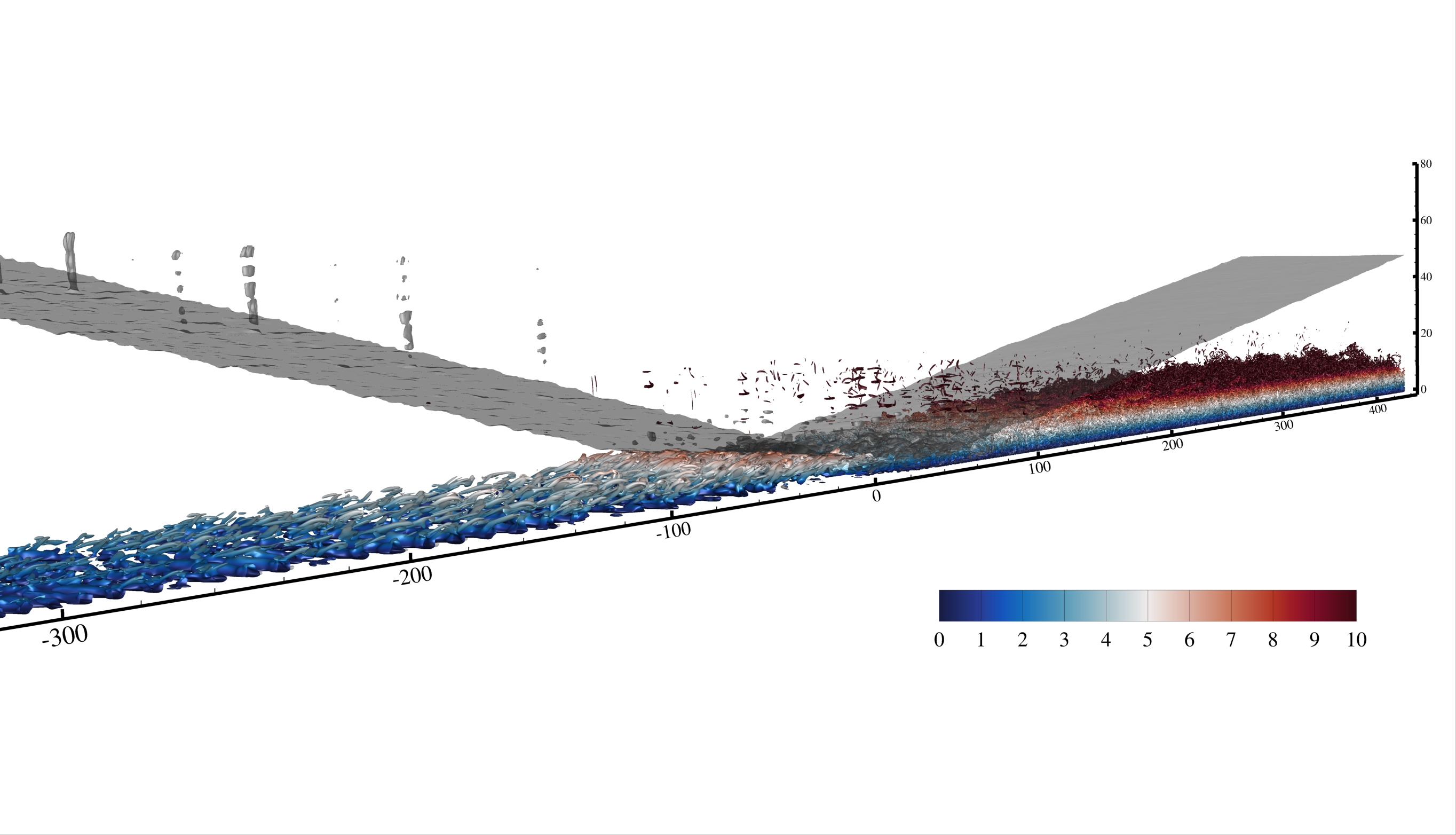}};
  \begin{scope}[x={(image.south east)},y={(image.north west)}]
     \node [align=center] at (0.50,0.85) {\scriptsize Free-stream disturbances};
     \draw [->] (0.50, 0.82) -- (0.29, 0.75);
     \draw [->] (0.50, 0.82) -- (0.55, 0.65);
     \node [align=center,rotate= 8] at (0.10,0.38) {\scriptsize Rope-like structures};
     \draw [->] (0.10, 0.36) -- (0.107, 0.26);
     \node [align=center,rotate= 8] at (0.35,0.46) {\scriptsize Hairpins};
     \draw [->] (0.35, 0.44) -- (0.42, 0.40);
     \node [align=center,rotate=10] at (0.35,0.20) {\scriptsize 2D rolls};
     \draw [->] (0.35, 0.22) -- (0.322, 0.270);
     \node [align=center,rotate=10] at (0.60,0.25) {\scriptsize Average bubble};
     \draw [->] (0.60, 0.28) -- (0.58, 0.37);
     \draw [red,rotate=10] (0.58,0.08) ellipse (0.45cm and 0.06cm);
     \node [align=center,rotate=10] at (0.80,0.42) {\scriptsize Turbulent flow region};
     \node [align=center,rotate=10] at (0.55,0.32) {$\hat{x}$};
     \node [align=center] at (1.00,0.80) {$\frac{y}{\delta^\star_\text{in}}$};
     \node [align=center] at (0.62,0.11) {$\frac{y}{\delta^\star_\text{in}}$};
  \end{scope}
  \end{tikzpicture}
 \caption{Isosurfaces of the $Q$-criterion coloured with the wall distance for an instantaneous snapshot. The red circle denotes the location of the average recirculation bubble.}\label{fig:qcrit}
\end{figure}

\begin{figure}
\centering
 \begin{tikzpicture}
   \node[anchor=south west,inner sep=0] (a) at (0,0) {\includegraphics[width=0.49\textwidth]{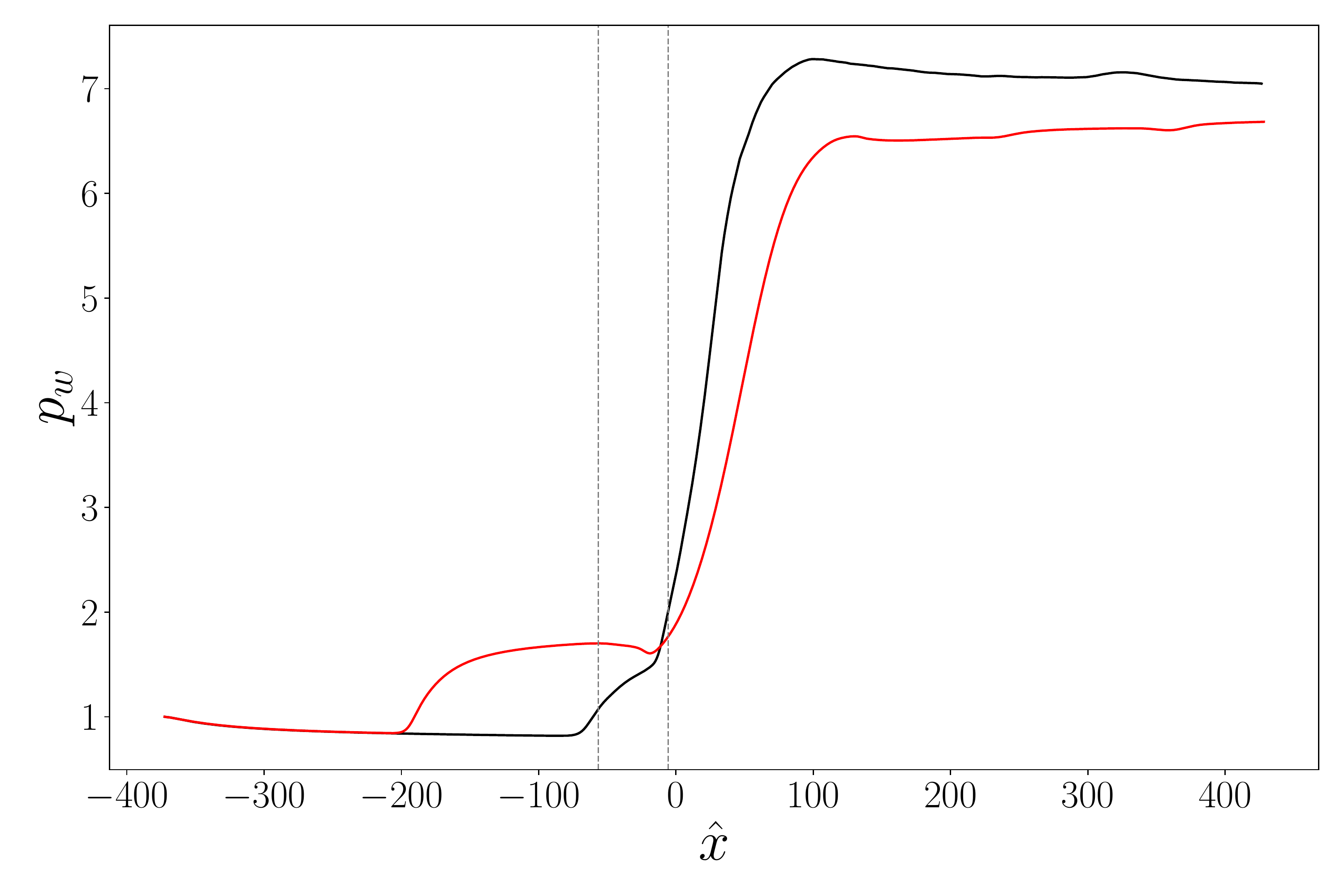}};
   \begin{scope}[x={(a.south east)},y={(a.north west)}] 
     \node [align=center] at (0.03,0.95) {(a)};
   \end{scope}
 \end{tikzpicture}
 \begin{tikzpicture}
   \node[anchor=south west,inner sep=0] (a) at (0,0) {\includegraphics[width=0.49\textwidth]{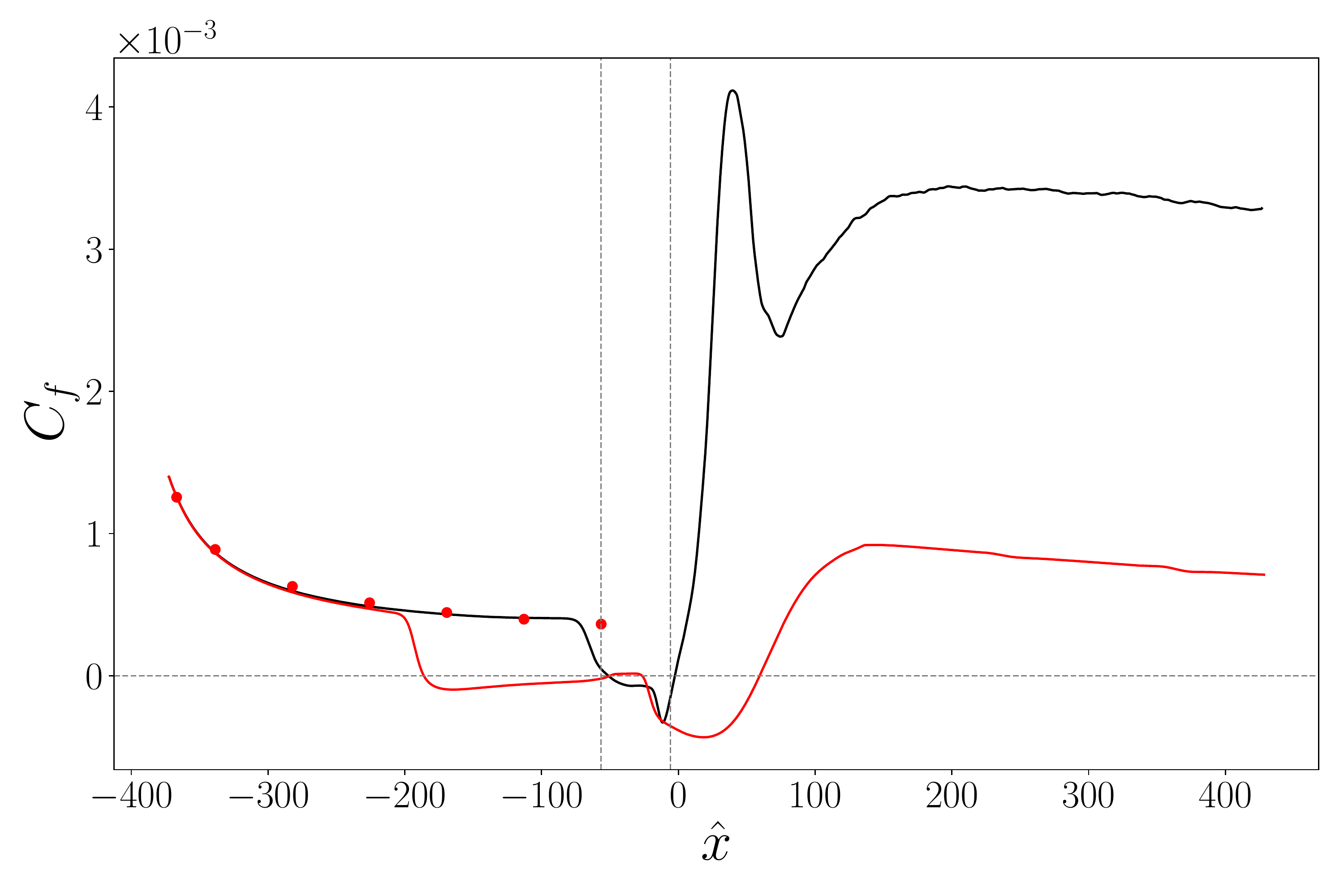}};
   \begin{scope}[x={(a.south east)},y={(a.north west)}]
     \node [align=center] at (0.03,0.95) {(b)};
       \end{scope}
 \end{tikzpicture}
 \caption{Evolution of normalized wall pressure $p_w$ (a) and skin friction coefficient (b). Red lines represent the evolution of the same quantities in the base flow computation; vertical lines stands for the beginning and the end of the separation zone in the 3D computation, whereas the horizontal line in panel (b) denotes the $C_f=0$ isoline. Symbols in panel (b) mark the results of the locally self-similar theory.}\label{fig:wall_quantities}
 \end{figure}

\begin{figure}
  \centering
  \begin{tikzpicture}
   \node[anchor=south west,inner sep=0] (a) at (0,0) {\includegraphics[width=0.49\textwidth]{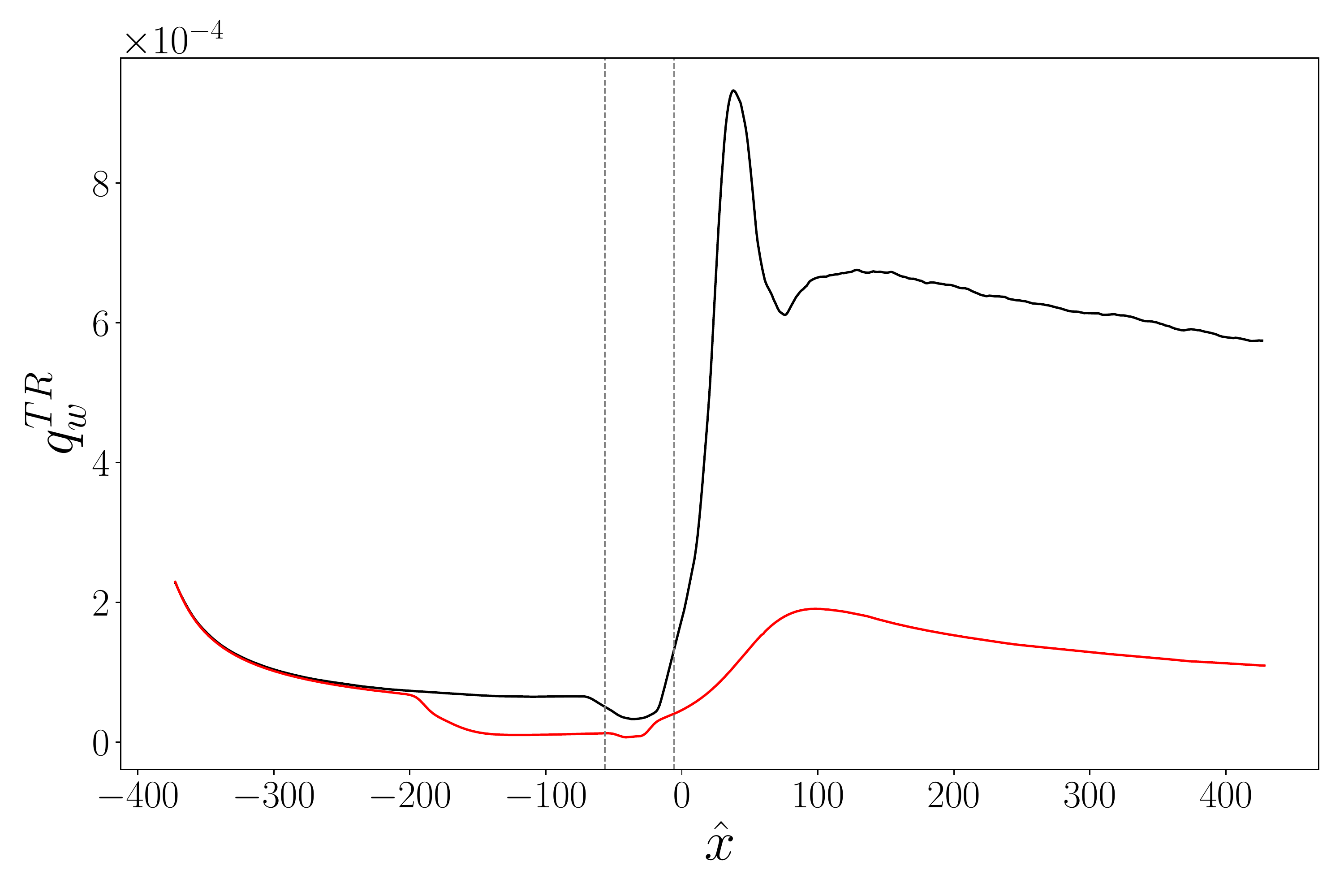}};
   \begin{scope}[x={(a.south east)},y={(a.north west)}]
     \node [align=center] at (0.03,0.95) {(a)};
       \end{scope}
 \end{tikzpicture}
  \begin{tikzpicture}
   \node[anchor=south west,inner sep=0] (a) at (0,0) {\includegraphics[width=0.49\textwidth]{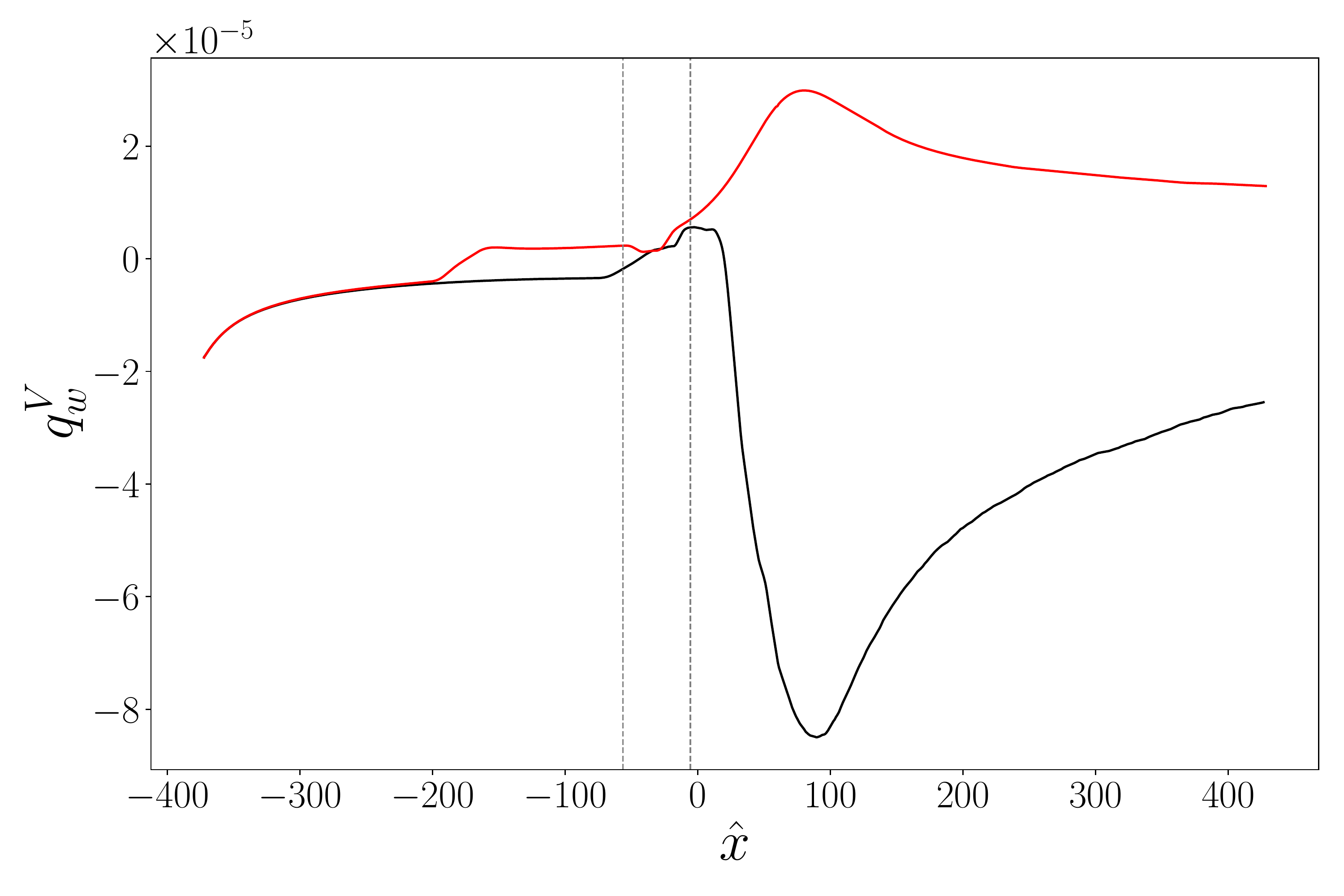}};
   \begin{scope}[x={(a.south east)},y={(a.north west)}]
     \node [align=center] at (0.03,0.95) {(b)};
       \end{scope}
 \end{tikzpicture}

 \begin{tikzpicture}
   \node[anchor=south west,inner sep=0] (a) at (0,0) {\includegraphics[width=0.49\textwidth]{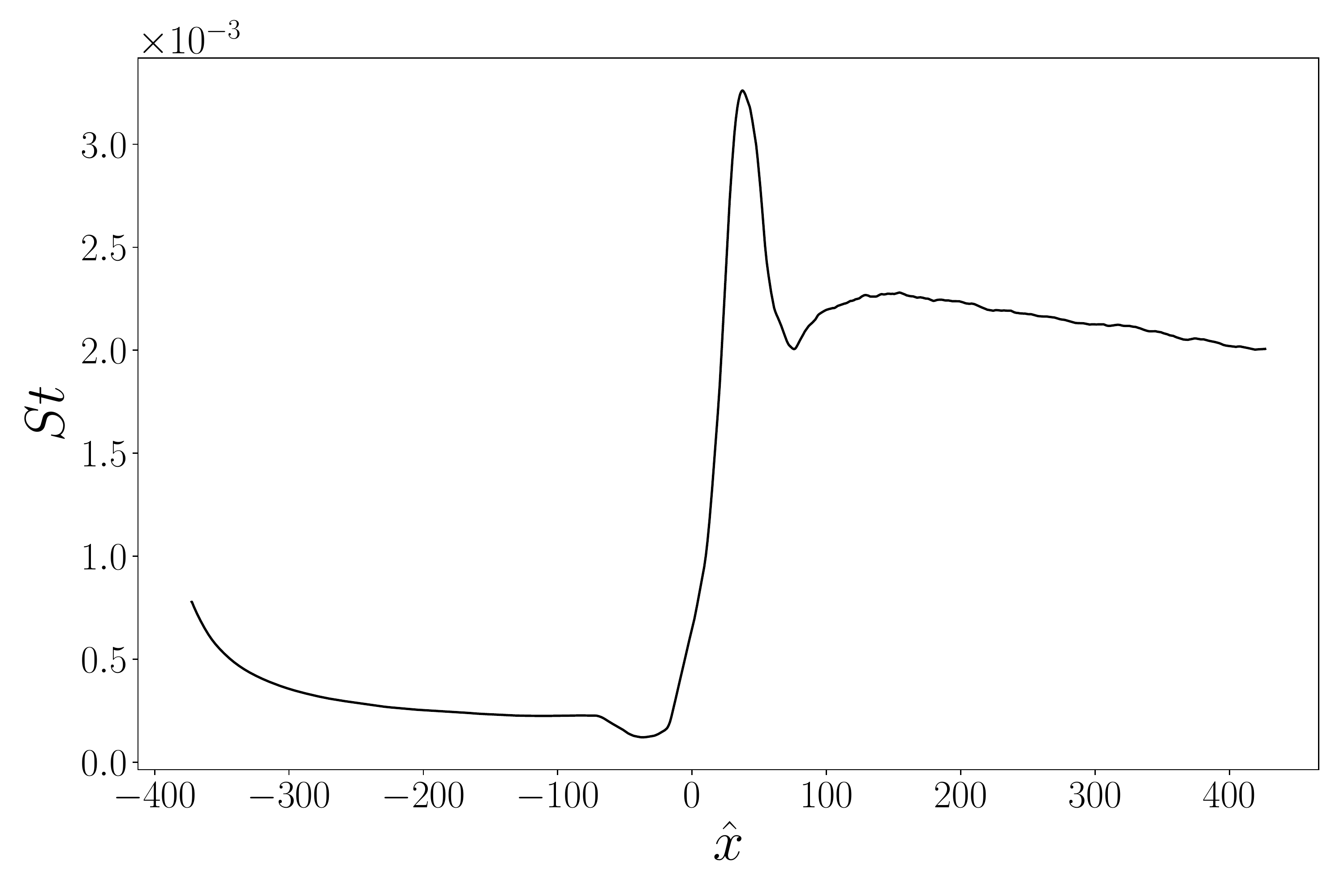}};
   \begin{scope}[x={(a.south east)},y={(a.north west)}]
     \node [align=center] at (0.03,0.95) {(c)};
       \end{scope}
 \end{tikzpicture}
   \begin{tikzpicture}
   \node[anchor=south west,inner sep=0] (a) at (0,0) {\includegraphics[width=0.49\textwidth]{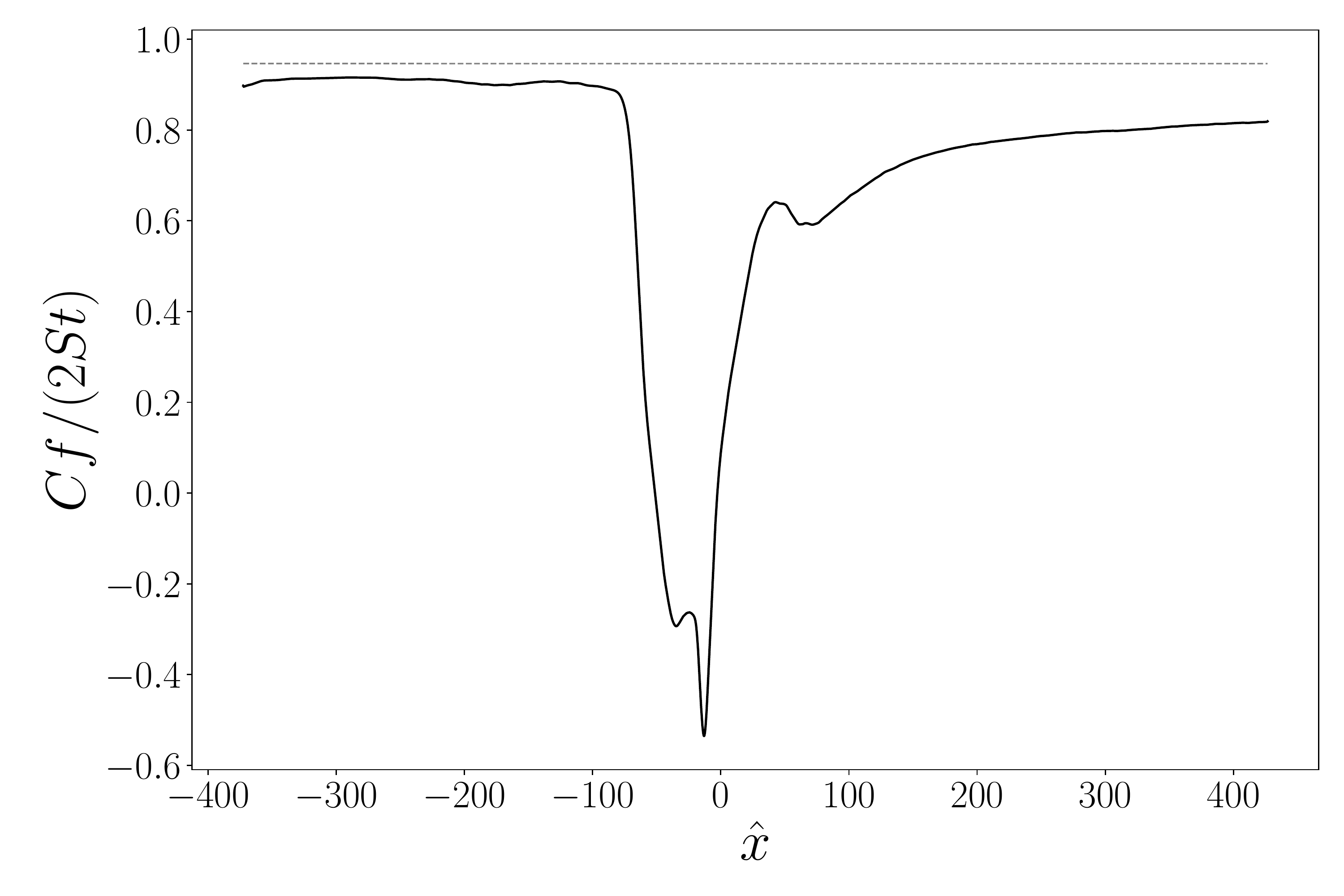}};
   \begin{scope}[x={(a.south east)},y={(a.north west)}]
     \node [align=center] at (0.03,0.95) {(d)};
       \end{scope}
 \end{tikzpicture}
 \caption{Top: distribution of the translational-rotational contribution (a) and the vibrational contribution (b) of the normalized heat flux at the wall. Red lines denote the base flow evolution of the corresponding variables. Bottom: evolution of the Stanton number (c) and the Reynolds analogy between $C_f$ and $St$ (d). The horizontal dashed line in panel (d) represents the $Pr^{2/3}$ line.}
 \label{fig:wall_heat}
 \end{figure}
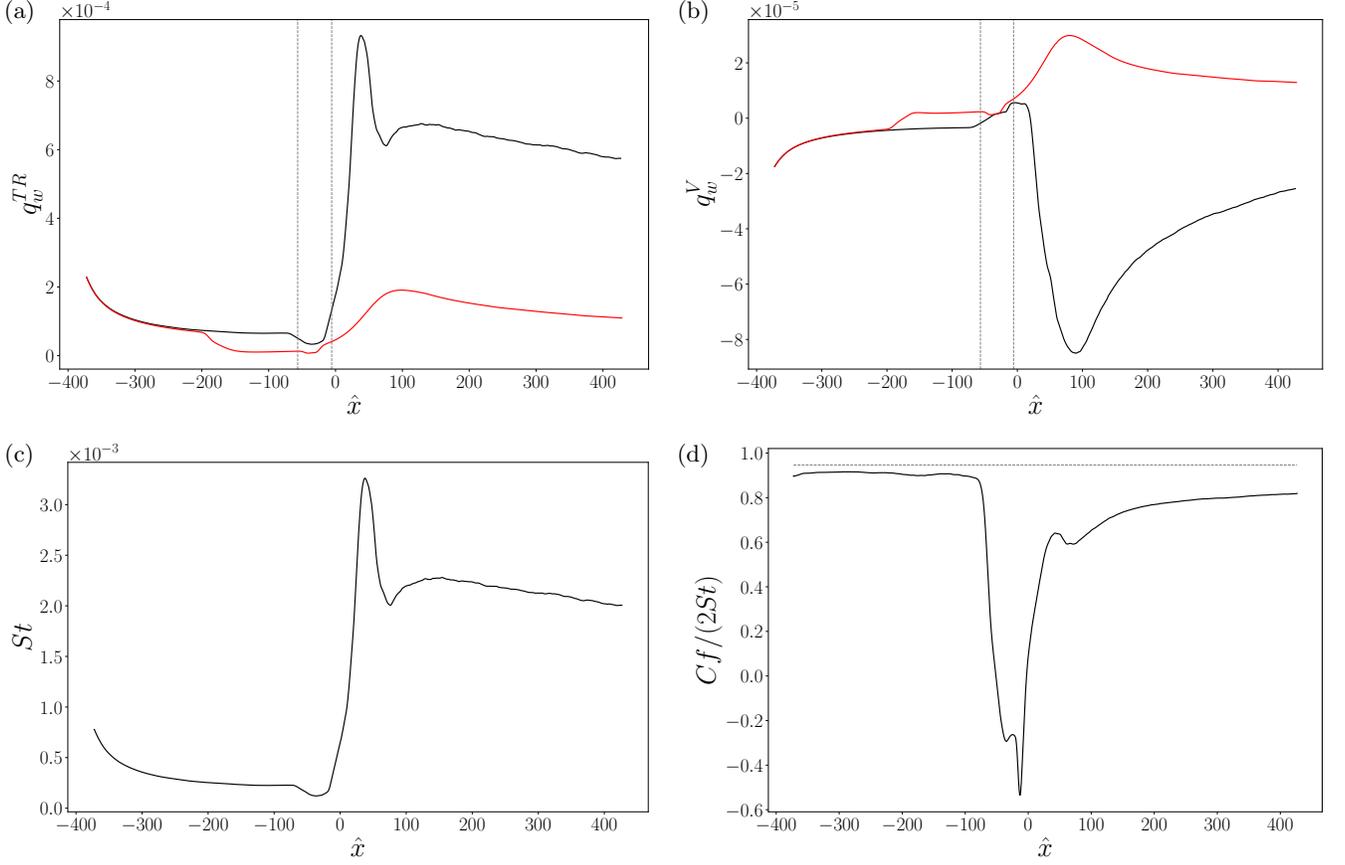

Figure~\ref{fig:qcrit} shows a 3D view of the flow structures for an instantaneous snapshot. Traces of the inflow perturbation are visible in the free-stream after the impinging shock. The disturbances interact with the incoming laminar boundary layer and destabilize it, leading to the generation of coherent structures upstream of the interaction region. 2D near-wall rolls are visible beneath rope-like structures; the latter evolve giving rise to hairpins ahead of the recirculation region that break down at the interaction. Despite the rich variety of flow structures populating the boundary layer, the analysis of wall quantities reveal that it essentially follows a laminar-like evolution almost up to the average separation point, as will be shown afterwards.
We start the analysis with the streamwise distributions of the normalized wall pressure $p_w=\overline{p}/p_\infty$ and the skin friction coefficient $C_f =\frac{2 \overline{\tau}_w}{\rho_\infty u_\infty^2}$, reported in figure~\ref{fig:wall_quantities}(a) and figure~\ref{fig:wall_quantities}(b), respectively. The vertical lines represent the beginning and the end of the separation zone, computed by intersecting the $C_f{=}0$ isoline with the evolution of the statistically-averaged skin friction coefficient. On the same figures, the laminar base flow solution is reported with red lines. In the laminar case, the pressure jump across the interaction zone is $\approx 6$, whereas it reaches $\approx 7$ in the three-dimensional transitional case. As already observed by Sandham \textit{et al.}\cite{sandham2014transitional} for cold SWBLI at $M_\infty=6$, there is a significant reduction of the bubble length with respect to the laminar case, the separation region $L_\text{sep}$ being equal to $48 \delta^\star_\text{in}$. The behavior may be attributed to the freestream perturbation causing a reduction of the shock strength, but also to the different nature of the incoming boundary layer which causes increased mixing due to velocity fluctuations \cite[see e.g.,][]{quadros2018numerical}. Of particular interest is the different trend of the skin friction coefficient with respect to the base flow, even before the shock impingement. In the region $-200 < \hat{x} < -80$, the $C_f$ follows the laminar distributions, deviating from the base flow profile for which the recirculation zone begins at $\hat{x} \approx -200$. Of note, the evolution in the upstream zone is in perfect agreement with the results estimated by the locally self-similar theory (represented by the red dots in figure~\ref{fig:wall_quantities}b). In the fully-turbulent region downstream of the interaction, $C_f$ values are approximately four times larger than those registered in the laminar case, similarly to \cite{sandham2014transitional}. On the other hand, its evolution in the interaction region is rather different. The increase observed after reaching the global minimum, at $\hat{x} \approx 0$, is attributed to the reattachment. As the $C_f$ experiences a ramp-like increase, at $\hat{x} \approx 40$, the incident shock penetrates the boundary layer and reaches the wall, causing a sudden increase of wall friction and heating.\\
The evolution of the two contributions of the normalized wall heat flux, defined as
\begin{equation}
   q_{w}^{\text{TR}} =\frac{\overline{\lambda_{TR} \partial T/\partial y}}{\rho_\infty u_\infty^3}, \qquad
   q_{w}^{\text{V}}  =\frac{\overline{\lambda_V \partial T_V/\partial y}}{\rho_\infty u_\infty^3}.
\end{equation}
are reported in figure~\ref{fig:wall_heat}(a)-(b). The results are also compared with the corresponding laminar evolutions of the same quantities. The rototranslational heat flux, $q_{w}^{\text{TR}}$, follows essentially the $C_f$ distribution, with a minimum in the separation zone, a peak of almost $10^{-4}$ and a significant overheating in the fully turbulent region with respect to the case without perturbation. Of particular interest is the trend of the vibrational heat flux. As already observed in the flat-plate boundary layer configuration investigated by Passiatore \textit{et al.} \cite{passiatore2022thermochemical}, the latter is one order of magnitude smaller with respect to the translational-rotational one. However, thermal non-equilibrium before the interaction is so strong that the wall heats the flow from a vibrational energy standpoint (i.e., $q_{w}^{\text{V}}$ is negative and the profiles of $T_\text{V}$ are monotonic, as it will be shown in section~\ref{sec:tcne}). For the case without perturbation, the vibrational heat flux switches to positive values in the recirculation bubble, then increases in the reattachment region and relaxes to the post shock conditions while keeping positive values.
On the other hand, when perturbations are added, $q_{w}^{\text{V}}$ keeps negative values almost everywhere, except in the small separation bubble. From the reattachment region onwards, its streamwise evolution is opposed to the one obtained in the laminar regime. The global increase of temperature due to the shock impingement causes strong aerodynamic heating, transferring a considerable amount of kinetic energy into internal energy, which is distributed across all the energetic modes. In the same figure, we also report the evolution of the total Stanton number, defined as:
\begin{equation}
   St  =\frac{q_w}{\rho_\infty u_\infty (h_{aw}-h_\infty)},
\end{equation}
where $q_w = q_{w}^{\text{TR}} + q_{w}^{\text{V}}$ and $h_{aw} = h_\infty + \frac{1}{2}r u_\infty^2$, with $r=0.9$. Note that the recovery factor has the same value previously used for high-enthalpy configurations \cite{duan2011direct4,passiatore2022thermochemical}. The evolution of the Stanton number is in accordance with the trend of the translational contribution of the wall heat flux. The orders of magnitude after the breakdown are in accordance with results for calorically-perfect gases \cite{sandham2014transitional,volpiani2018effects} and also with high-enthalpy thermally-equilibrated boundary layers \cite{duan2011direct4}. Therefore, the small vibrational heat flux contribution does not affect the Stanton number distribution even in the present strong thermochemical non-equilibrium conditions. In figure~\ref{fig:wall_heat}(d) we assess the validity of the Reynolds analogy relating $C_f$ and $St$. The ratio $C_f/(2St)$ is expected to vary as $Pr^{2/3}$ (with $Pr = \mu c_p /\lambda$), which amounts to $\approx 0.85$ for classical values of the Prandtl number. In the present case, the mean Prandtl number reaches $\approx 0.9$ in the near wall region and displays a nearly constant streamwise evolution (as shown in figure~\ref{fig:wall_heat}d), with variations in the recirculation region less than $1\%$ with respect to the turbulent zone. As previously observed for other SWBLI configurations in the literature, the relation performs poorly in the interaction region and seems to slowly relax back to the expected trend afterwards. It is reasonable to suppose that $C_f/(2St)$ tends asymptotically to $Pr^{2/3}$, albeit longer computational domains would be needed to confirm its validity \cite{roy2006review}.\\
%

%
\subsection{Flow field near the interaction}
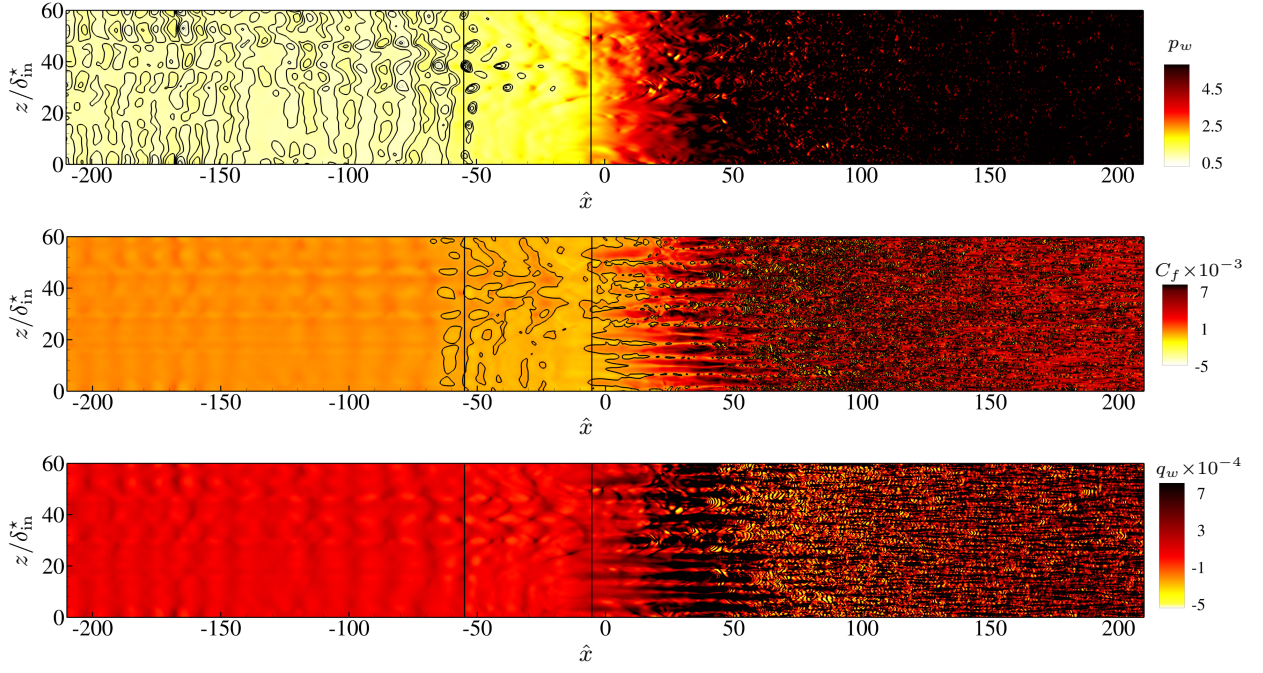
\begin{figure}
 \centering
  \begin{tikzpicture}
   \node[anchor=south west,inner sep=0] (image) at (0,0) {
   \includegraphics[width=0.9\columnwidth,trim={10 10 10 200},clip]{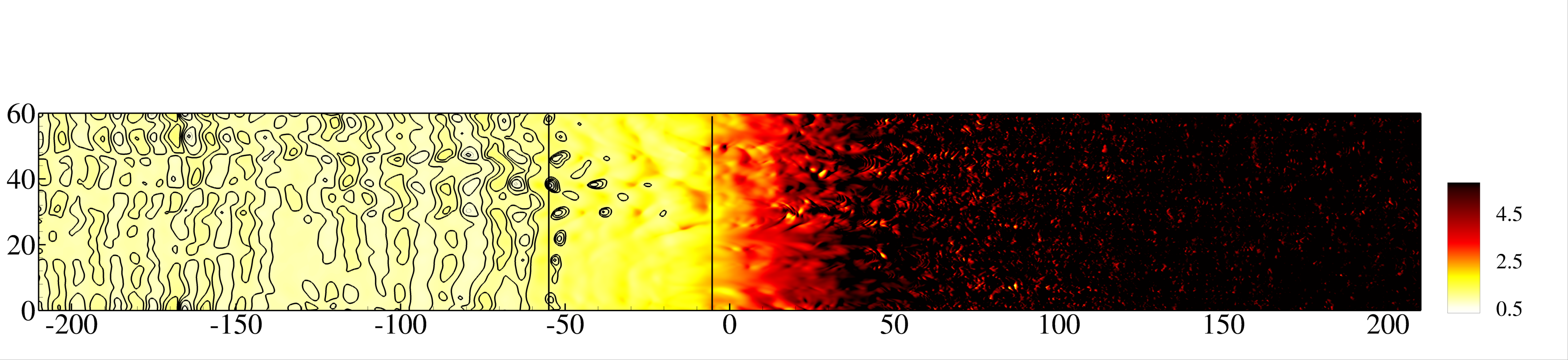}};
   \begin{scope}[x={(image.south east)},y={(image.north west)}]
    \node [anchor=center] at (0.45,0.0) {\small $\hat{x}$};
    \node [anchor=center] at (0.94,0.74) {\scriptsize $p_w$};
    \node [anchor=center,rotate=90] at (-0.015,0.54) {\small $z/\delta^\star_\text{in}$};
   \end{scope}
   \end{tikzpicture}
   \begin{tikzpicture}
   \node[anchor=south west,inner sep=0] (image) at (0,0) {
   \includegraphics[width=0.9\columnwidth,trim={10 10 10 200},clip]{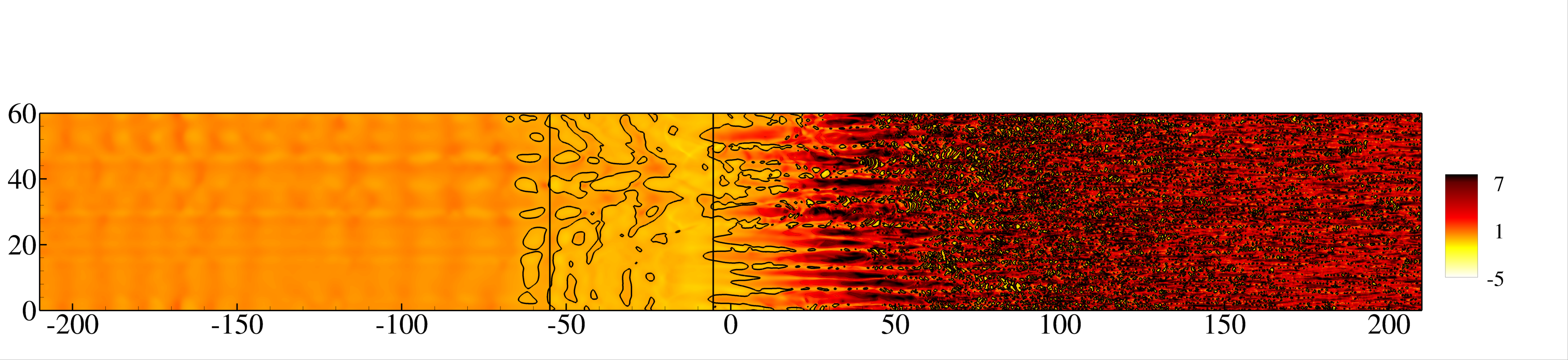}};
   \begin{scope}[x={(image.south east)},y={(image.north west)}]
    \node [anchor=center] at (0.955,0.75) {\scriptsize $C_f {\times} 10^{-3}$};
    \node [anchor=center] at (0.45,0.0) {\small $\hat{x}$};
    \node [anchor=center,rotate=90] at (-0.015,0.54) {\small $z/\delta^\star_\text{in}$};
   \end{scope}
   \end{tikzpicture}
  \begin{tikzpicture}
   \node[anchor=south west,inner sep=0] (image) at (0,0) {
   \includegraphics[width=0.9\columnwidth,trim={10 10 10 200},clip]{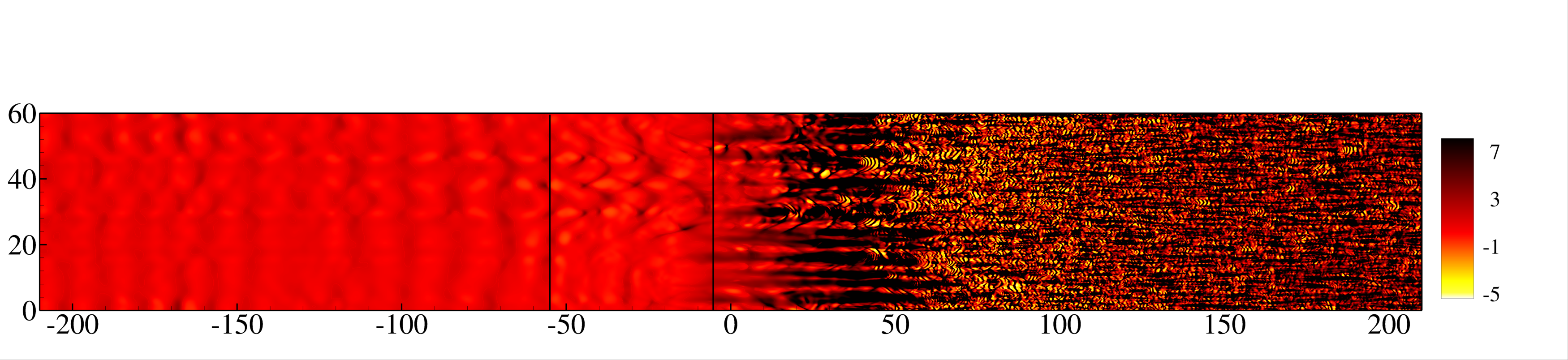}};
   \begin{scope}[x={(image.south east)},y={(image.north west)}]
    \node [anchor=center] at (0.45,0.0) {\small $\hat{x}$};
    \node [anchor=center] at (0.955,0.89) {\scriptsize $q_w {\times} 10^{-4}$};
    \node [anchor=center,rotate=90] at (-0.015,0.54) {\small $z/\delta^\star_\text{in}$};
   \end{scope}
   \end{tikzpicture}
 \caption{Top: isocontours of normalized wall pressure $p_w$ and pressure isolines; middle: isocontours of the instantaneous skin friction coefficient $C_f$, along with $C_f=0$ isolines; bottom: isocontours of instantaneous normalized total wall heat flux, $q_w$. The vertical black lines in each subfigure mark the beginning and the end of the separation bubble.}
\label{fig:wall_inst}
\end{figure}

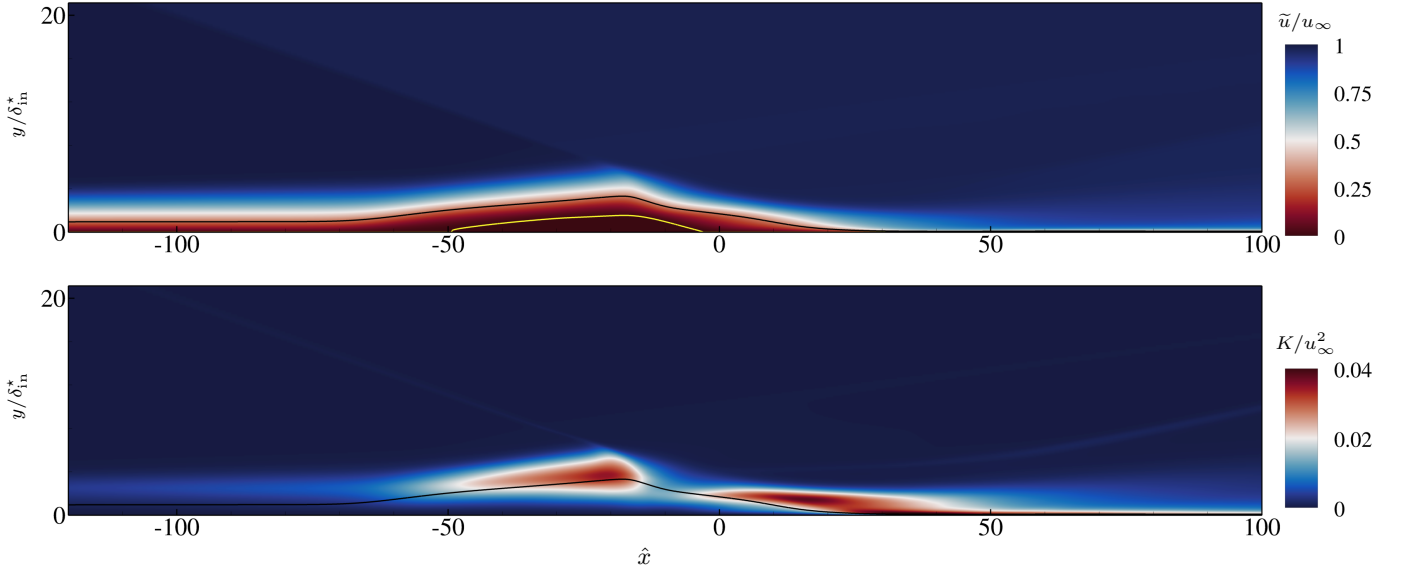
\begin{figure}
 \centering
 \begin{tikzpicture}
   \node[anchor=south west,inner sep=0] (a) at (0,0) {\includegraphics[width=1\textwidth, trim={5 5 5 35}, clip]{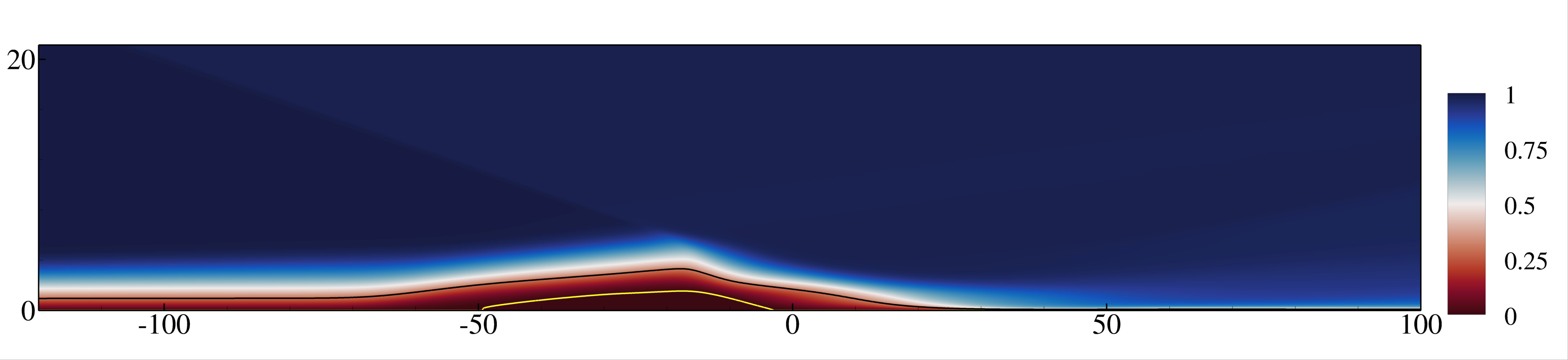}};
   \begin{scope}[x={(a.south east)},y={(a.north west)}]
     \node [align=center] at (0.94,0.85) {\scriptsize $\widetilde{u}/u_\infty$};
     \node [align=center,rotate=90] at (-0.015,0.54) {\scriptsize $y/\delta^\star_\text{in}$};
       \end{scope}
 \end{tikzpicture}\\[-0.2cm]
 \begin{tikzpicture}
   \node[anchor=south west,inner sep=0] (a) at (0,0) {\includegraphics[width=1\textwidth, trim={5 5 5 35}, clip]{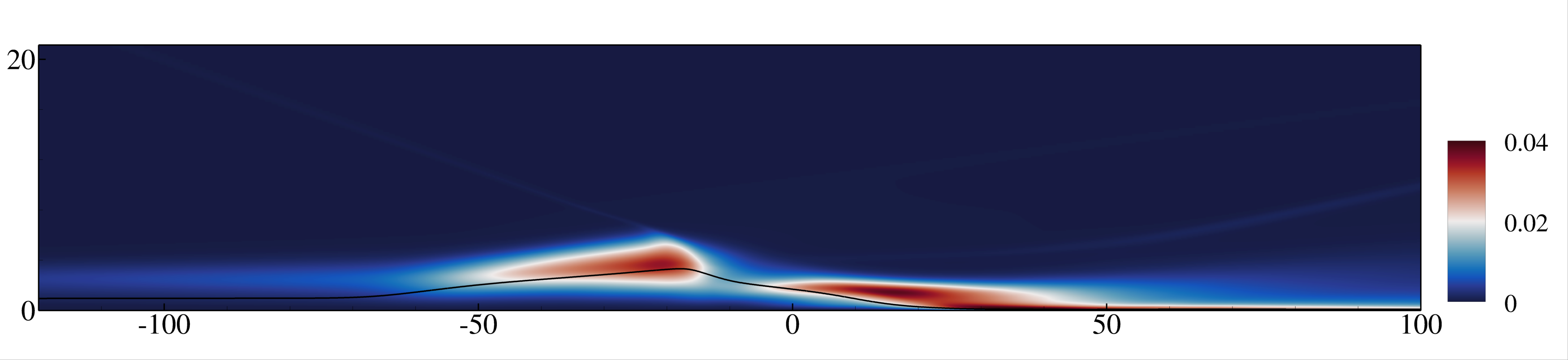}};
   \begin{scope}[x={(a.south east)},y={(a.north west)}]
     \node [align=center] at (0.45,0.0) {\small $\hat{x}$};
     \node [align=center] at (0.94,0.72) {\scriptsize $K/u_\infty^2$};
     \node [align=center,rotate=90] at (-0.015,0.54) {\scriptsize $y/\delta^\star_\text{in}$};
       \end{scope}
 \end{tikzpicture}
 \caption{Visualization of normalized mean streamwise velocity (top panel) and normalized fluctuating kinetic energy (bottom panel).}\label{fig:stats_uK}
 \end{figure}

Figure~\ref{fig:wall_inst} shows instantaneous views of the wall pressure (top), skin friction coefficient (middle) and total wall heat flux (bottom) in the neighbourhood of the shock impingement region. The average reversed flow region is also marked with isolines. In the three figures, the footprint of the freestream perturbations is visible in the form of wave packets, corresponding to pressure isolines smaller than $p_\infty$ in the top figure. These wave packets remain coherent up to the separation point and start to destabilize in the bubble at $z/\delta^\star_\text{in} \approx 40$, albeit they do not completely break. After the reattachment point, breakdown to turbulence begins and $C_f$ rapidly increases towards its post-shock value. For the conditions under investigation, the breakdown is particularly sharp as witnessed by the large overshoot of $C_f$. This is related to the emergence of energetic structures at $\hat{x} \approx 40$ visible in the middle and bottom panels of figure~\ref{fig:wall_inst}. Further downstream, the skin friction reaches a local minimum at $\hat{x} \approx 50$ and the flow relaxes to equilibrium turbulence starting from $\hat{x} \approx 190$. Despite the statistically-averaged separation region is small, relatively large structures of negative skin friction coefficient (middle panel of figure~\ref{fig:wall_inst}) are visible, showing that instantaneous separation exists well upstream and downstream of the average bubble.\\
%
%
The topology of the separation bubble along an $xy$ plane cut is highlighted in the mean streamwise velocity plot, displayed in the top panel of figure~\ref{fig:stats_uK}. In the same figure, the yellow and black lines represent the $\widetilde{u}=0$ and the sonic line, respectively. The recirculation bubble is found to be very flat and close to the sonic line. The width of the bubble depends on many parameters, such as the shock angle $\beta$, the amount of wall cooling and the Reynolds number at the impinging station; further investigations will be performed in the future to assess the sensitivity of the bubble size to them. Even if the flow regime in the interaction region is still transitional, another variable worth investigating is the fluctuating kinetic energy $K=\widetilde{u''_i u''_i}/2$. Inspection of figure~\ref{fig:stats_uK}(bottom) illustrates that there are two peaks of kinetic energy emerging in the proximity of the interaction region. The first peak of $K$ corresponds to the shear layer at the interface with the recirculation bubble, as already pointed out, for instance, by Volpiani \textit{et al.}\cite{volpiani2020effects}. The second peak is shifted towards the wall and corresponds to the transitional structures observed in figure~\ref{fig:wall_inst}; it is therefore peculiar to the specific case dynamics. It is also possible to note the increase of $K$ in the near wall region at $\hat{x} \approx 40$, due to the impact of the shock foot on the wall.
 %
\subsubsection{Instability mechanisms}
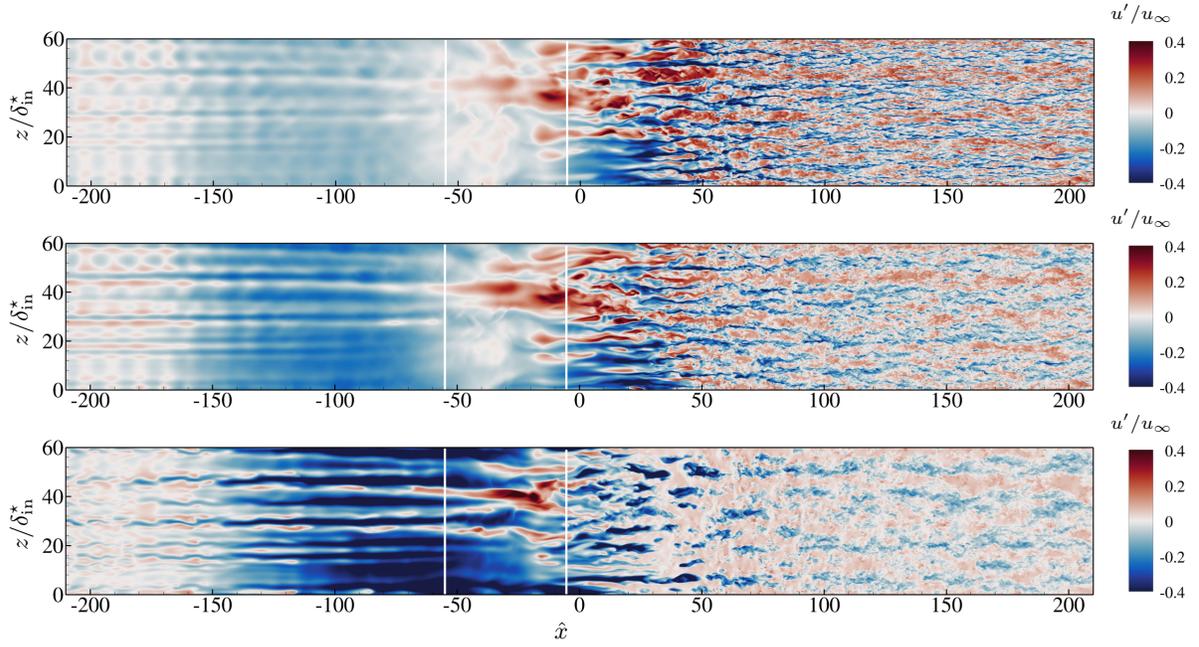
\begin{figure}
 \centering
    \begin{tikzpicture}
   \node[anchor=south west,inner sep=0] (image) at (0,0) {
   \includegraphics[width=0.86\columnwidth,trim={5 3 7 150},clip]{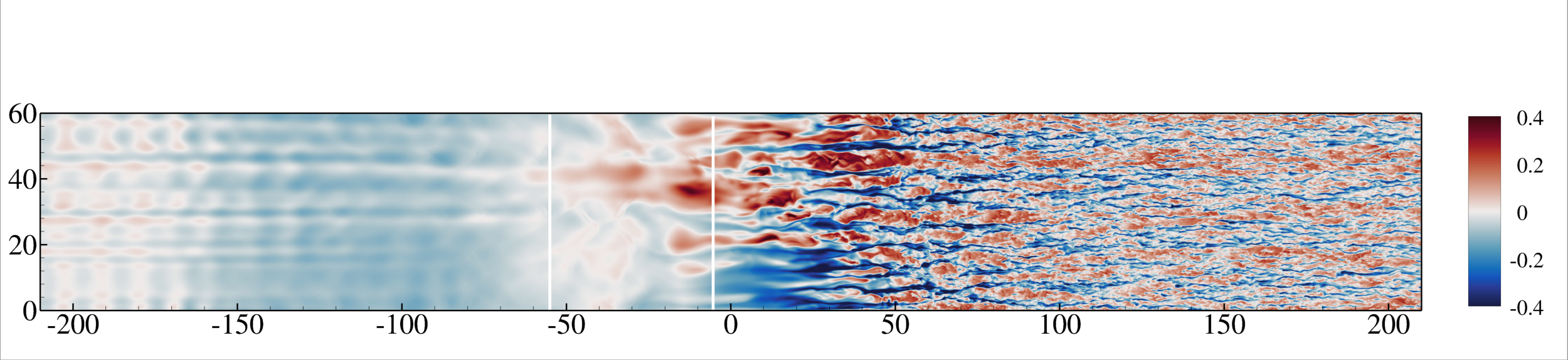}};
   \begin{scope}[x={(image.south east)},y={(image.north west)}]
    \node [align=center,rotate=90] at (-0.015,0.54) {\small $z/\delta^\star_\text{in}$};
    \node [align=center] at (0.95,1.05) {\scriptsize $u'/u_\infty$};
   \end{scope}
  \end{tikzpicture}\\[-0.3cm]
   \centering
      \begin{tikzpicture}
   \node[anchor=south west,inner sep=0] (image) at (0,0) {
   \includegraphics[width=0.86\columnwidth,trim={5 3 7 150},clip]{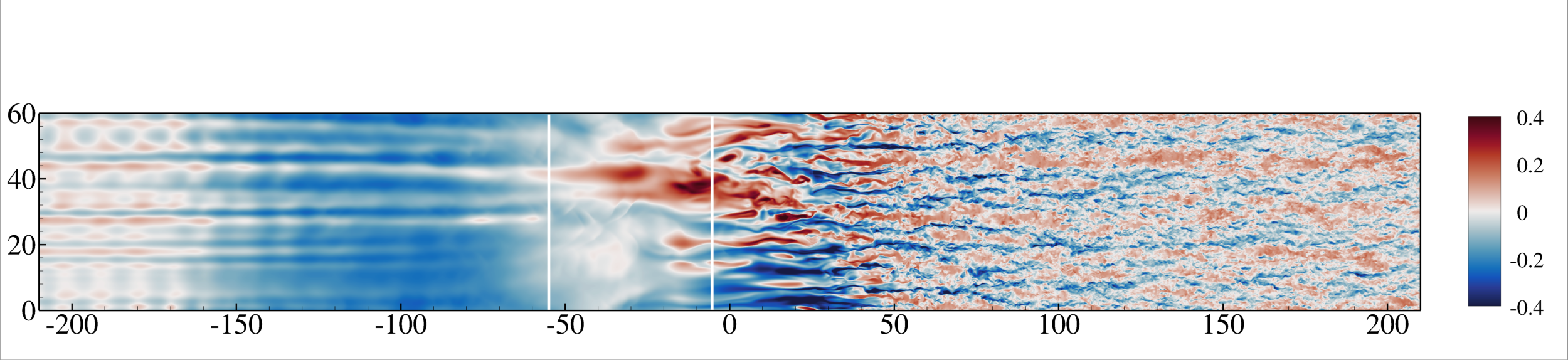}};
   \begin{scope}[x={(image.south east)},y={(image.north west)}]
    \node [align=center,rotate=90] at (-0.015,0.54) {\small $z/\delta^\star_\text{in}$};
    \node [align=center] at (0.95,1.05) {\scriptsize $u'/u_\infty$};
   \end{scope}
  \end{tikzpicture}\\[-0.3cm]
   \centering
      \begin{tikzpicture}
   \node[anchor=south west,inner sep=0] (image) at (0,0) {
   \includegraphics[width=0.86\columnwidth,trim={5 3 7 150},clip]{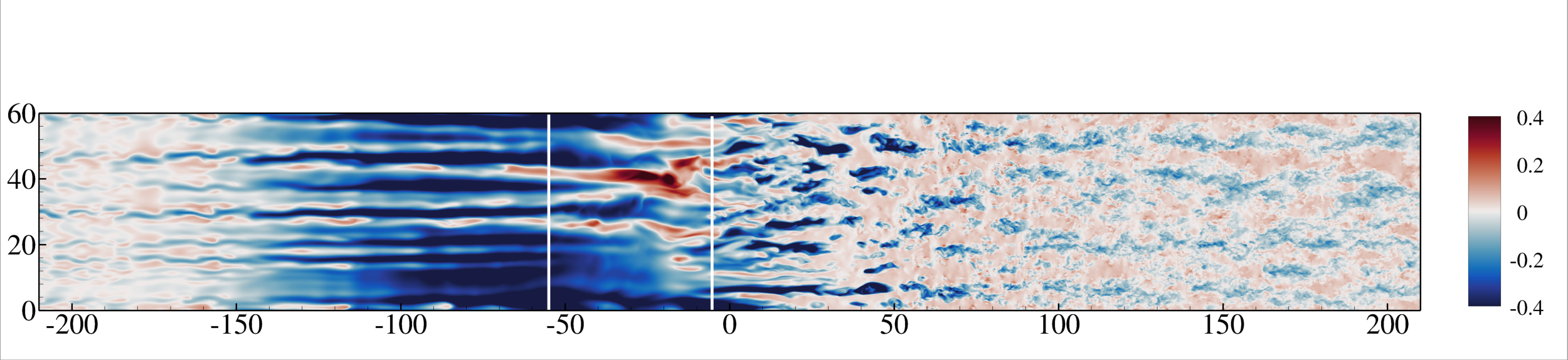}};
   \begin{scope}[x={(image.south east)},y={(image.north west)}]
     \node [align=center] at (0.45,0.0) {\small $\hat{x}$};
    \node [align=center,rotate=90] at (-0.015,0.54) {\small $z/\delta^\star_\text{in}$};
    \node [align=center] at (0.95,1.05) {\scriptsize $u'/u_\infty$};
   \end{scope}
  \end{tikzpicture}
 \caption{Isocontours of streamwise velocity perturbation in $xz$-slices, extracted at $y/\delta^\star_\text{in} = 0.5$ (top), $y/\delta^\star_\text{in} = 1$ (middle) and  $y/\delta^\star_\text{in} = 3$ (bottom). White vertical lines denote the beginning and the end of the recirculation zone.}
\label{fig:streaks}
\end{figure}

\begin{figure}
\centering
 \begin{tikzpicture}
   \node[anchor=south west,inner sep=0] (a) at (0,0) {\includegraphics[width=0.8\textwidth, trim={10 0 10 0}, clip]{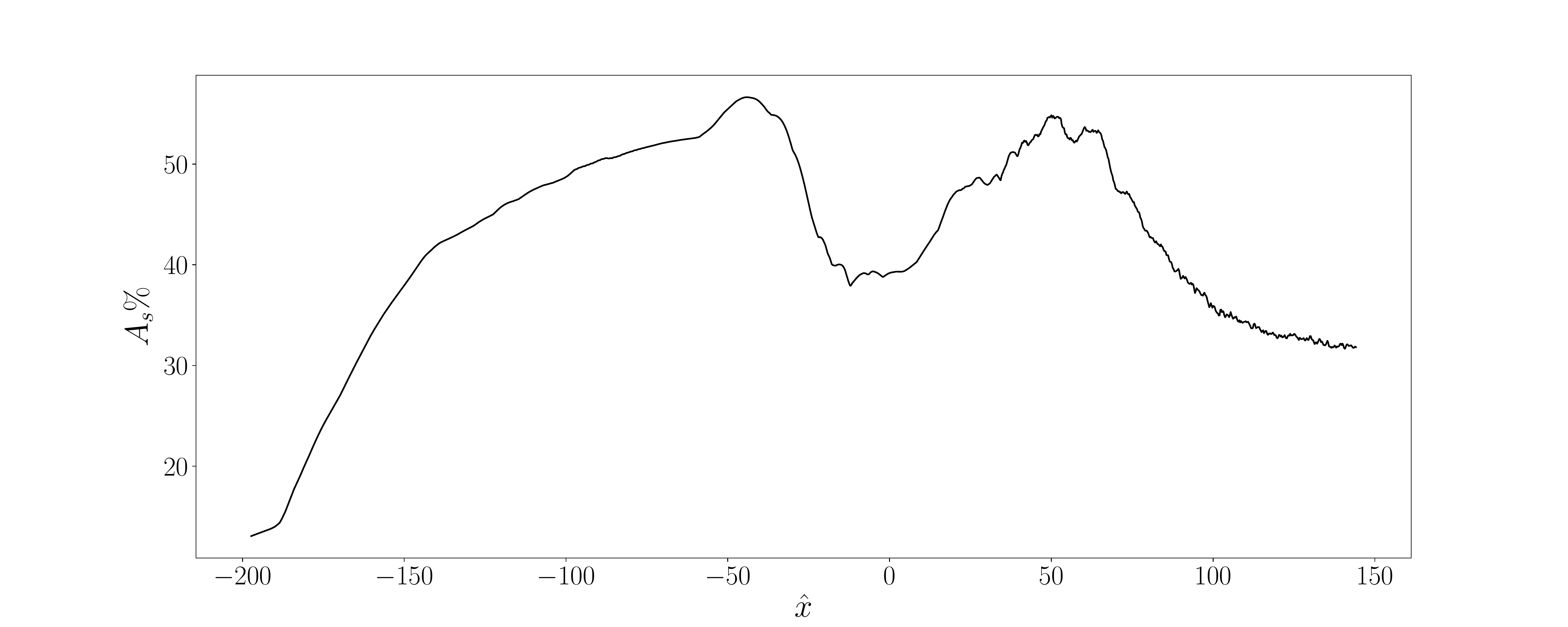}};
   \begin{scope}[x={(a.south east)},y={(a.north west)}] 
      \end{scope}
 \end{tikzpicture}
 \caption{Streamwise evolution of streaks amplitude, computed as in equation~\eqref{eq:ampl_streaks}.}
\label{fig:ampl_streaks}
 \end{figure}

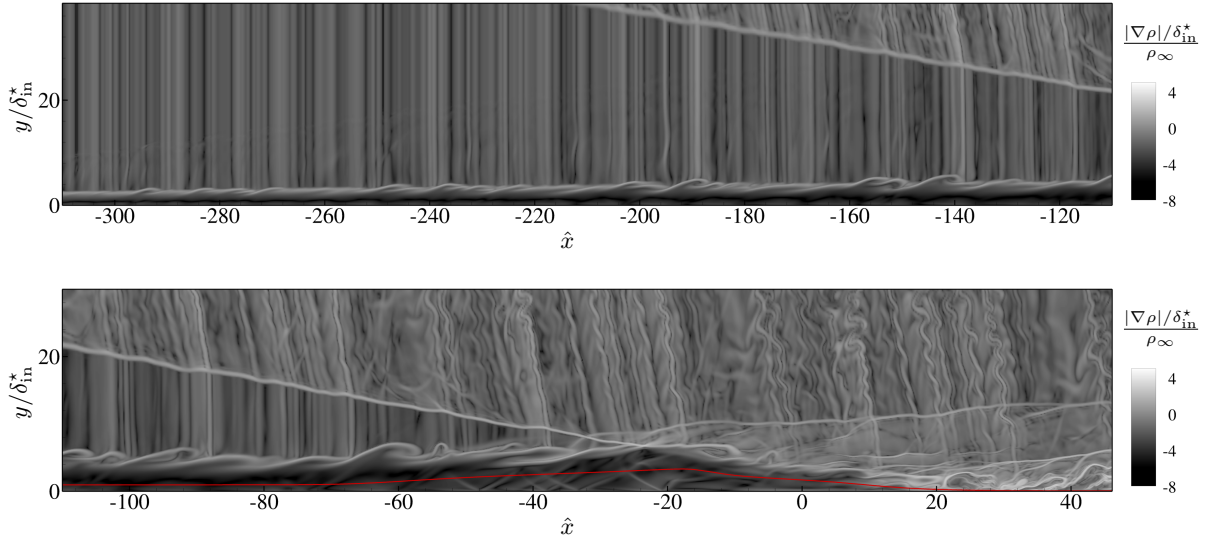
\begin{figure}
 \centering
    \begin{tikzpicture}
   \node[anchor=south west,inner sep=0] (image) at (0,0) {
   \includegraphics[width=0.88\columnwidth,trim={0 10 10 10},clip]{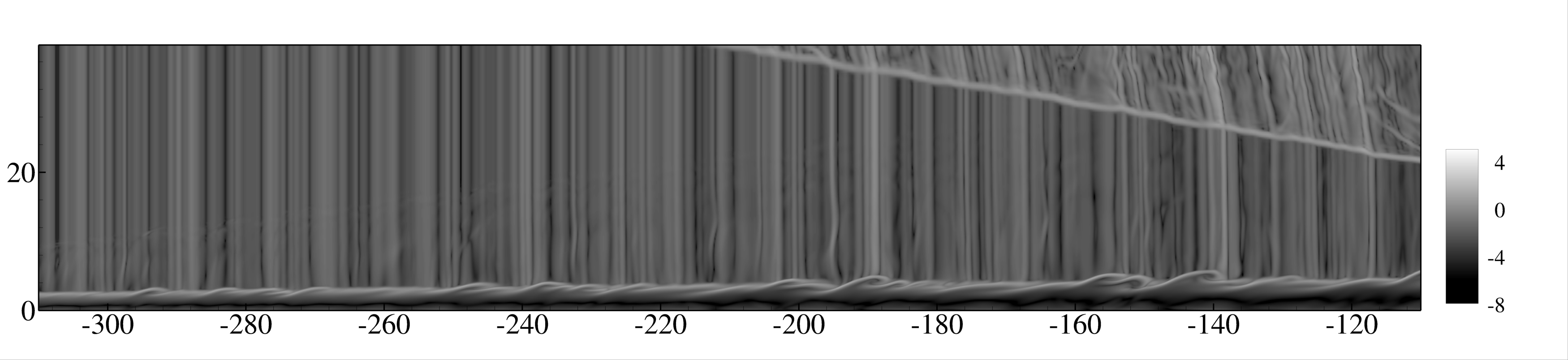}};
   \begin{scope}[x={(image.south east)},y={(image.north west)}]
       \node [align=center] at (0.95,0.74) {\small $\frac{|\nabla \rho|/\delta^\star_\text{in}}{\rho_\infty}$};
   \node [align=center, rotate=90] at (-0.01,0.5) {$y/\delta^\star_\text{in}$};
       \node [align=center] at (0.45,0) {$\hat{x}$};
   \end{scope}
  \end{tikzpicture}
 \centering
   \begin{tikzpicture}
   \node[anchor=south west,inner sep=0] (image) at (0,0) {
   \includegraphics[width=0.88\columnwidth,trim={0 10 10 10},clip]{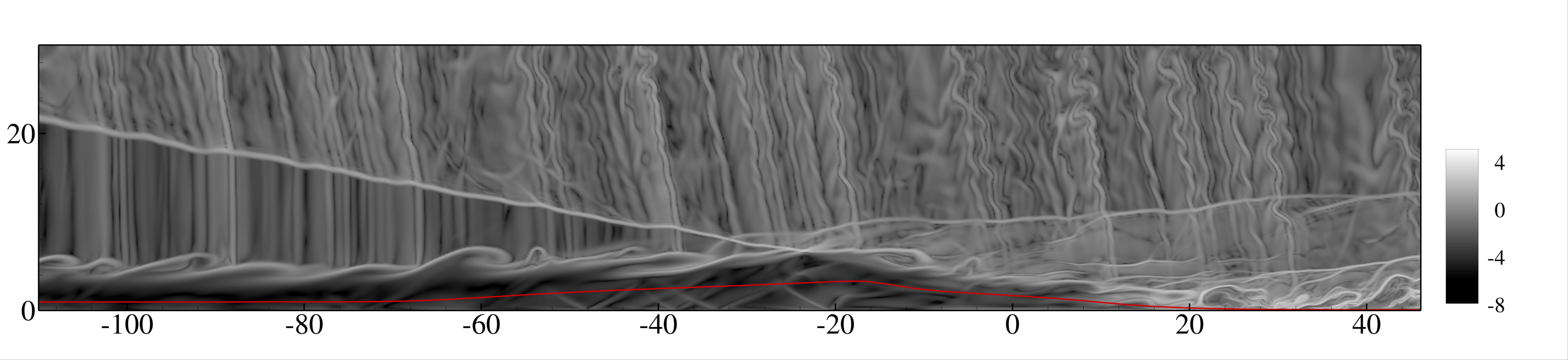}};
   \begin{scope}[x={(image.south east)},y={(image.north west)}]
       \node [align=center] at (0.95,0.74) {\small $\frac{|\nabla \rho|/\delta^\star_\text{in}}{\rho_\infty}$};
   \node [align=center, rotate=90] at (-0.01,0.5) {$y/\delta^\star_\text{in}$};
    \node [align=center] at (0.45,0) {$\hat{x}$};
   \end{scope}
  \end{tikzpicture}
 \caption{Isocontours of normalized density gradient magnitude in an instantaneous $xy$ slice of the computational domain.}
\label{fig:inst_Mack}
\end{figure}

\begin{figure}
 \centering
   \begin{tikzpicture}
   \node[anchor=south west,inner sep=0] (image) at (0,0) {
   \includegraphics[width=0.6\columnwidth,trim={5 5 5 5 },clip]{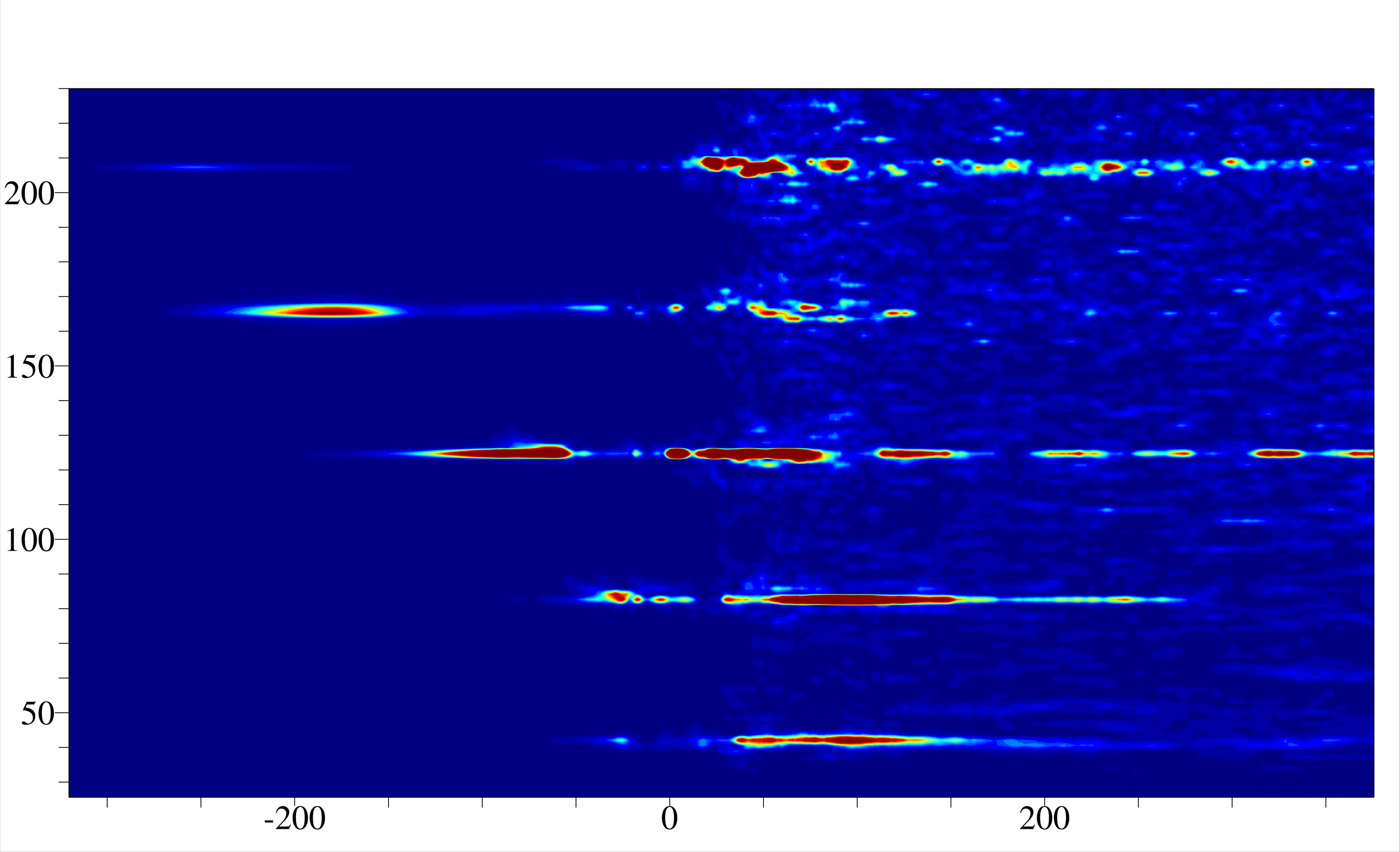}};
   \begin{scope}[x={(image.south east)},y={(image.north west)}]
   \node [align=center, rotate=90] at (-0.02,0.5) {\small $f$[kHz]};
   \node [align=center] at (0.55,0.) {$\hat{x}$};
  \end{scope}
   \end{tikzpicture}
   \caption{Premultiplied power spectral density of the wall pressure in the interaction region.}
\label{fig:PSD}
\end{figure}

The strong adverse pressure gradient associated with the incident shock tends to amplify the boundary layer perturbations injected by the inflow disturbances, inducing transition to turbulence. Nevertheless, the standalone perturbation has a large influence on the overall flow dynamics, even before the shock impingement station and the recirculation bubble. Figure~\ref{fig:streaks} reports the isocontours of the streamwise velocity perturbation at $y/\delta^\star_\text{in} \approx 0.5$ (top panel), $y/\delta^\star_\text{in} \approx 1$ (middle panel) and $y/\delta^\star_\text{in} \approx 3$ (bottom panel). The emergence of intense streaky structures is visible starting from $\hat{x} \approx -150$. These structures develop with different characteristic dimensions and intensity moving away from the wall; they are disrupted by the shock impingement and disappear in the recirculation bubble. When the flow reattaches, other less-coherent streaky structures appear and finally break down and lead to transition to turbulence. The observed dynamics further confirms that breakdown to turbulence occurs after the reattachment, coherently with the streamwise evolution of the skin friction coefficient.
In the attempt to quantify the breakdown, we follow the procedure of Andersson \textit{et al.} \cite{andersson2001breakdown} for incompressible flows and we estimate the streaks amplitude as:
\begin{equation}\label{eq:ampl_streaks}
\text{A}_s = \frac{1}{2 u_\infty} \left[ \max_{y,z}\left(\overline{u}-u_\text{BF}\right)- \min_{y,z}\left(\overline{u}-u_\text{BF}\right) \right]
\end{equation}
where $u_\text{BF}$ stays for the velocity of the base flow. Evaluating their amplitude on the current highly-compressible non-equilibrium flow is of course made difficult by the fact that quantitative criteria in the literature exist only for incompressible flows; however, the streamwise evolution of streaks amplitude may help understanding their role in the transition process. The analysis is performed over 300 three-dimensional subdomains collected in runtime, spanning the extent $-200\leq \hat{x} \leq 150$, $0\leq y/\delta^\star_\text{in} \leq 20$ and $0\leq z/\delta^\star_\text{in} \leq 60$. Figure~\ref{fig:ampl_streaks} shows the streamwise evolution of $A_s$, in the region $-200 < \hat{x} < 150$.
As already observed from the instantaneous slices, the streaks amplitude grows significantly well upstream of shock impingement, in a region subjected to zero or even slightly favourable pressure gradient; It is therefore to be uniquely ascribed to the growth of inflow perturbations. Afterwards, the impinging shock disrupts the growing flow structures and generates a recirculation region, which results in a sharp decrease of the estimated streak amplitude. Downstream of $\hat{x}=0$, the structures grow again up to $\hat{x} \approx 60$,  i.e. shortly downstream of the location where the skin friction and the heat flux peak. Then, coherence is lost as the flow transitions and the amplitude drops again.\\
Although the previous procedure allows describing the evolution of such stationary structures, the exact transition process is more difficult to detect. The coexistence of the perturbation and the incident shock makes it more difficult to distinguish the different mechanisms that non-linearly combine to induce breakdown. The instability that often dominates transition in the hypersonic flow regime is the one related to second (or Mack) mode. This two-dimensional inviscid instability arises when a region of the mean flow becomes supersonic relative to the phase speed of the instability, and is characterized by higher frequencies with respect to the first mode. In the past, both linear and weakly-nonlinear stability studies \cite{fedorov2001prehistory,bitter2015stability} have pointed out that wall cooling tends to stabilize first-mode instability while destabilizing the second mode, which may even become the most unstable one at lower Mach numbers. Recently, many authors have observed the presence of such instability in high-enthalpy flows as well \cite{miro2019high,salemi2018synchronization,zanus2020parabolized,chen2021secondary,passiatore2021simulations}. Following the trend of the skin friction coefficient in figure~\ref{fig:wall_quantities}, the flow behavior starts to deviate from the base flow self-similar solution at $\hat{x} \approx -50$. Therefore, in the upstream flow, existing considerations corroborated for flat-plate boundary layers can be extended to the present study. The instantaneous density gradient $|\nabla \rho/|\delta^\star_\text{in}/\rho_\infty$ is shown in figure~\ref{fig:inst_Mack}. Rope-like structures can be observed in the top panel, focussing on the region upstream of the interaction zone. Such structures are shown not be as regular and pronounced as those found for cones configurations in Zhu \textit{et al.} \cite{zhu2018newly} and \cite{chen2021secondary}, albeit this difference can be attributed to the specific perturbations used in the present study. Concurrently, the bottom panel of figure~\ref{fig:inst_Mack} shows acoustic waves that are trapped and reflected by the wall and near the sonic line, marked in red in the figure. Visualizations of streamwise and wall-normal velocity fields (not shown) have also revealed the presence of distinguishable structures having a wavelength of almost twice the local boundary layer thickness, a common characteristic of second Mack modes. Moreover, two dimensional modes represented by roll-like shape structures near the wall are clearly visible in figure~\ref{fig:qcrit}, extending up to the interaction zone.\\
From a more quantitative standpoint, the frequency of Mack mode can be roughly estimated as $f_{M} = u_\delta/(2 \delta_l)$ for low-enthalpy boundary layers \cite{sandham2014transitional}, with $\delta_l$ the laminar boundary layer thickness. In the region of interest, comprised in $-200 \lessapprox \hat{x} \lessapprox -50$, such an estimation results in frequencies ranging between approximately 200 and \SI{100}{kHz}. Mack's mode frequency basically decreases with increasing boundary layer thickness and, consequently, with increasing streamwise coordinate. It is evident that this estimate loses its validity when a clear boundary layer thickness cannot be distinguished anymore, such as in the interaction region. To analyse the frequency content of the flow, we show in figure~\ref{fig:PSD} the premultiplied Power Spectral Density (PSD) of the wall pressure for the streamwise region of interest. The predominant modes are explicitly excited by equation~\eqref{eq:forcing}, the flow in this portion of the computational domain being still strongly influenced by the inflow forcing. Nevertheless, among all the frequencies excited through the perturbation function, specific frequencies emerge in the range comprised approximately between \SI{50}{kHz} and \SI{200}{kHz}. Before the impingement, a $\approx \SI{160}{kHz}$ frequency is significantly predominant. Consistently with the boundary layer thickening downstream, the peak is then shifted to lower frequencies ($\approx$ \SI{125}{kHz}, \SI{82}{kHz}). These results are in agreement with the $f_M$ estimate; despite the region being still biased by the imposed disturbances, the closest frequency to $f_M$ is naturally selected at each streamwise location and prevail among the others. When the boundary layer is subjected to the impinging shock, other mechanisms emerge and the spectrum starts to fill up, both sustaining the previously emerged frequencies and highlighting new ones. The large computational cost of the simulation limits the temporal window of sample collection, and therefore no information about low-frequency unsteadiness can be provided at the moment. Whether the bubble breathing phenomenon is present even when the shock impinges on a high-enthalpy boundary layer will be the subject of future works.
%
%
\subsection{Turbulent statistics}
\begin{figure}
   \centering
\begin{tikzpicture}
   \node[anchor=south west,inner sep=0] (image) at (0,0) {
   \includegraphics[width=0.49\textwidth,trim={0 0 0 0},clip]{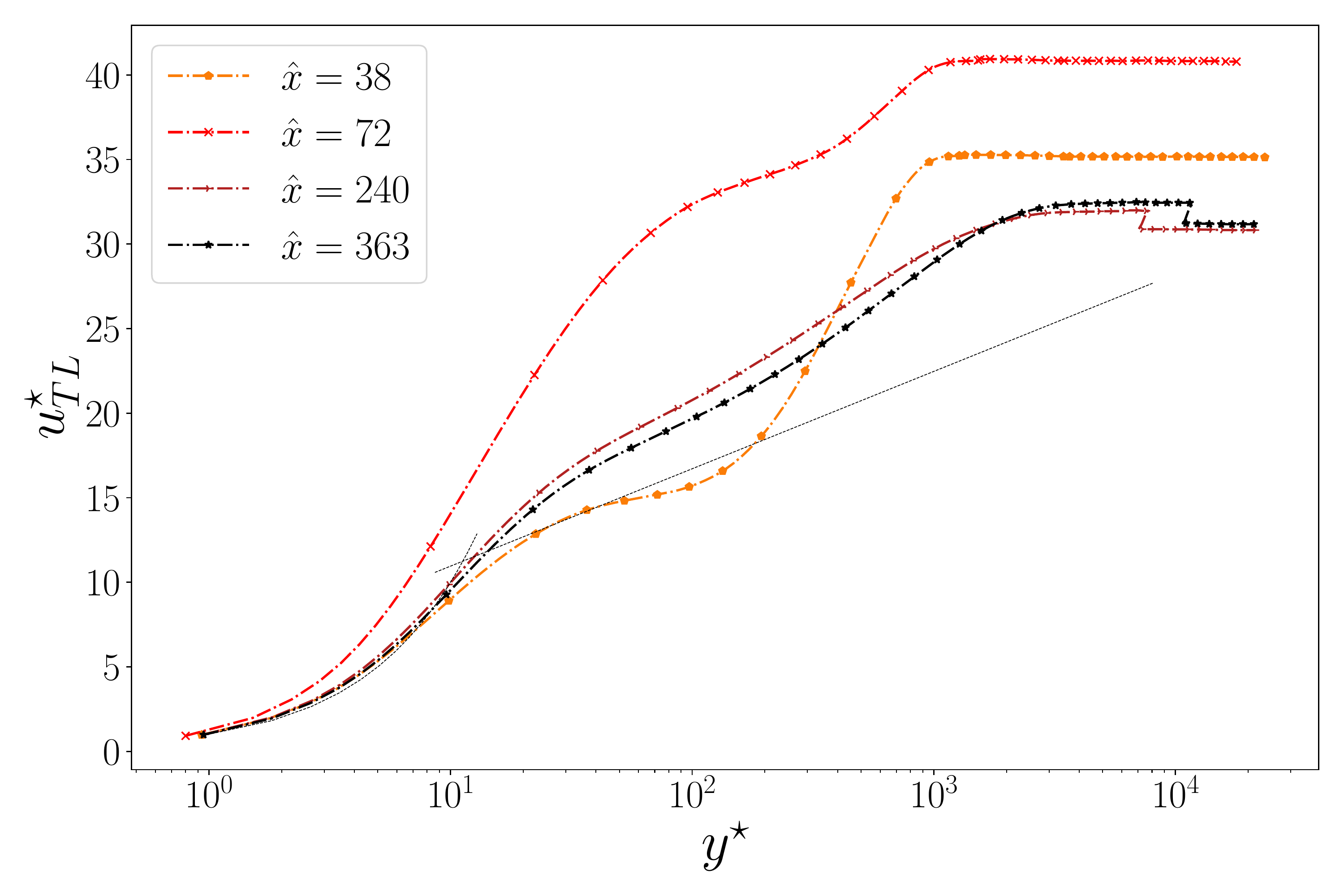}};
   \begin{scope}[x={(image.south east)},y={(image.north west)}]
    \node [align=center] at (0.03,0.95) {(a)};
      \end{scope}
  \end{tikzpicture}
  \begin{tikzpicture}
   \node[anchor=south west,inner sep=0] (image) at (0,0) {
   \includegraphics[width=0.49\textwidth,trim={0 0 0 0},clip]{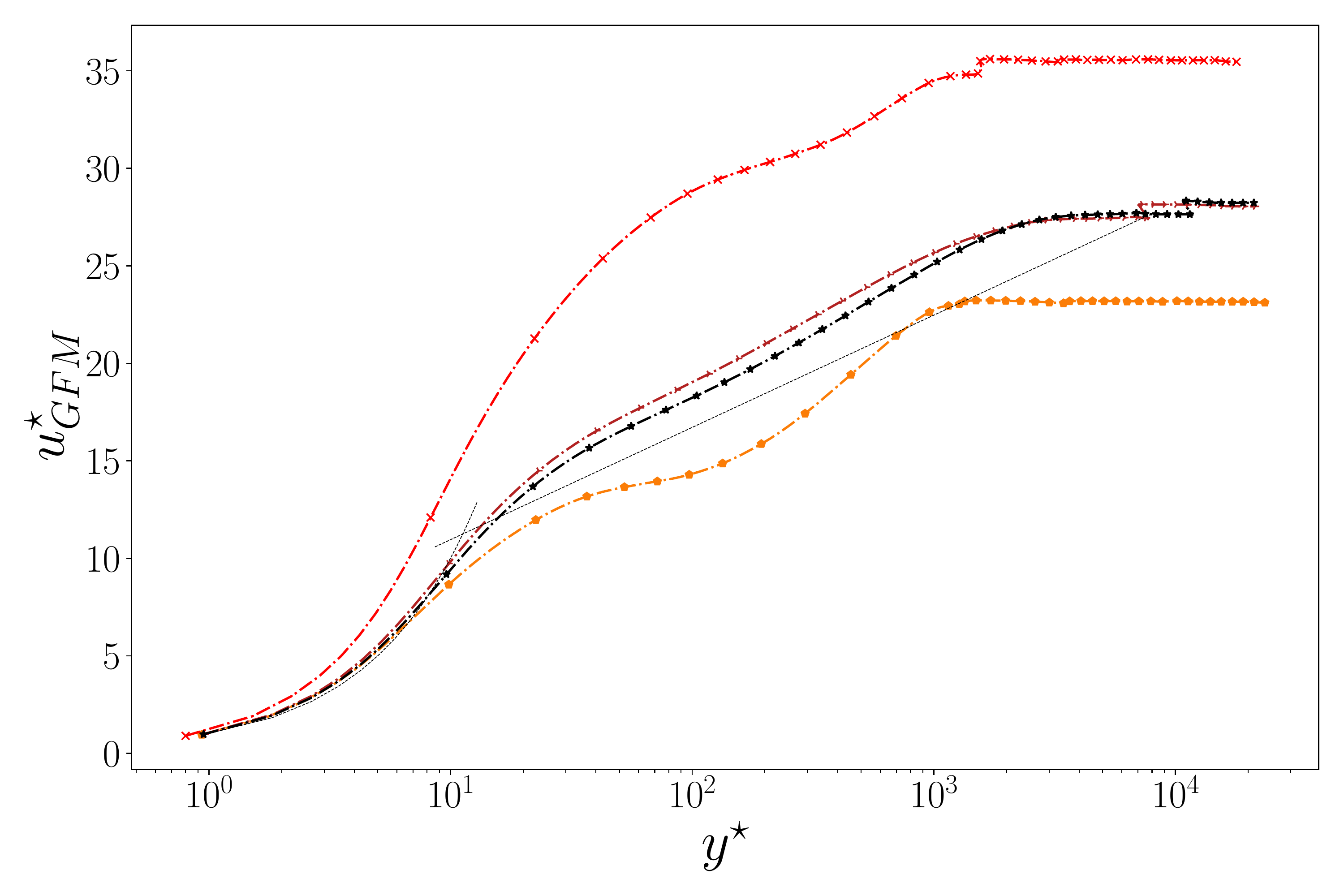}};
   \begin{scope}[x={(image.south east)},y={(image.north west)}]
    \node [align=center] at (0.03,0.95) {(b)};
    \end{scope}
  \end{tikzpicture}
 \caption{Streamwise velocity scalings. Trettel \& Larsson transformation (a) and Griffin-Fu-Moin transformation (b).}
\label{fig:velocity_scaling}
\end{figure}

\begin{figure}
 \centering
   \begin{tikzpicture}
   \node[anchor=south west,inner sep=0] (image) at (0,0) {
   \includegraphics[width=0.49\textwidth,trim={0 0 0 0},clip]{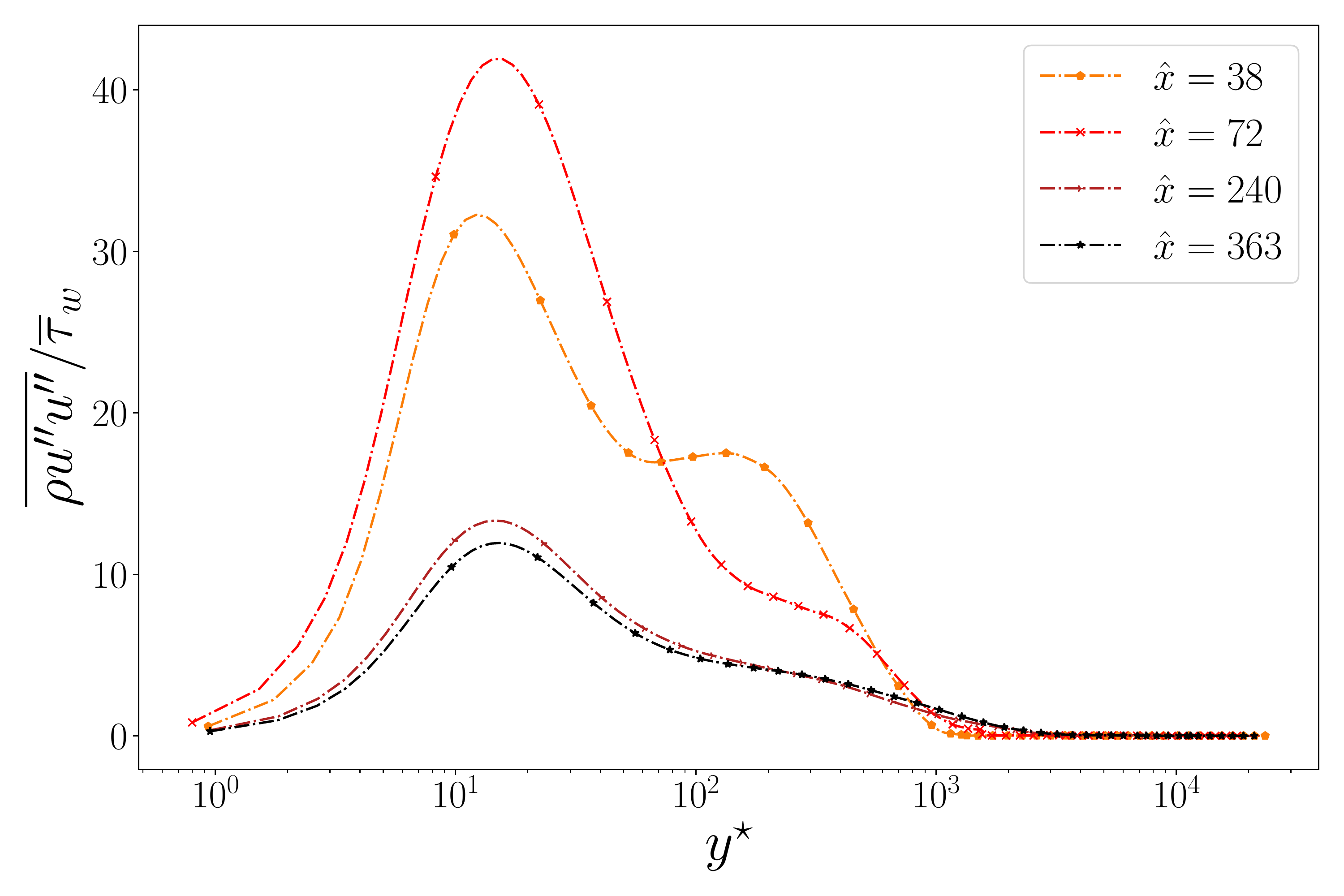}};
   \begin{scope}[x={(image.south east)},y={(image.north west)}]
    \node [align=center] at (0.03,0.95) {(a)};
   \end{scope}
  \end{tikzpicture}
\begin{tikzpicture}
   \node[anchor=south west,inner sep=0] (image) at (0,0) {
   \includegraphics[width=0.49\textwidth,trim={0 0 0 0},clip]{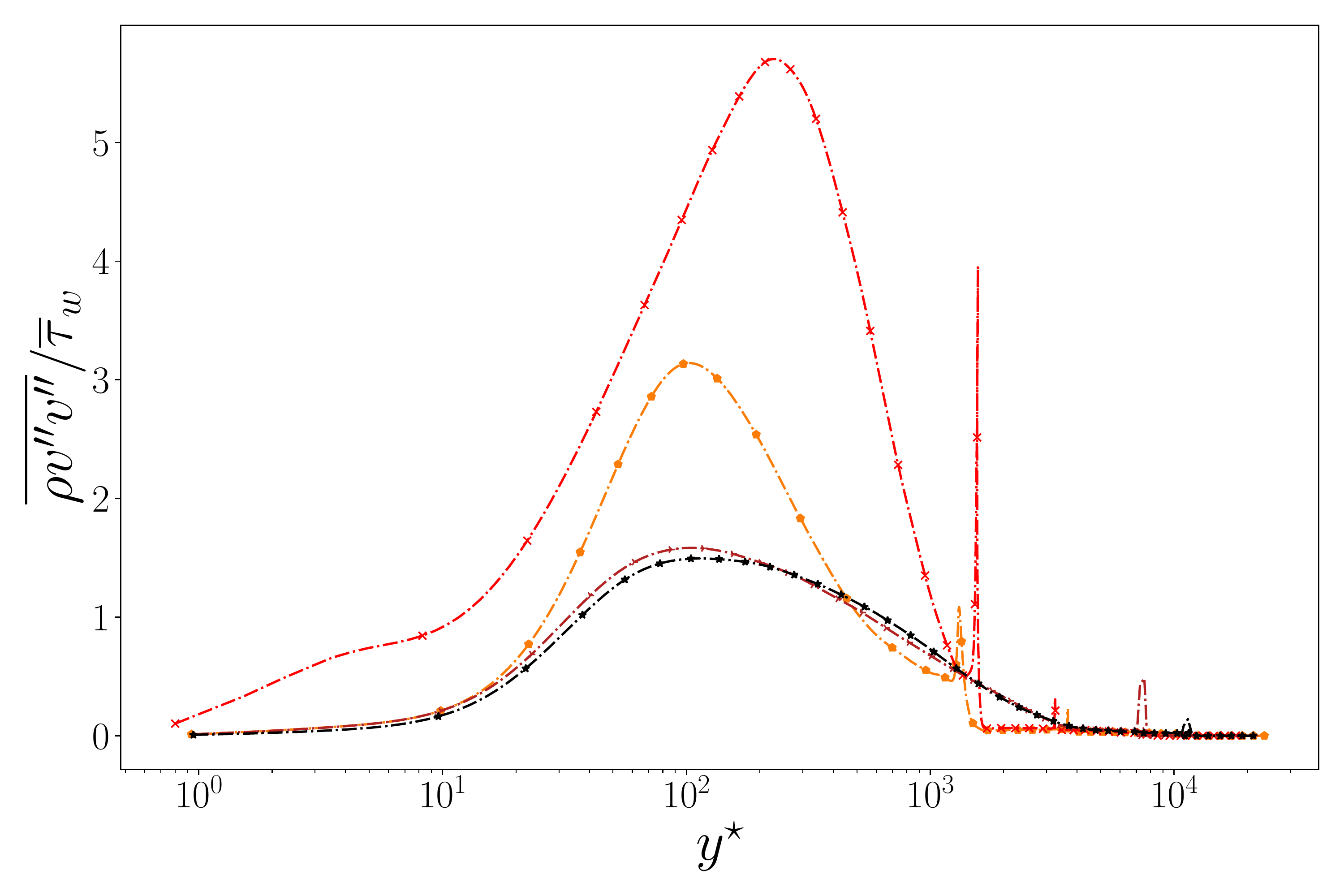}};
   \begin{scope}[x={(image.south east)},y={(image.north west)}]
    \node [align=center] at (0.03,0.95) {(b)};
   \end{scope}
  \end{tikzpicture}

 \begin{tikzpicture}
   \node[anchor=south west,inner sep=0] (image) at (0,0) {
   \includegraphics[width=0.49\textwidth,trim={0 0 0 0},clip]{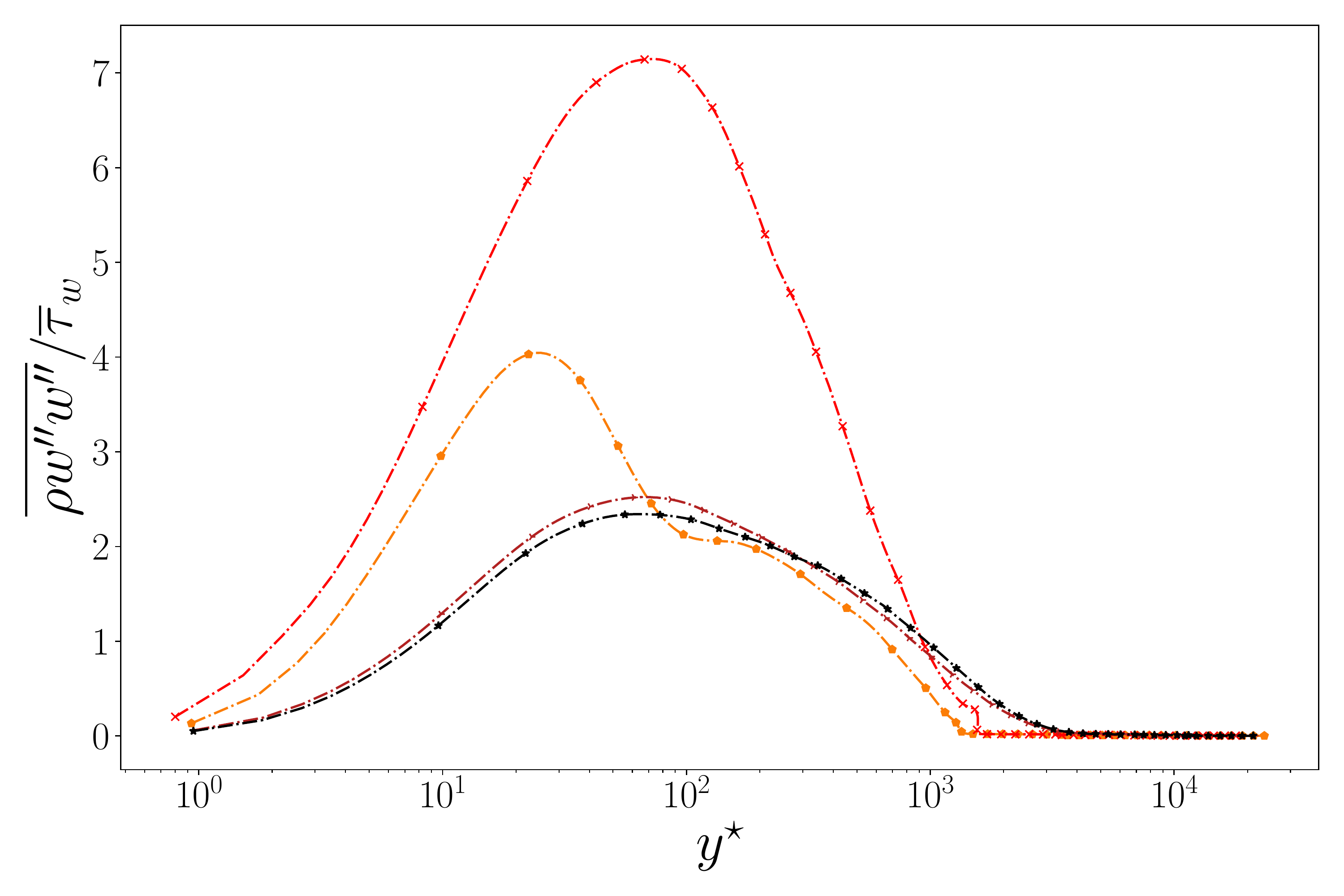}};
   \begin{scope}[x={(image.south east)},y={(image.north west)}]
    \node [align=center] at (0.03,0.95) {(c)};
   \end{scope}
  \end{tikzpicture}
   \begin{tikzpicture}
   \node[anchor=south west,inner sep=0] (image) at (0,0) {
   \includegraphics[width=0.49\textwidth,trim={0 0 0 0},clip]{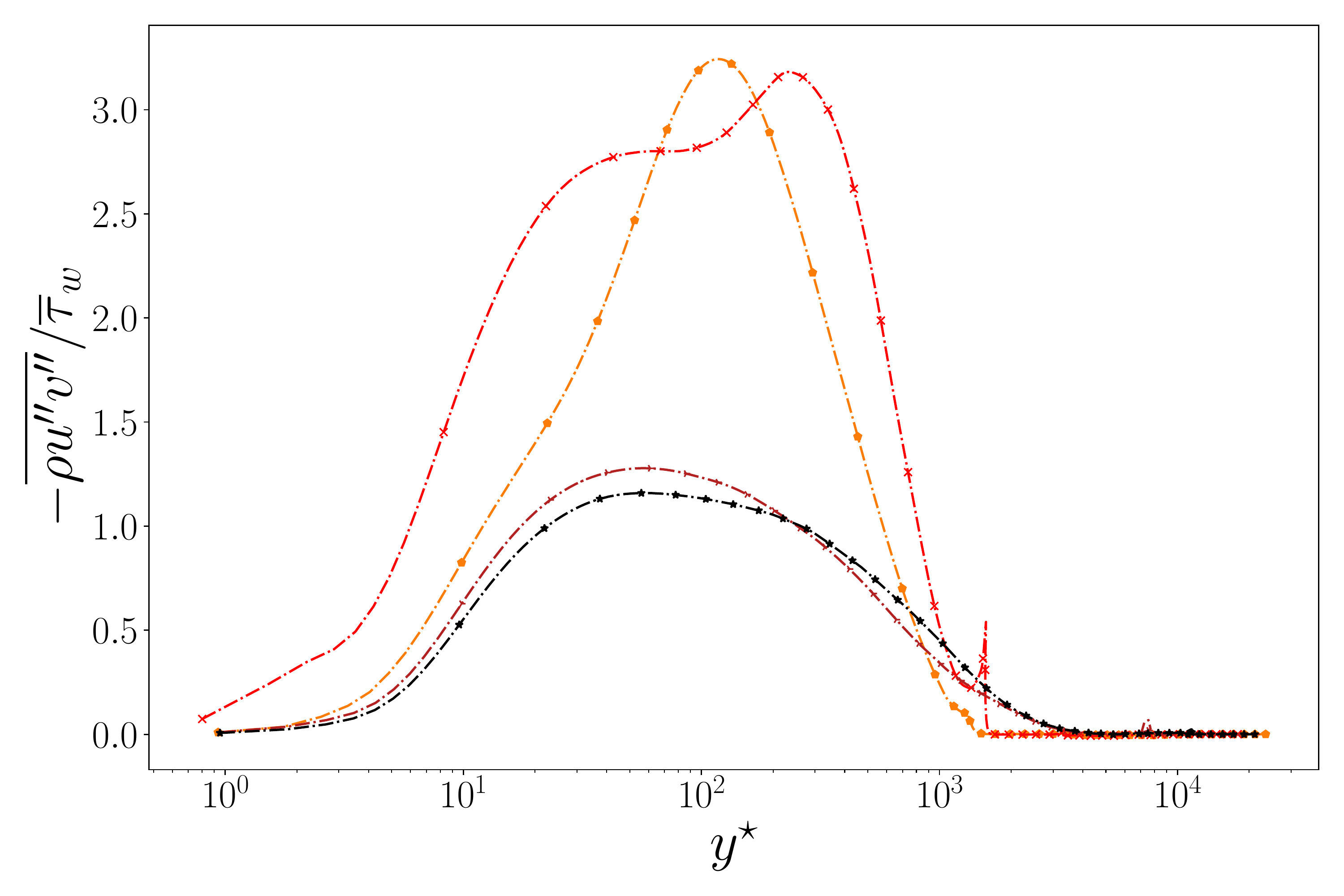}};
   \begin{scope}[x={(image.south east)},y={(image.north west)}]
    \node [align=center] at (0.03,0.95) {(d)};
   \end{scope}
  \end{tikzpicture}
 \caption{Wall-normal profiles of the Reynolds stresses for the streamwise (a), wall-normal (b), spanwise (c) and shear (d) components.}
\label{fig:reynolds_stresses}
\end{figure}
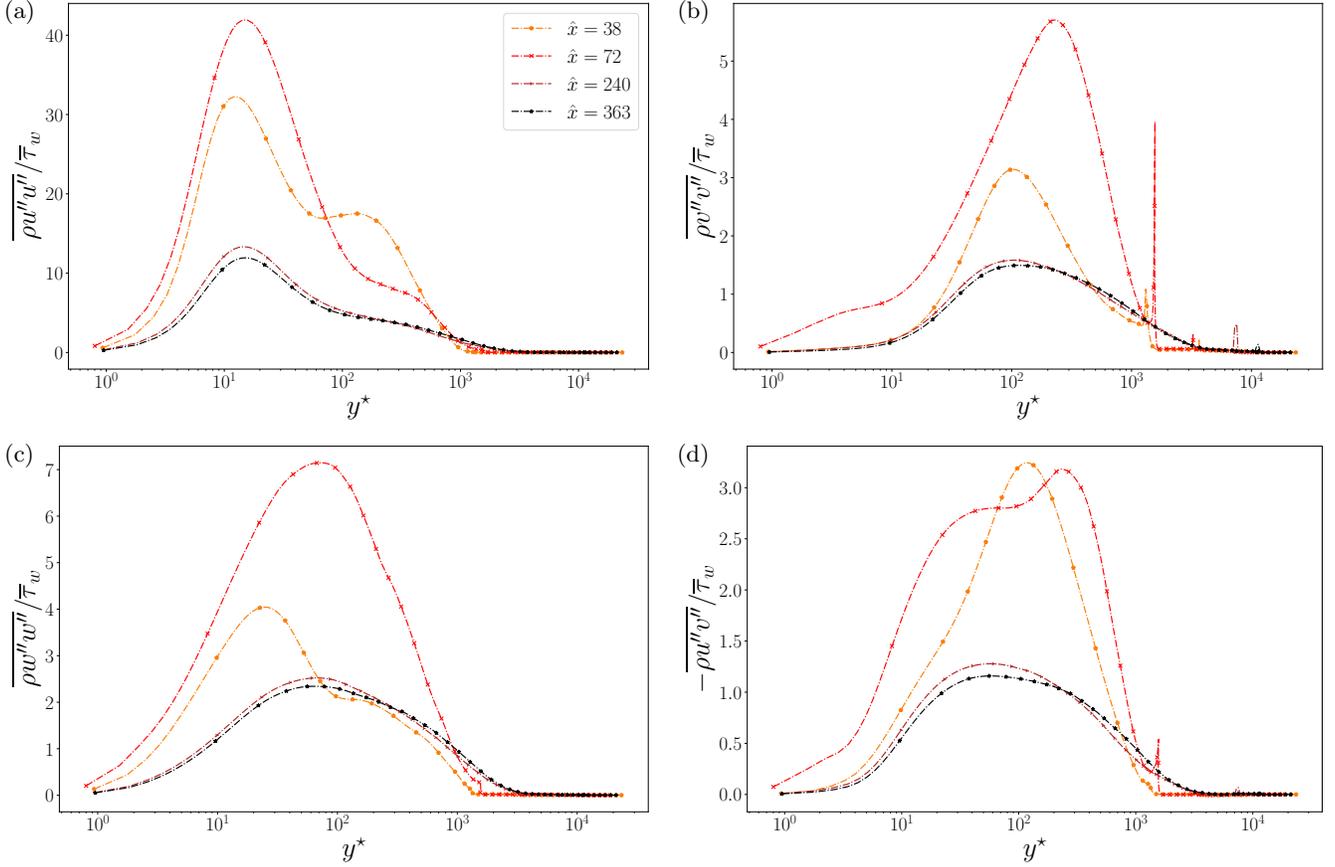

\begin{figure}
 \centering
   \begin{tikzpicture}
   \node[anchor=south west,inner sep=0] (image) at (0,0) {
   \includegraphics[width=0.49\textwidth,trim={0 0 0 0},clip]{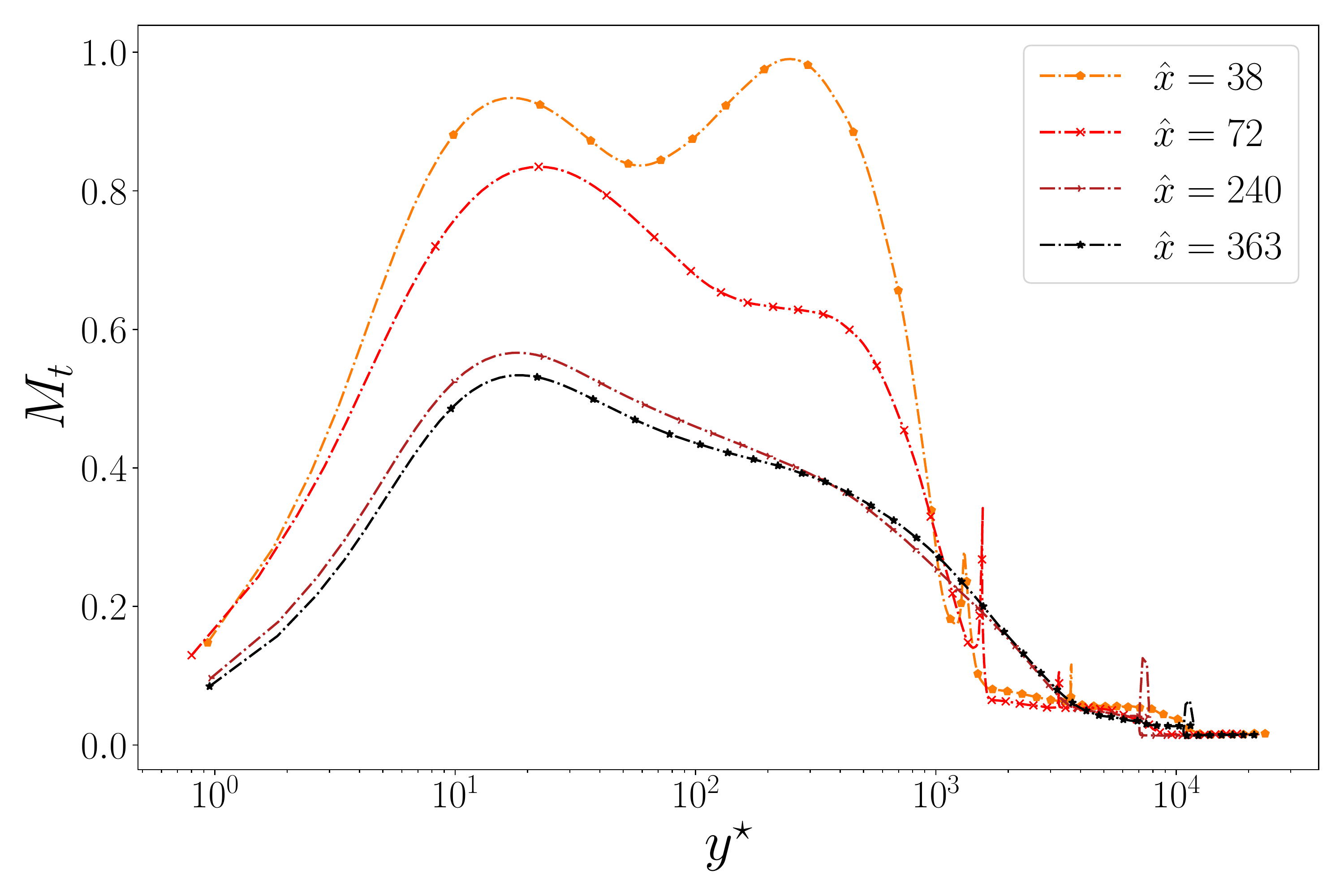}};
   \begin{scope}[x={(image.south east)},y={(image.north west)}]
    \node [align=center] at (0.03,0.95) {(a)};
    \end{scope}
  \end{tikzpicture}
   \begin{tikzpicture}
   \node[anchor=south west,inner sep=0] (image) at (0,0) {
   \includegraphics[width=0.49\textwidth,trim={0 0 0 0},clip]{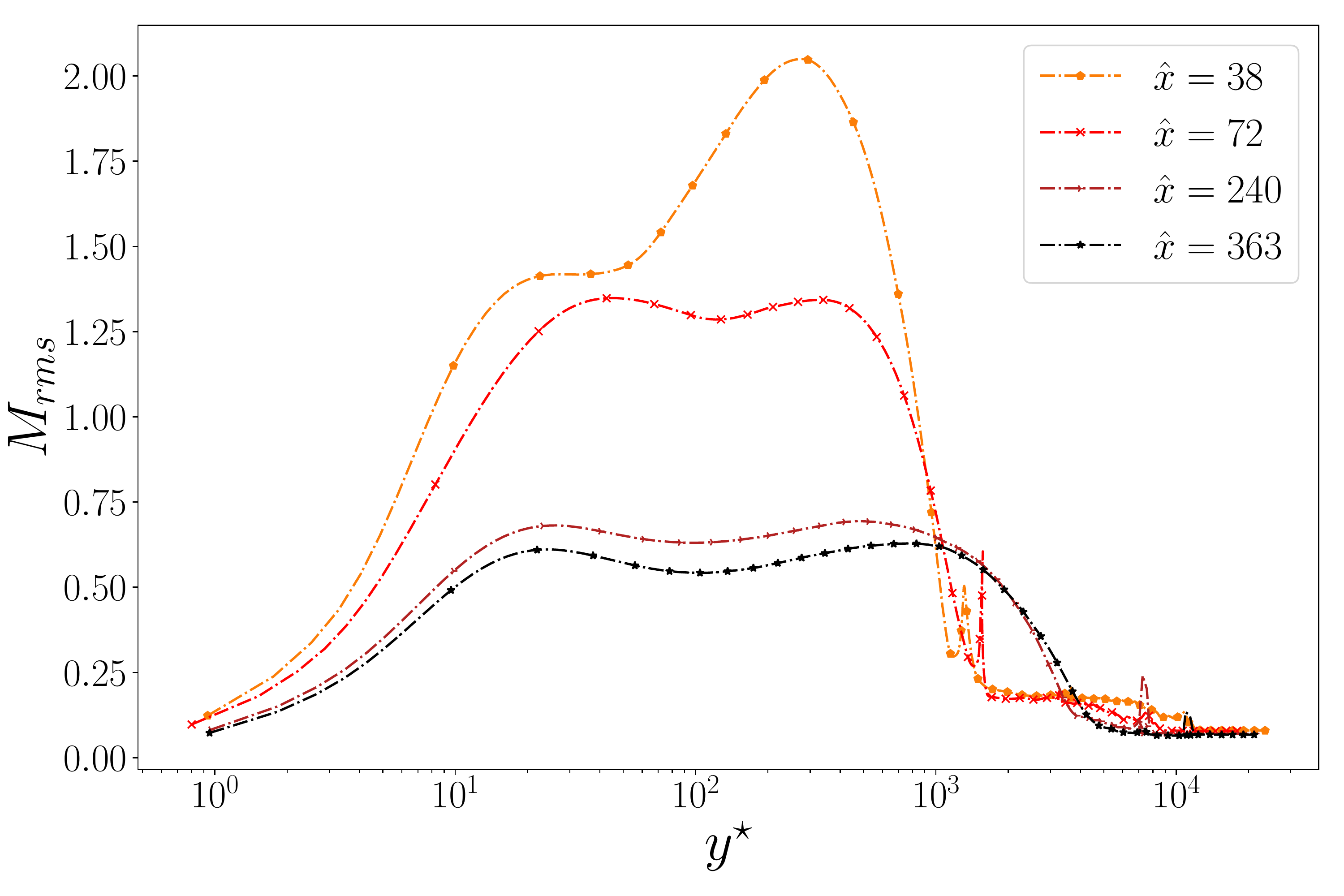}};
   \begin{scope}[x={(image.south east)},y={(image.north west)}]
    \node [align=center] at (0.03,0.95) {(b)};
    \end{scope}
  \end{tikzpicture}
\caption{Wall-normal profiles of turbulent Mach number (a) and rms Mach number (b) for the last four stations of table~\ref{tab:stations}.}
\label{fig:mach_comp}
\end{figure}

We present hereafter an overview of the main turbulent statistics at the last four stations of table~\ref{tab:stations}. Note that an in-depth analysis of the turbulent flow over a thermochemical out-of-equilibrium boundary layer was already performed by Passiatore \textit{et al.} \cite{passiatore2022thermochemical}; the results here presented share similar trends. First, the transformations of Van Driest \cite{van1956problem}, Trettel \& Larsson \cite{trettel2016mean} and Griffin \textit{et al.} \cite{griffin2021velocity} for the averaged streamwise velocity are applied to the transitional and fully turbulent stations. Figure~\ref{fig:velocity_scaling} shows the results only for the last two scalings, both providing better predictions than the Van Driest one.
%
The collapse with the classical logarithmic profile is very poor for $\hat{x} = 38$ and $\hat{x} =72$, confirming the purely transitional state of the boundary layer in this region. For the two last stations, these transformations fail to collapse the mean velocity profiles on the incompressible logarithmic law, as already observed by many authors \cite{zhang2018direct,fu2021shock,passiatore2022thermochemical}. It is of common agreement that the nominal K\`arm\`an constant should be smaller than the classical value of $\approx 0.4$ and the intercept should be greater than 5.2 at least for cooled boundary layers \cite{duan2011direct4,li2022wall}. Reasonable self-similarity can be observed for the last two stations, the better collapse being obtained by the \cite{griffin2021velocity} transform. The Reynolds stresses, shown in figure~\ref{fig:reynolds_stresses}, exhibit a reasonable collapse when plotted in semi-local units, also due to the small changes in $Re_\tau$ for the two last streamwise positions (less than 10\%). At the transitional stations, the flow is subjected to massive velocity and pressure fluctuations, two to three times larger than in the turbulent region. These lead to very large values for the turbulent and rms Mach numbers (figure~\ref{fig:mach_comp}a and b, respectively), similar to those obtained by Passiatore \textit{et al.} \cite{passiatore2022thermochemical} at much larger friction Reynolds numbers.
%



\subsection{Thermochemical non-equilibrium}\label{sec:tcne}
\begin{figure}
 \centering
 \begin{tikzpicture}
   \node[anchor=south west,inner sep=0] (a) at (0,0) {\includegraphics[width=0.49\textwidth]{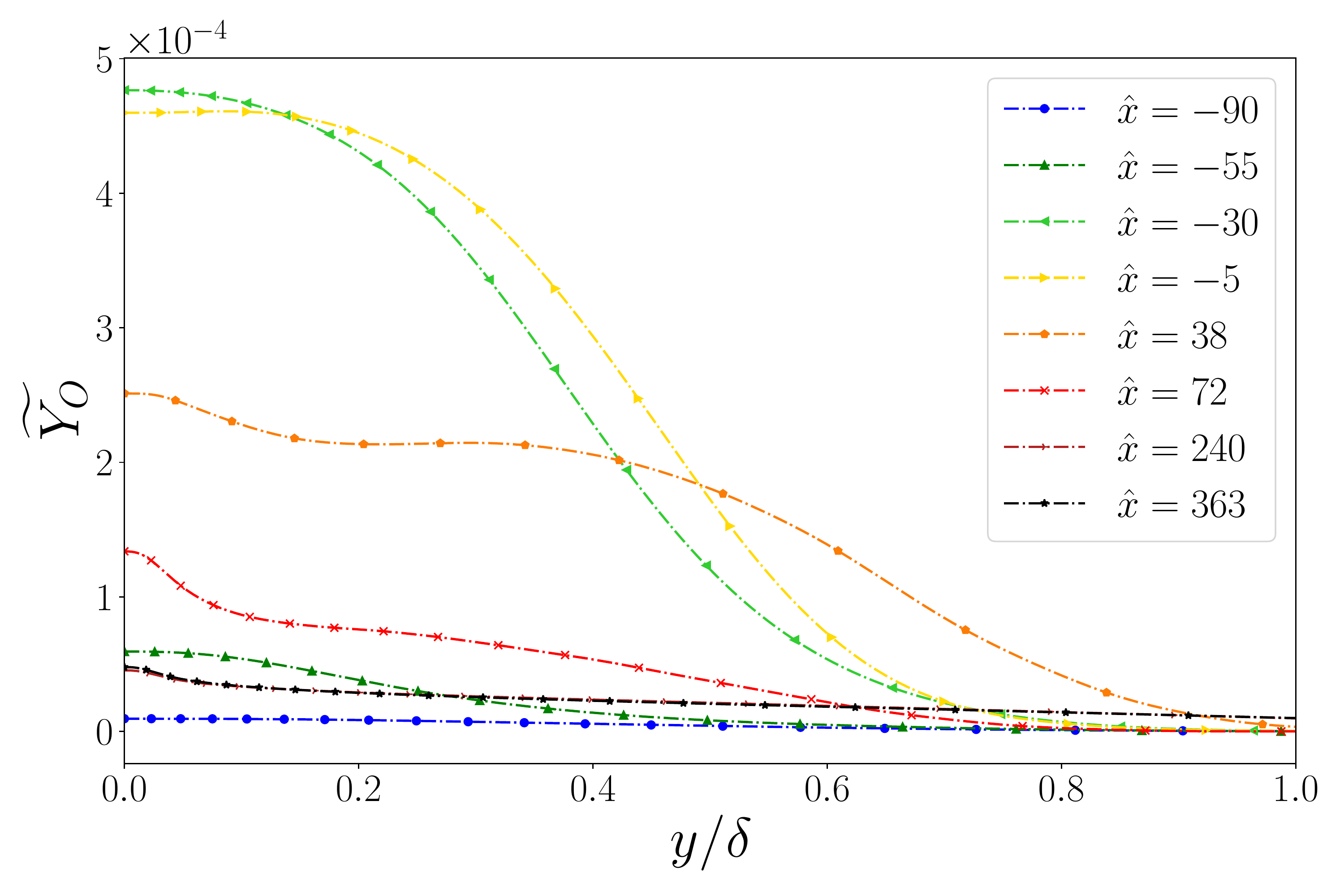}};
   \begin{scope}[x={(a.south east)},y={(a.north west)}] 
    \node [align=center] at (0.03,0.95) {(a)};
    \end{scope}
 \end{tikzpicture}
 \begin{tikzpicture}
   \node[anchor=south west,inner sep=0] (a) at (0,0) {\includegraphics[width=0.49\textwidth]{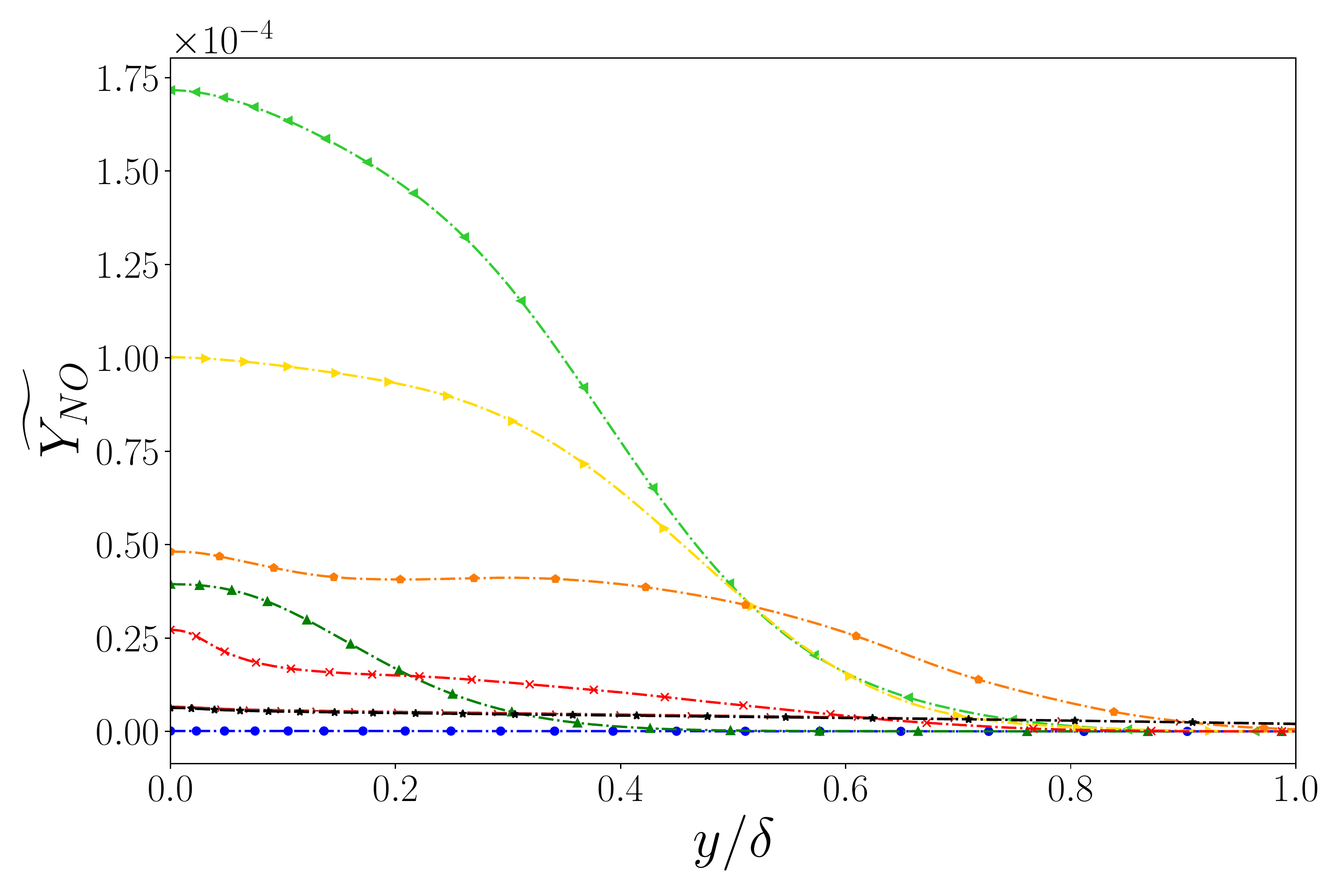}};
   \begin{scope}[x={(a.south east)},y={(a.north west)}]
    \node [align=center] at (0.03,0.95) {(b)};
    \end{scope}
 \end{tikzpicture}
 \caption{Wall-normal distributions of atomic oxygen (a) and nitric oxide (b) for the different streamwise stations reported in table~\ref{tab:stations}.}\label{fig:mean_species}
\end{figure}

Before focusing the attention on vibrational excitation, we provide some insights on chemical activity. As already observed in previous works for cooled flat-plate boundary layers at similar freestream Mach numbers \cite{direnzo2021direct,passiatore2022thermochemical}, chemical activity is relatively weak and scarcely influences flow dynamics. A more significant effect can be appreciated for pseudo-adiabatic \cite{passiatore2021finite} and adiabatic walls. Alternatively, when the freestream temperature is important \cite[e.g.,][]{li2022wall} the influence of chemical activity is far from being marginal, the heat flux increasing due to species mass fraction fluctuations. For the present configuration, chemical dissociation is quite faint, reaching a peak of atomic oxygen mass fraction of $Y_\text{O} \approx 5 \times 10^{-4}$, despite the relatively high temperatures registered.
The wall-normal distributions of the chemical products at the different stations are shown in figure~\ref{fig:mean_species}(a)-(b) for O and NO, respectively. Note that production of atomic nitrogen is negligible and will not be considered in the discussion. Notwithstanding the cooled boundary condition and the resulting non-monotonic temperature profiles, the largest amount of chemical products is encountered at the wall in each region of the computational domain.
The concentrations of species products increase passing from the laminar region to the interaction zone; when entering the recirculation bubble, where the temperature values are the largest, a rise up to an order of magnitude is registered, whereas their amount decreases in the transitional region. The distributions for the last two stations in the fully-turbulent regimes are almost perfectly superposed because of the extremely low Damk{\"o}hler numbers achieved here, which highlighting a large difference between the characteristic timescales of turbulent motions and chemical activity. Such a behavior is distinct from the one shown by Volpiani \cite{volpiani2021numerical}, in which chemical dissociation progressively increases reaching a peak downstream of the interaction region. The difference can be ascribed to the much larger temperature values assigned at the wall and in the free-stream in that study.

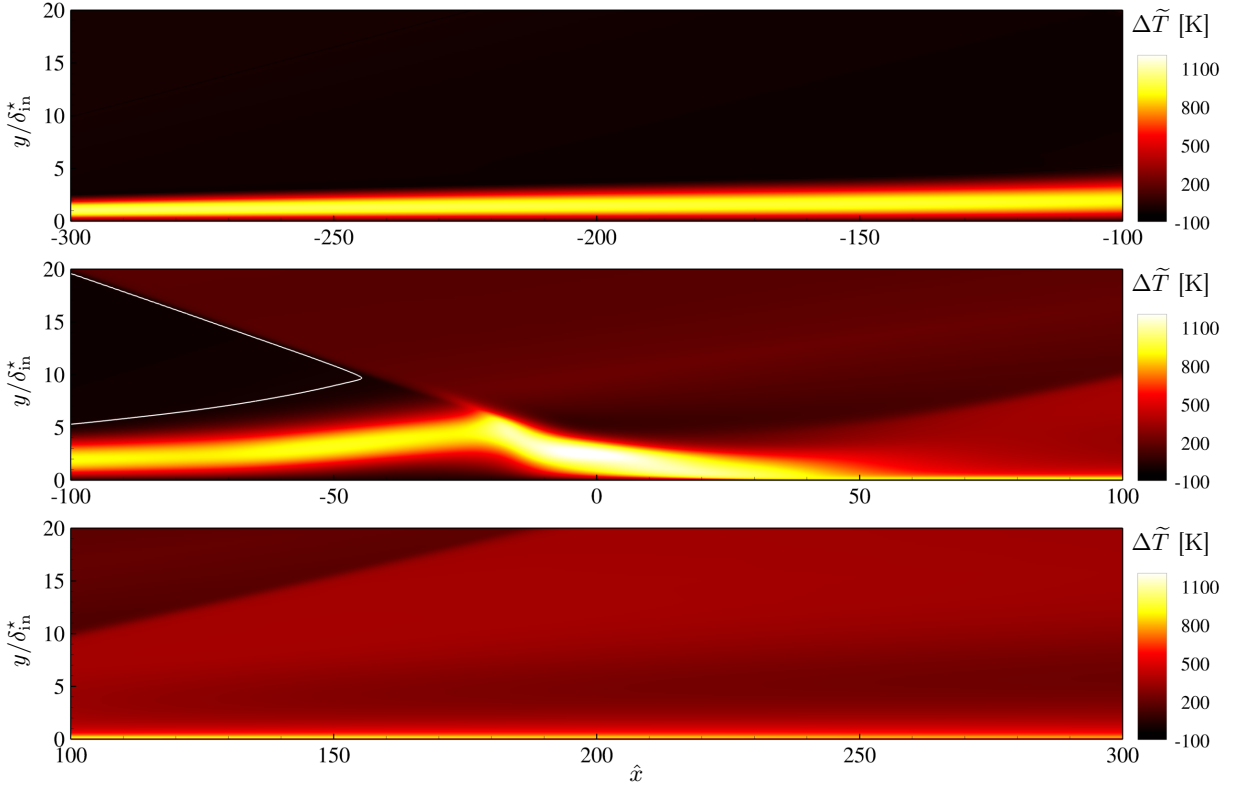
\begin{figure}
 \centering
    \begin{tikzpicture}
   \node[anchor=south west,inner sep=0] (a) at (0,0) {\includegraphics[width=0.88\textwidth, trim={4 10 10 10}, clip]{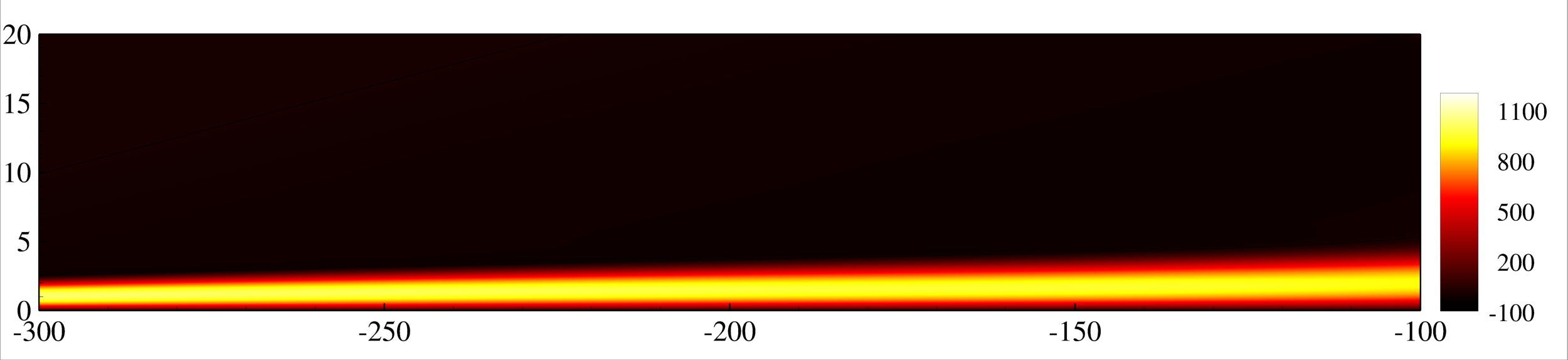}};
   \begin{scope}[x={(a.south east)},y={(a.north west)}]
     \node [align=center] at (0.95,0.87) {$\Delta \widetilde{T}$ [K]};
     \node [align=center,rotate=90] at (-0.02,0.5) {$y/\delta^\star_\text{in}$};
       \end{scope}
 \end{tikzpicture}\\[-0.1cm]
 \centering
 \begin{tikzpicture}
   \node[anchor=south west,inner sep=0] (a) at (0,0) {\includegraphics[width=0.88\textwidth, trim={4 10 10 10}, clip]{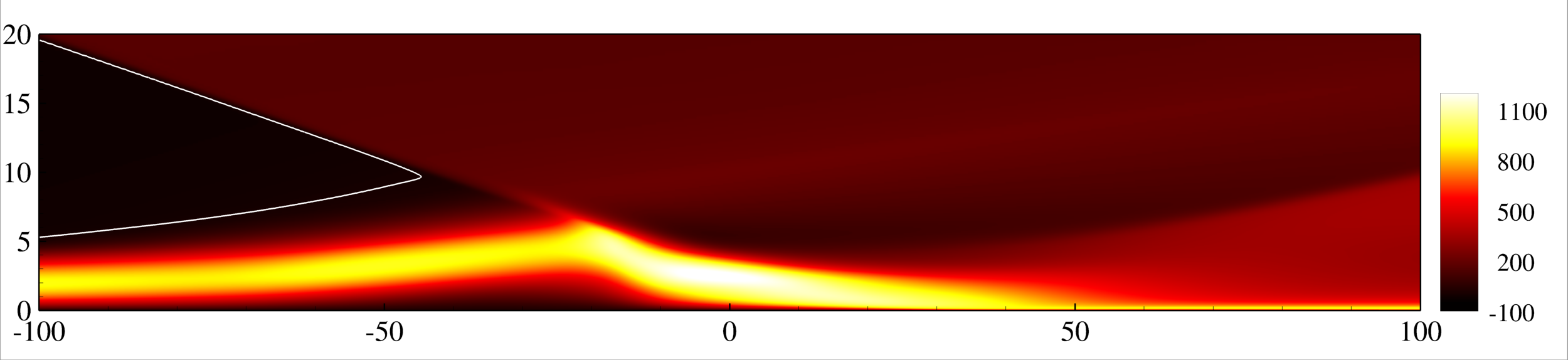}};
   \begin{scope}[x={(a.south east)},y={(a.north west)}]
     \node [align=center] at (0.95,0.87) {$\Delta \widetilde{T}$ [K]};
     \node [align=center,rotate=90] at (-0.02,0.5) {$y/\delta^\star_\text{in}$};
     \end{scope}
 \end{tikzpicture}\\[-0.1cm]
 \centering
 \begin{tikzpicture}
   \node[anchor=south west,inner sep=0] (a) at (0,0) {\includegraphics[width=0.88\textwidth, trim={4 10 10 10}, clip]{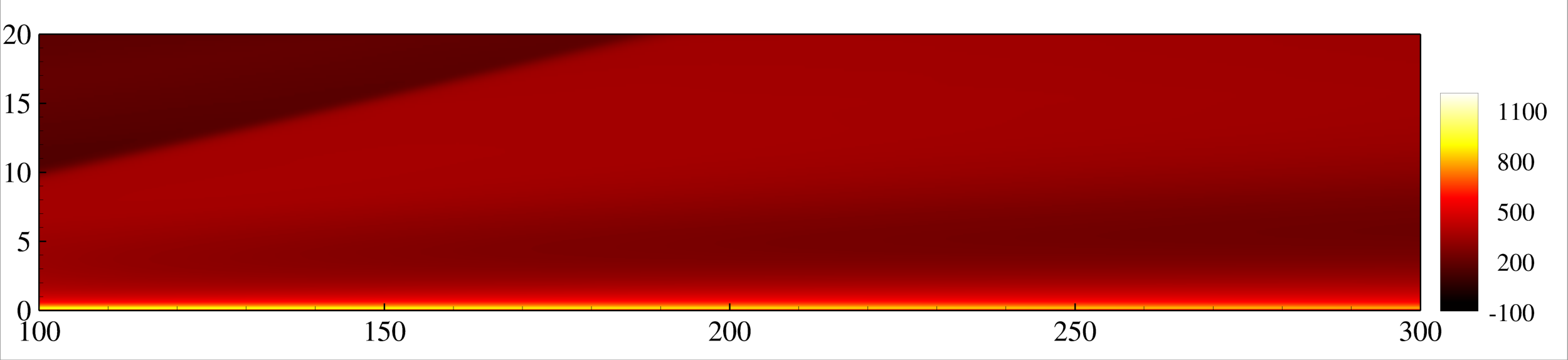}};
   \begin{scope}[x={(a.south east)},y={(a.north west)}]
     \node [align=center] at (0.5,0.) {$\hat{x}$};
     \node [align=center] at (0.95,0.87) {$\Delta \widetilde{T}$ [K]};
     \node [align=center,rotate=90] at (-0.02,0.5) {$y/\delta^\star_\text{in}$};
     \end{scope}
 \end{tikzpicture}
 \caption{Isocontours of the Favre-averaged dimensional temperature difference $\Delta \widetilde{T}$. The white line denotes the $\Delta \widetilde{T} = 0$ isoline. The $y$ axis has been stretched for better visualization.}\label{fig:deltaT}
\end{figure}

\begin{figure}
\centering
\begin{tikzpicture}
\node[anchor=south west,inner sep=0] (a) at (0,0) {\includegraphics[width=0.49\textwidth]{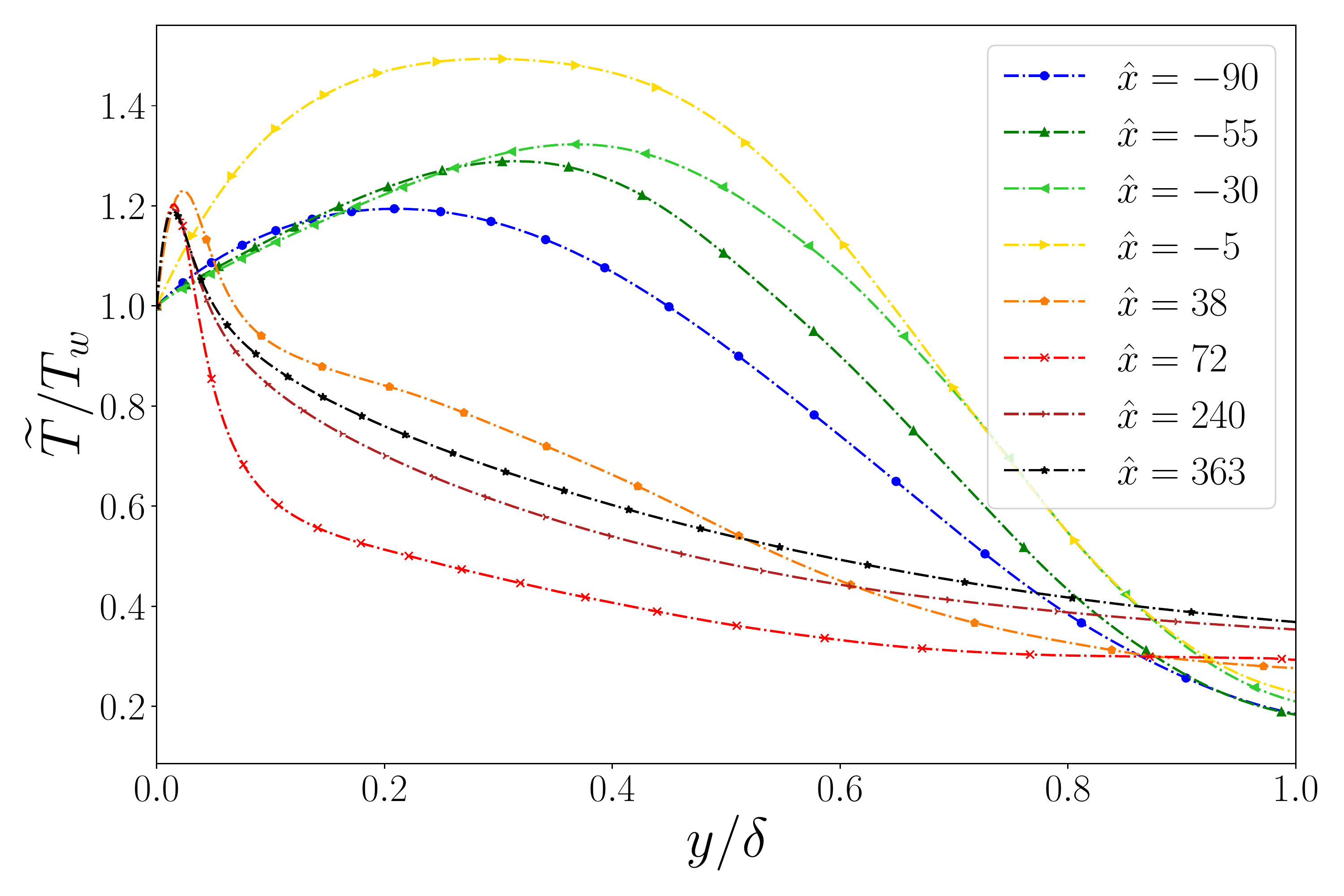}};
\begin{scope}[x={(a.south east)},y={(a.north west)}]
  \node [align=center] at (0.03,0.95) {(a)};
   \end{scope}
 \end{tikzpicture}
   \begin{tikzpicture}
   \node[anchor=south west,inner sep=0] (a) at (0,0) {\includegraphics[width=0.49\textwidth]{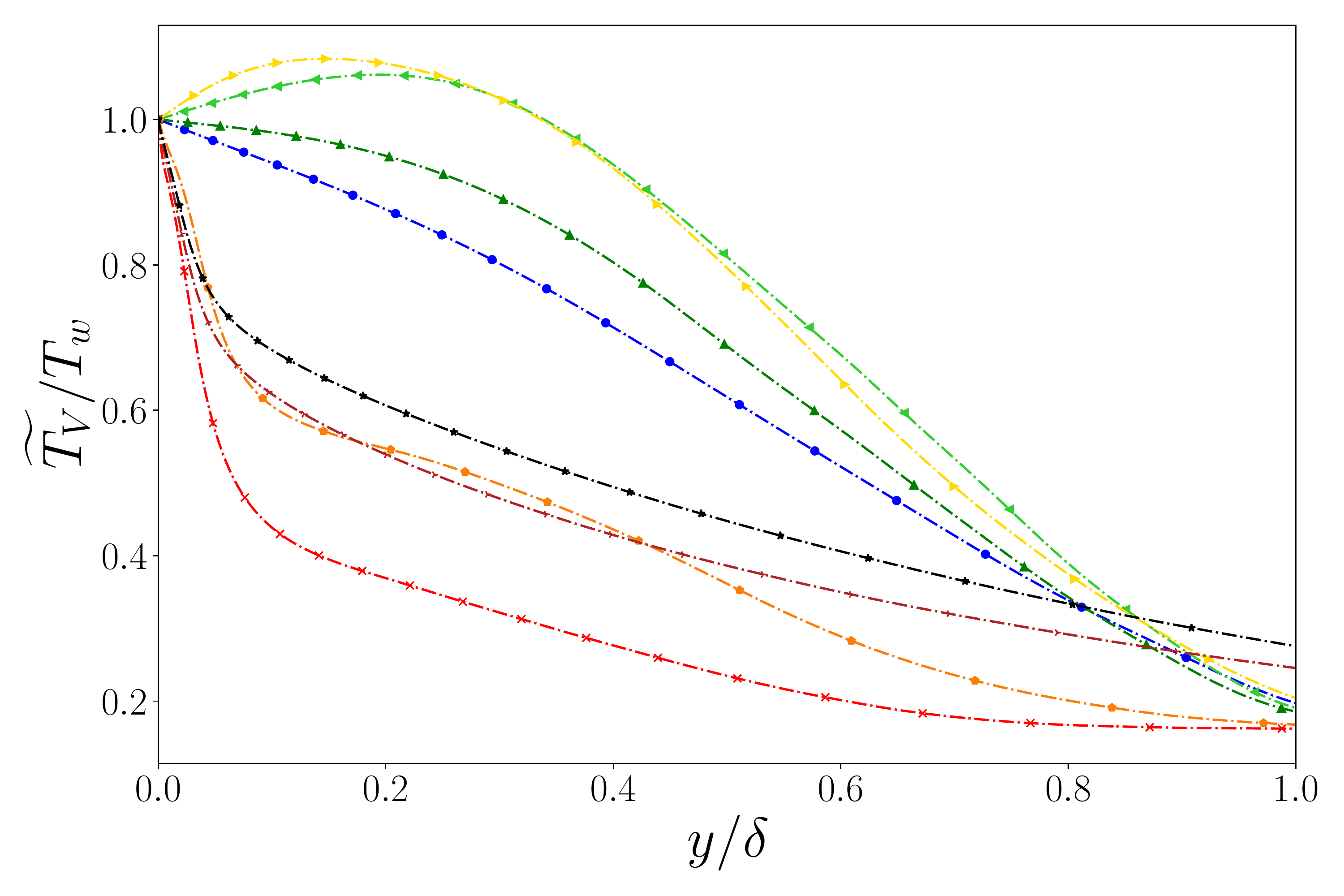}};
   \begin{scope}[x={(a.south east)},y={(a.north west)}]
     \node [align=center] at (0.03,0.95) {(b)};
       \end{scope}
 \end{tikzpicture}

  \begin{tikzpicture}
   \node[anchor=south west,inner sep=0] (a) at (0,0) {\includegraphics[width=0.49\textwidth]{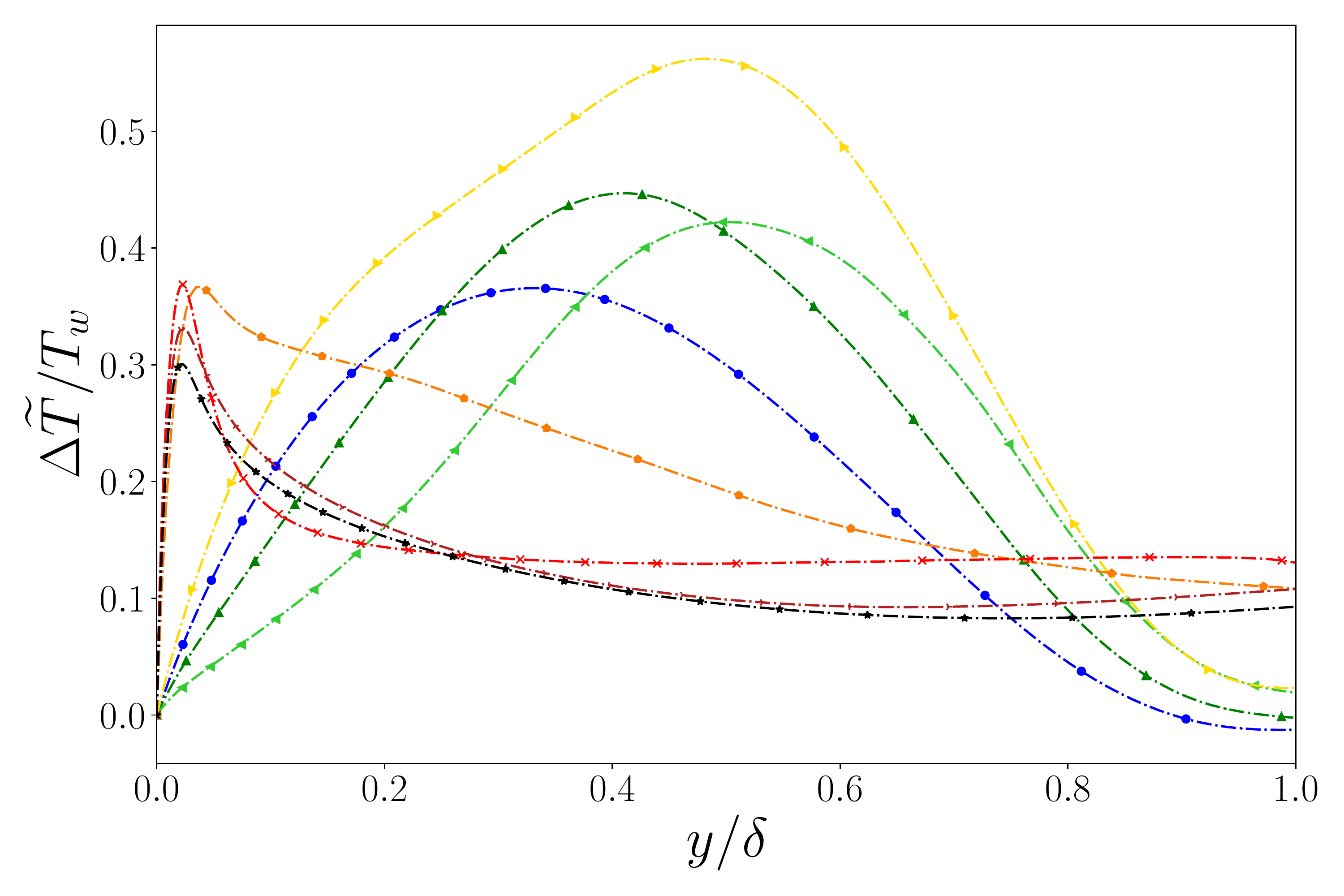}};
   \begin{scope}[x={(a.south east)},y={(a.north west)}]
     \node [align=center] at (0.03,0.95) {(c)};
       \end{scope}
 \end{tikzpicture}
 \caption{Evolution of normalized mean translational temperature (a), vibrational temperature (b) and temperatures difference (c) in outer scaling.}\label{fig:mean_temperatures}
 \end{figure}
Figure~\ref{fig:deltaT} reports the isocontours of the Favre-averaged dimensional temperature difference $\Delta \widetilde{T} = \widetilde{T} - \widetilde{T_V}$. The amount of thermal non-equilibrium is extremely high along the entire boundary layer, with maximum temperature differences larger than \SI{1000}{K} registered in the interaction zone. The non-equilibrium state is mostly vibrationally under-excited, with $\widetilde{T}_V$ lagging behind $\widetilde{T}$ almost everywhere apart from the pre-shock freestream region. The two small over-excited flow pockets inside the recirculation bubble shown in figure~\ref{fig:base_flow} for the base flow disappear, albeit small $\Delta T<0$ values are locally visible in the instantaneous flow field. To get further insights on the streamwise evolution of the thermal non-equilibrium, figure~\ref{fig:mean_temperatures} displays the wall-normal profiles of $\widetilde{T}$, $\widetilde{T_V}$ and $\Delta \widetilde{T}$ in panels (a), (b) and (c), respectively, at the selected stations.
Upstream of the impingement station, the profiles of the two temperatures keep distinct values due to the low Reynolds numbers. Owing to the presence of a cooled wall, the translational temperature always exhibits non-monotonic distributions. On the other hand, the profiles of $\widetilde{T}_V$ display a monotonic profile with a maximum at the wall which is therefore vibrationally heating the flow, coherently with the vibrational wall heat flux evolution. Such a behaviour is partially reversed in the bubble region, where the lower flow speed promote a temperature increase up to values higher than $T_w$. The largest absolute temperatures (as well as their difference) are indeed registered at the reattachment point, with peaks of $\approx \SI{3700}{K}$ and \SI{2700}{K} for $\widetilde{T}$ and $\widetilde{T_V}$, respectively. Here, the gap between the temperatures also reaches a maximum and the wall-normal location of its peak is the farthest from the wall ($y/\delta \approx 0.48$, compared to $y/\delta \approx 0.32$ in the laminar region). From this station on, the peak is rapidly shifted towards the wall in the range $0.02 < y/\delta < 0.03$, due to the sudden decrease of the boundary layer thickness before, and the increase in turbulent activity after. In the last two stations, the turbulent mixing efficiently redistribute the gas \cite[as shown in][]{passiatore2022thermochemical} such that the relaxation towards equilibrium of the vibrational modes is strongly delayed, resulting in a profoundly different dynamics with respect to the base flow predictions \cite{passiatore2022high}.
\begin{figure}
 \centering
     \begin{tikzpicture}
   \node[anchor=south west,inner sep=0] (image) at (0,0) {
   \includegraphics[width=0.49\textwidth,trim={0 0 0 0},clip]{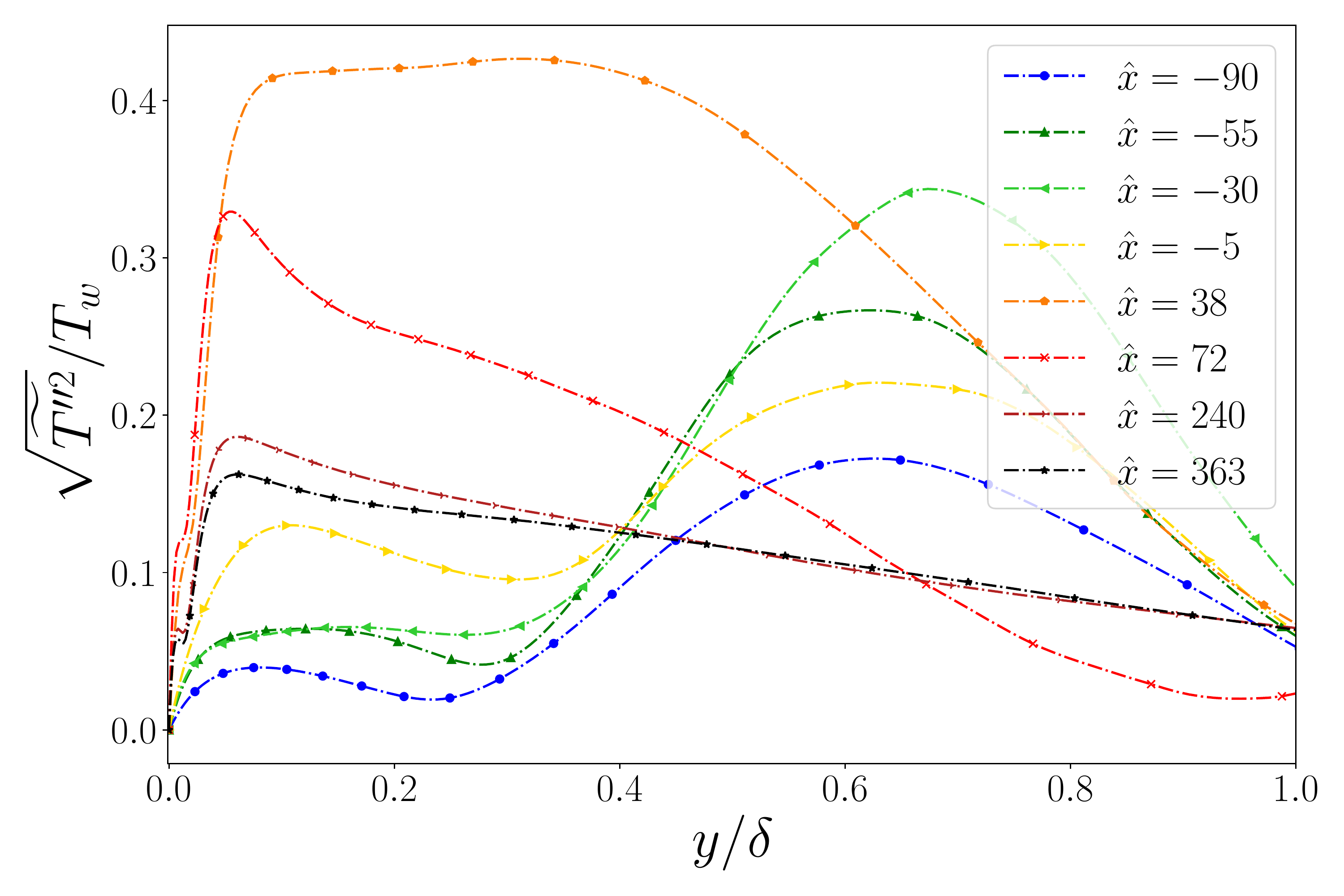}};
   \begin{scope}[x={(image.south east)},y={(image.north west)}]
    \node [align=center] at (0.03,0.95) {(a)};
   \end{scope}
  \end{tikzpicture}
  \begin{tikzpicture}
   \node[anchor=south west,inner sep=0] (image) at (0,0) {
   \includegraphics[width=0.49\textwidth,trim={0 0 0 0},clip]{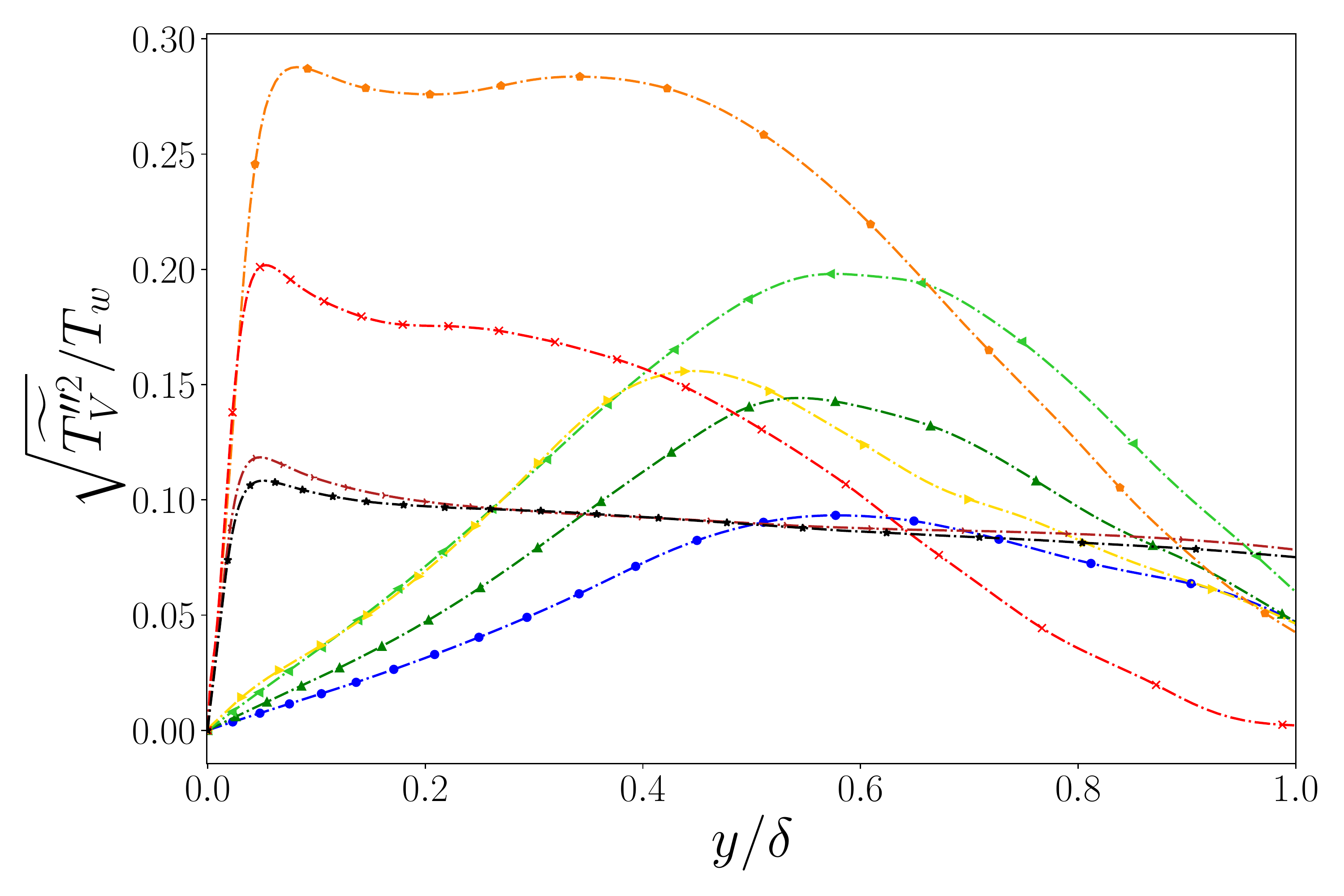}};
   \begin{scope}[x={(image.south east)},y={(image.north west)}]
    \node [align=center] at (0.03,0.95) {(b)};
   \end{scope}
  \end{tikzpicture}

  \begin{tikzpicture}
   \node[anchor=south west,inner sep=0] (image) at (0,0) {
   \includegraphics[width=0.49\textwidth,trim={0 0 0 0},clip]{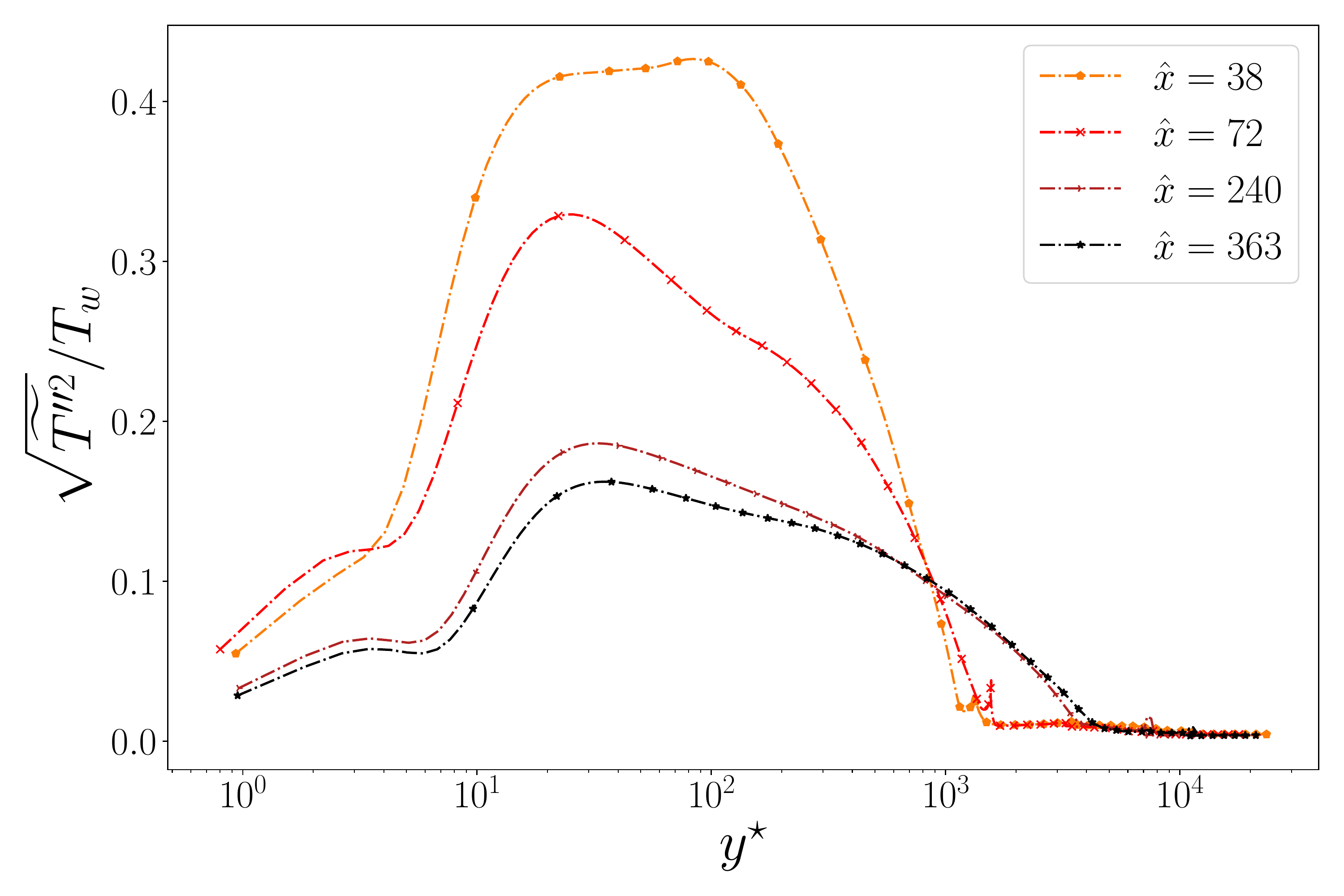}};
   \begin{scope}[x={(image.south east)},y={(image.north west)}]
    \node [align=center] at (0.03,0.95) {(c)};
    \end{scope}
  \end{tikzpicture}
  \begin{tikzpicture}
   \node[anchor=south west,inner sep=0] (image) at (0,0) {
   \includegraphics[width=0.49\textwidth,trim={0 0 0 0},clip]{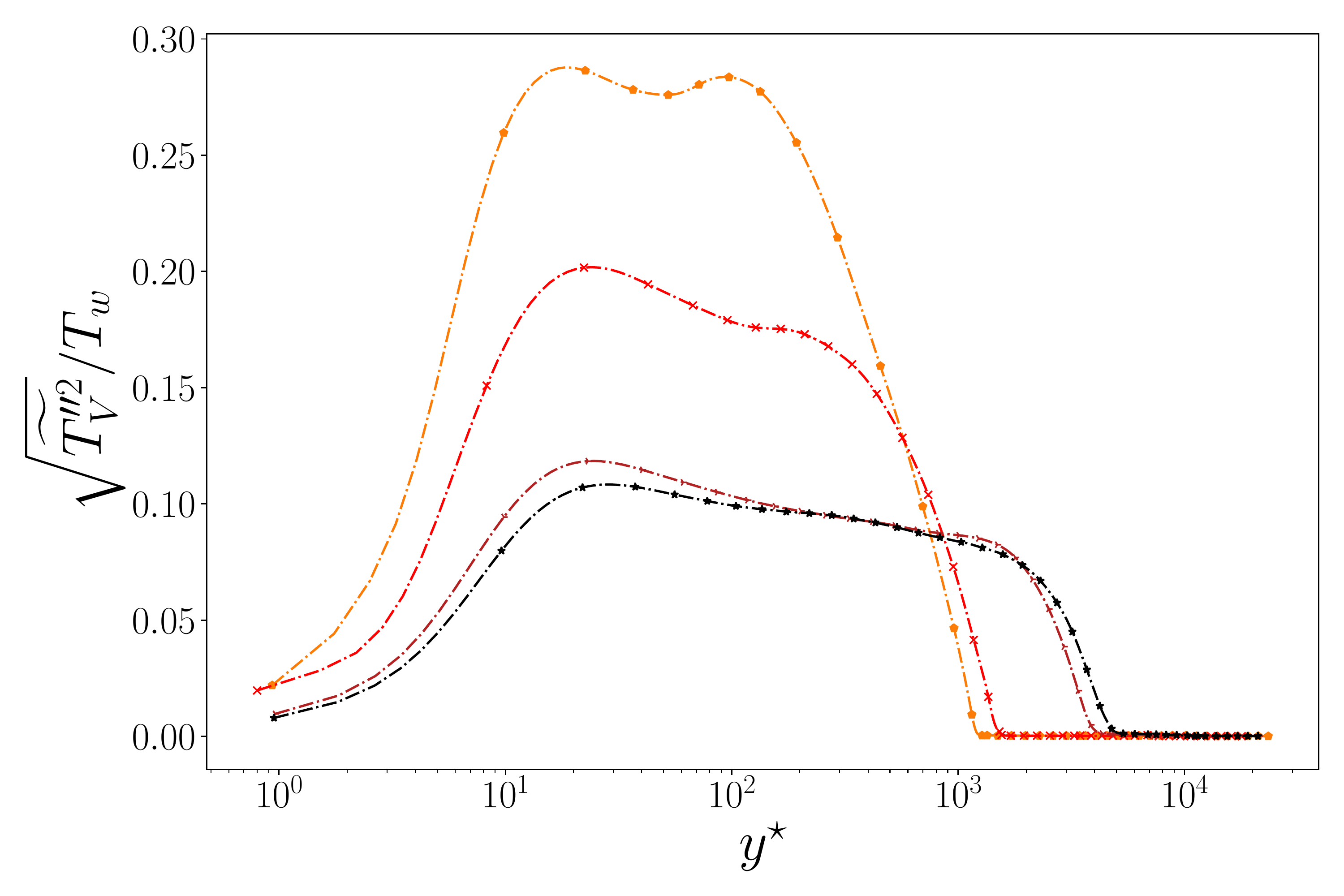}};
   \begin{scope}[x={(image.south east)},y={(image.north west)}]
    \node [align=center] at (0.03,0.95) {(d)};
   \end{scope}
  \end{tikzpicture}
 \caption{Evolution of translational temperatures fluctuations (a-c) and vibrational temperatures fluctuations (b-d), in outer scaling (a-b) and inner scaling (c-d).}
\label{fig:temp_rms}
\end{figure}

\begin{figure}
\centering
 \begin{tikzpicture}
   \node[anchor=south west,inner sep=0] (a) at (0,0) {\includegraphics[width=0.49\textwidth]{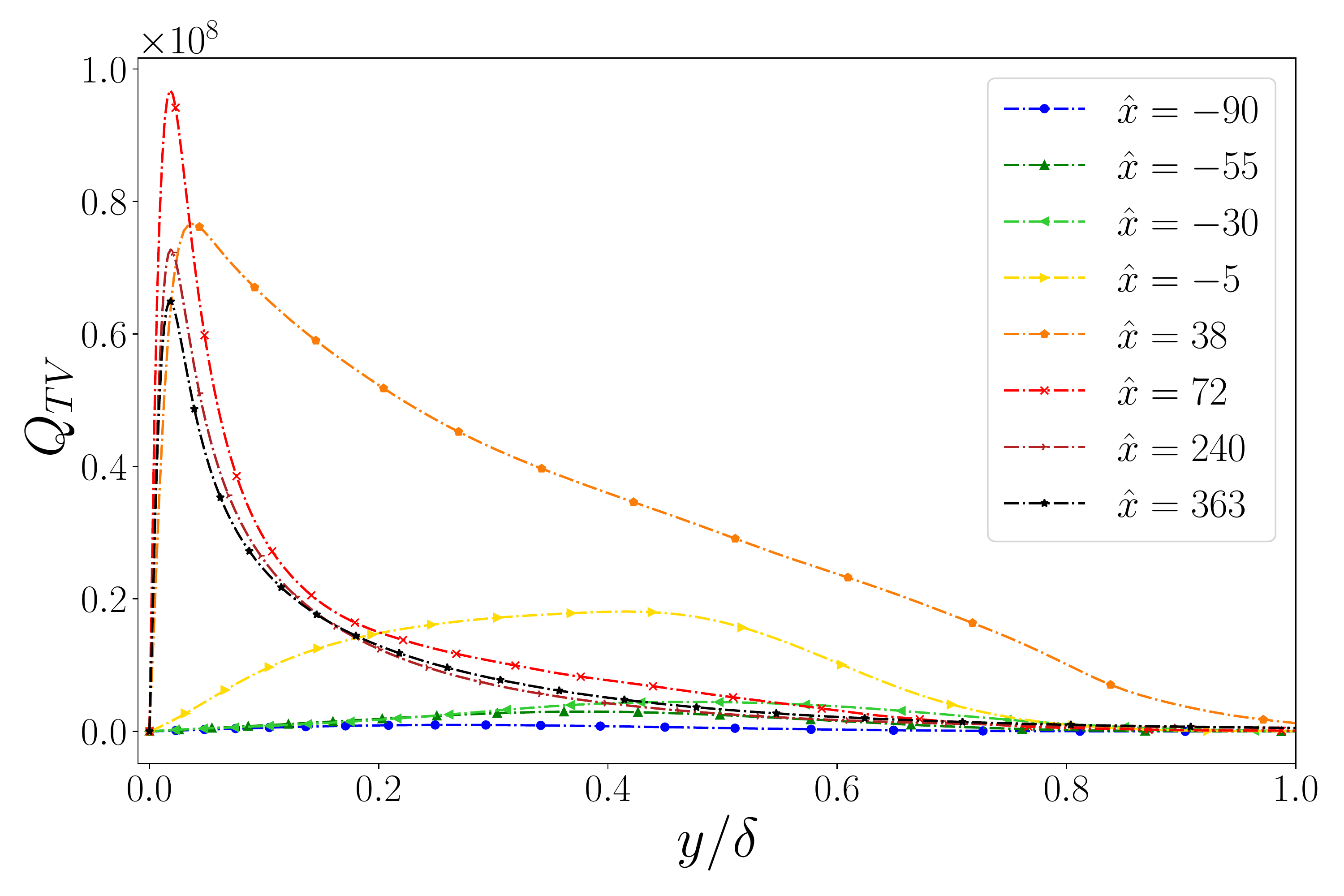}};
   \begin{scope}[x={(a.south east)},y={(a.north west)}] 
     \node [align=center] at (0.03,0.95) {(a)};
     \end{scope}
 \end{tikzpicture}
  \begin{tikzpicture}
   \node[anchor=south west,inner sep=0] (a) at (0,0) {\includegraphics[width=0.49\textwidth]{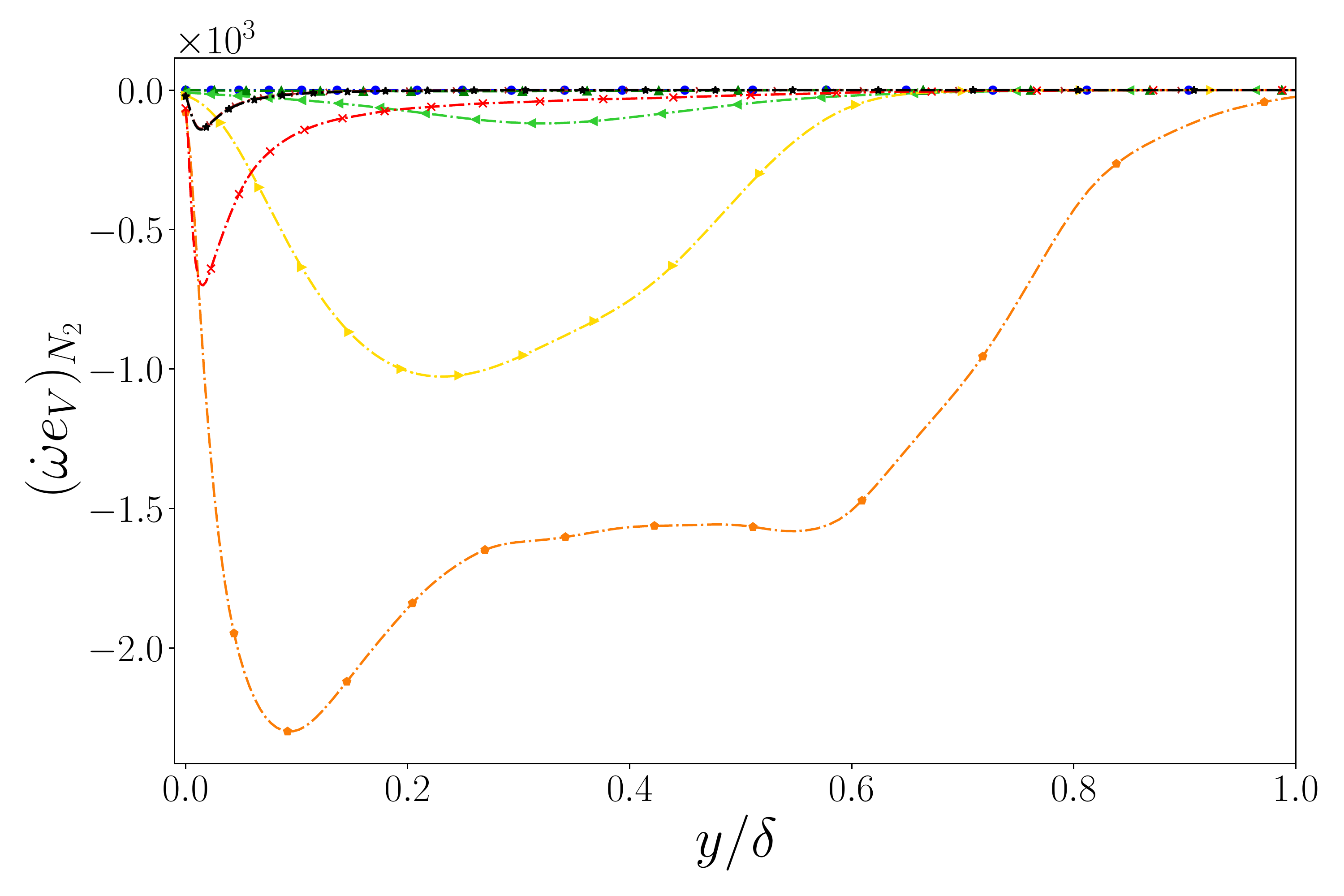}};
   \begin{scope}[x={(a.south east)},y={(a.north west)}]
     \node [align=center] at (0.03,0.95) {(b)};
       \end{scope}
 \end{tikzpicture}

   \begin{tikzpicture}
   \node[anchor=south west,inner sep=0] (a) at (0,0) {\includegraphics[width=0.49\textwidth]{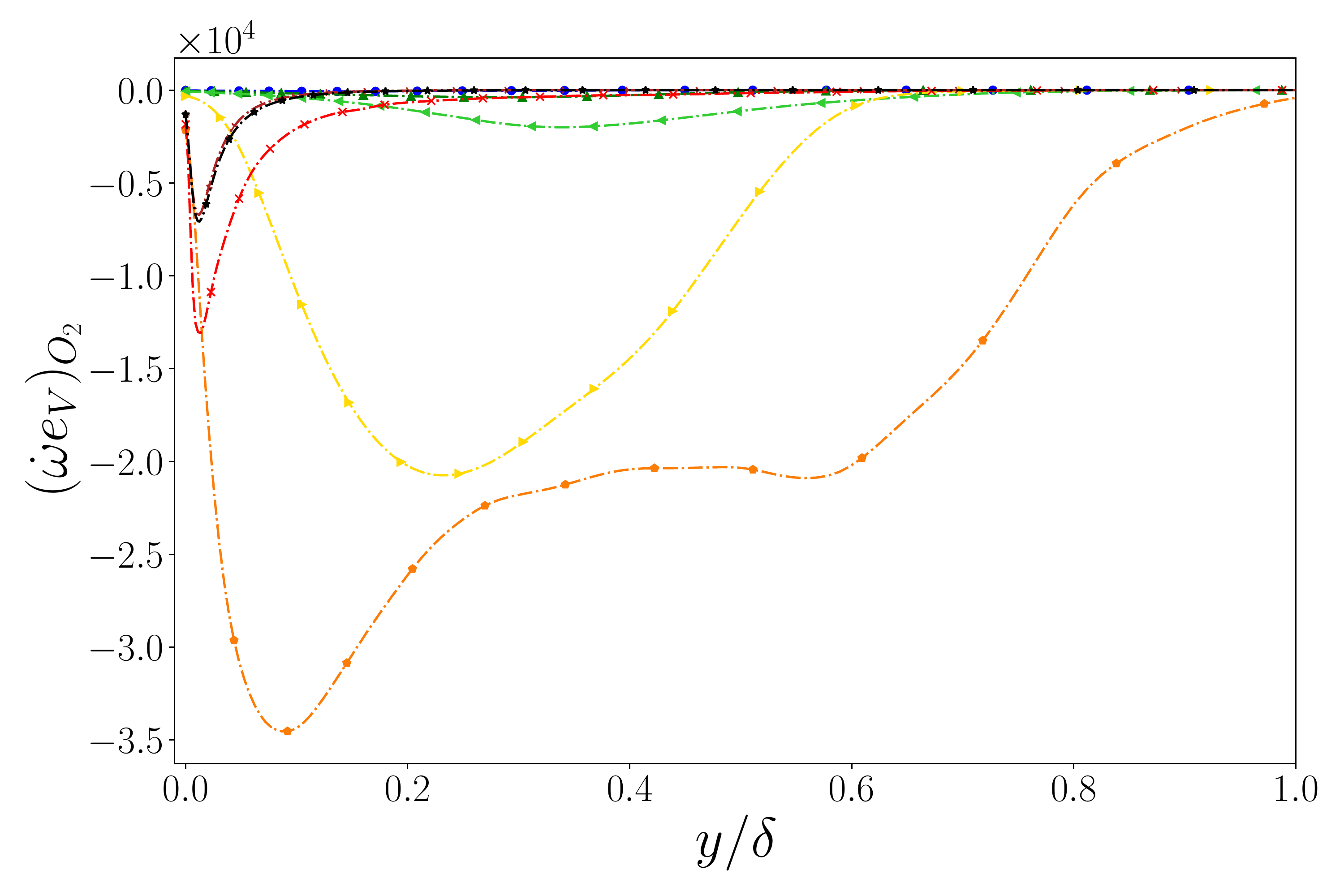}};
   \begin{scope}[x={(a.south east)},y={(a.north west)}]
     \node [align=center] at (0.03,0.95) {(c)};
       \end{scope}
 \end{tikzpicture}
   \begin{tikzpicture}
   \node[anchor=south west,inner sep=0] (a) at (0,0) {\includegraphics[width=0.49\textwidth]{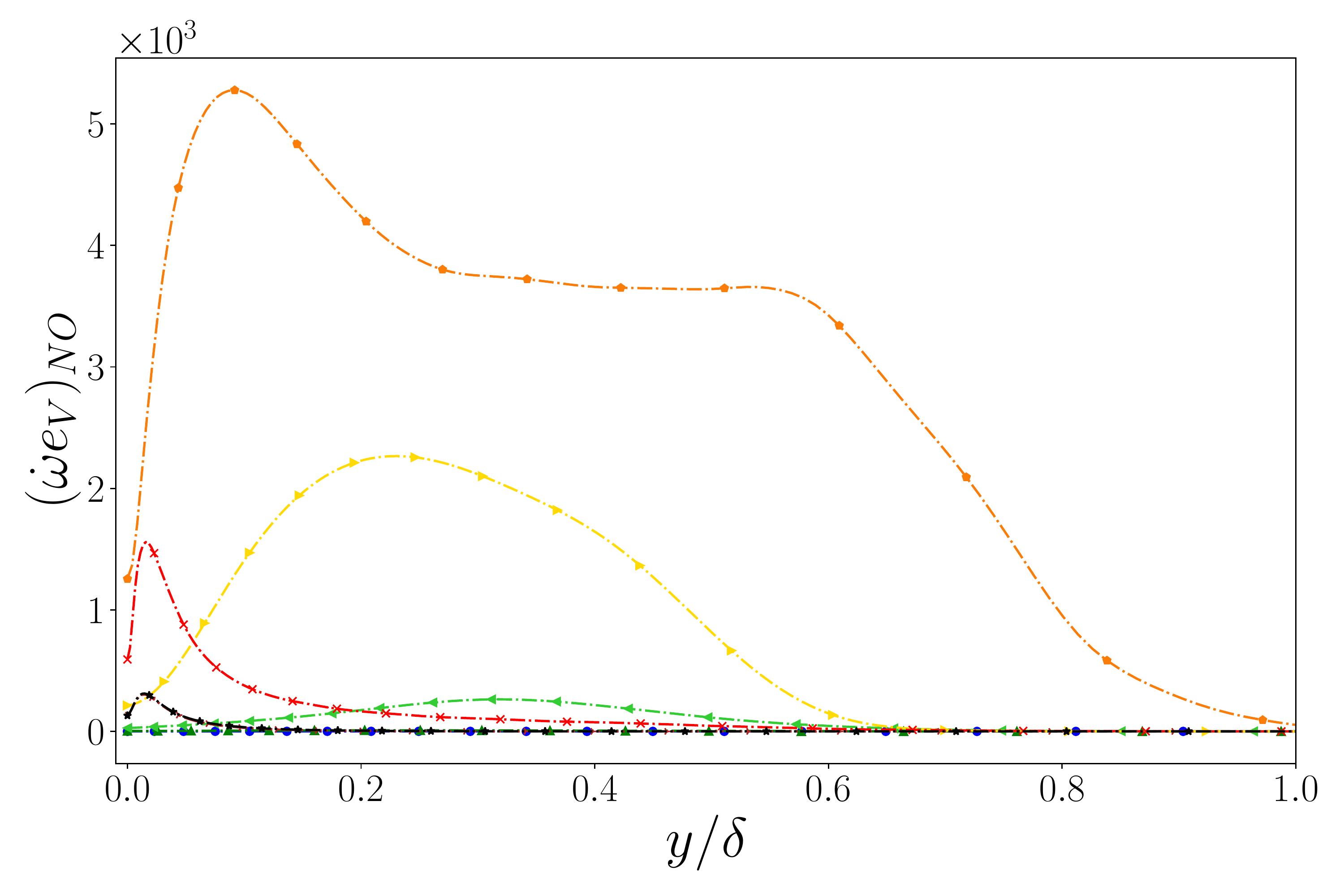}};
   \begin{scope}[x={(a.south east)},y={(a.north west)}]
     \node [align=center] at (0.03,0.95) {(d)};
       \end{scope}
 \end{tikzpicture}
  \centering
 \caption{Evolution of vibrational source terms. Translational-vibrational energy exchange (a), chemical source term of N$_2$ (b), chemical source term of O$_2$ (c), chemical source term of NO (d). The quantities are dimensional, in $\SI{}{W/m^3}$.}\label{fig:vibr_source_terms}
 \end{figure}

The evolution of the root mean square of the Favre fluctuations of rototranslational and vibrational temperatures is shown in figure~\ref{fig:temp_rms} for all the stations in outer scaling, and in inner scaling for the streamwise position after the interaction region. The level of turbulent fluctuations is high even before the shock impingement, especially for the translational temperature. The profiles of $\sqrt{\widetilde{T''^2}}$ in the fully turbulent region display two peaks, that are a local maximum at $y^\star < 10$ and a global maximum at $y^\star \approx 30$. These two peaks are more distinguishable when the temperature fluctuations are normalized with the corresponding Favre average \cite{direnzo2021direct,passiatore2022thermochemical}. The r.m.s. of the vibrational temperature present a radically different behavior, due to the particular trend of $\widetilde{T_V}$. The inner maxima registered for the rototranslational temperature at all the selected stations are smeared out for the vibrational one; moreover, vibrational fluctuations are less intense than their translational counterparts everywhere. This is a direct consequence of the under-excited non-equilibrium persisting throughout the boundary layer, with $T_V$ values lagging behind $T$ ones.
\begin{figure}
\centering
 \begin{tikzpicture}
   \node[anchor=south west,inner sep=0] (a) at (0,0) {\includegraphics[width=0.49\textwidth]{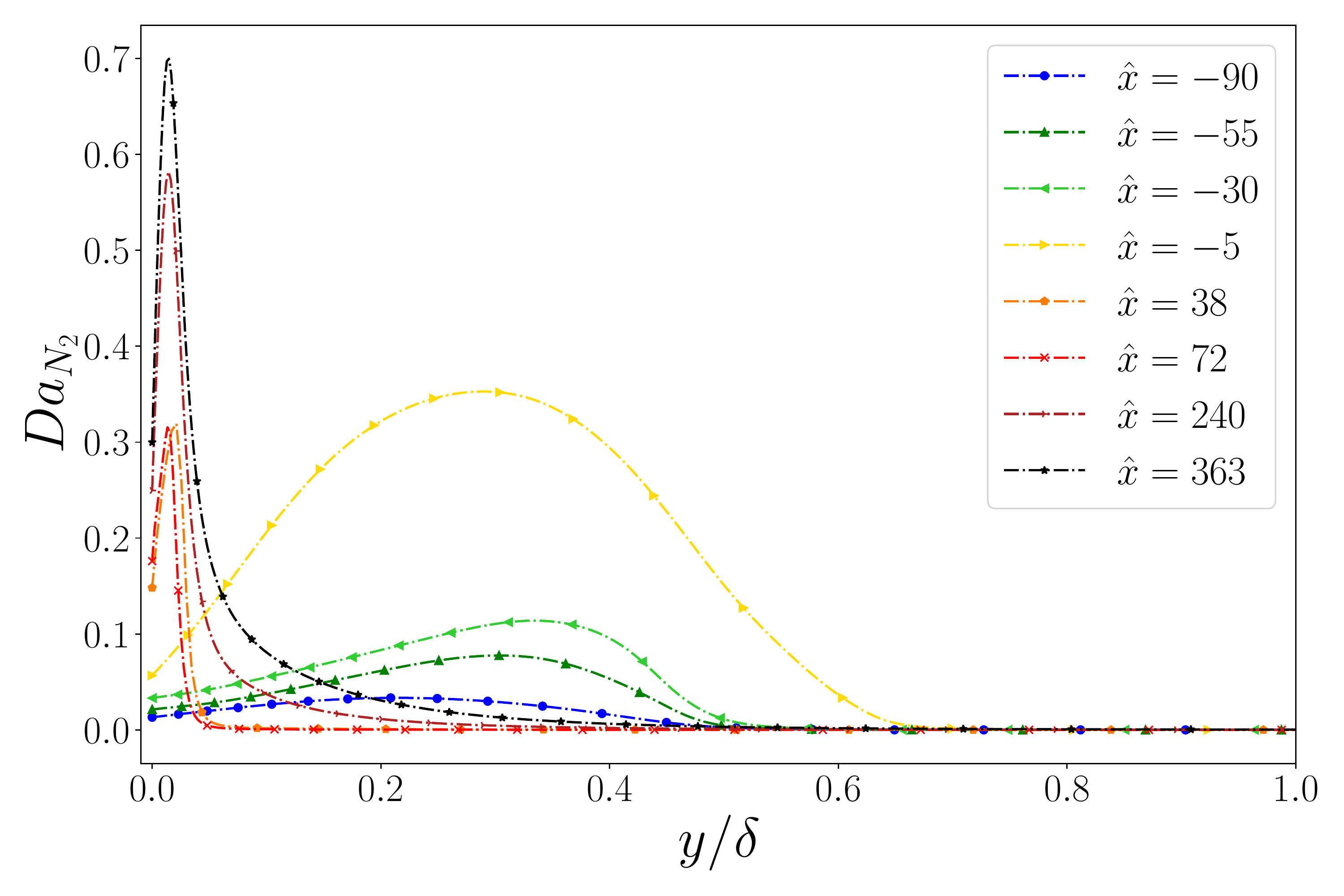}};
   \begin{scope}[x={(a.south east)},y={(a.north west)}]
     \node [align=center] at (0.03,0.95) {(a)};
       \end{scope}
 \end{tikzpicture}
  \begin{tikzpicture}
   \node[anchor=south west,inner sep=0] (a) at (0,0) {\includegraphics[width=0.49\textwidth]{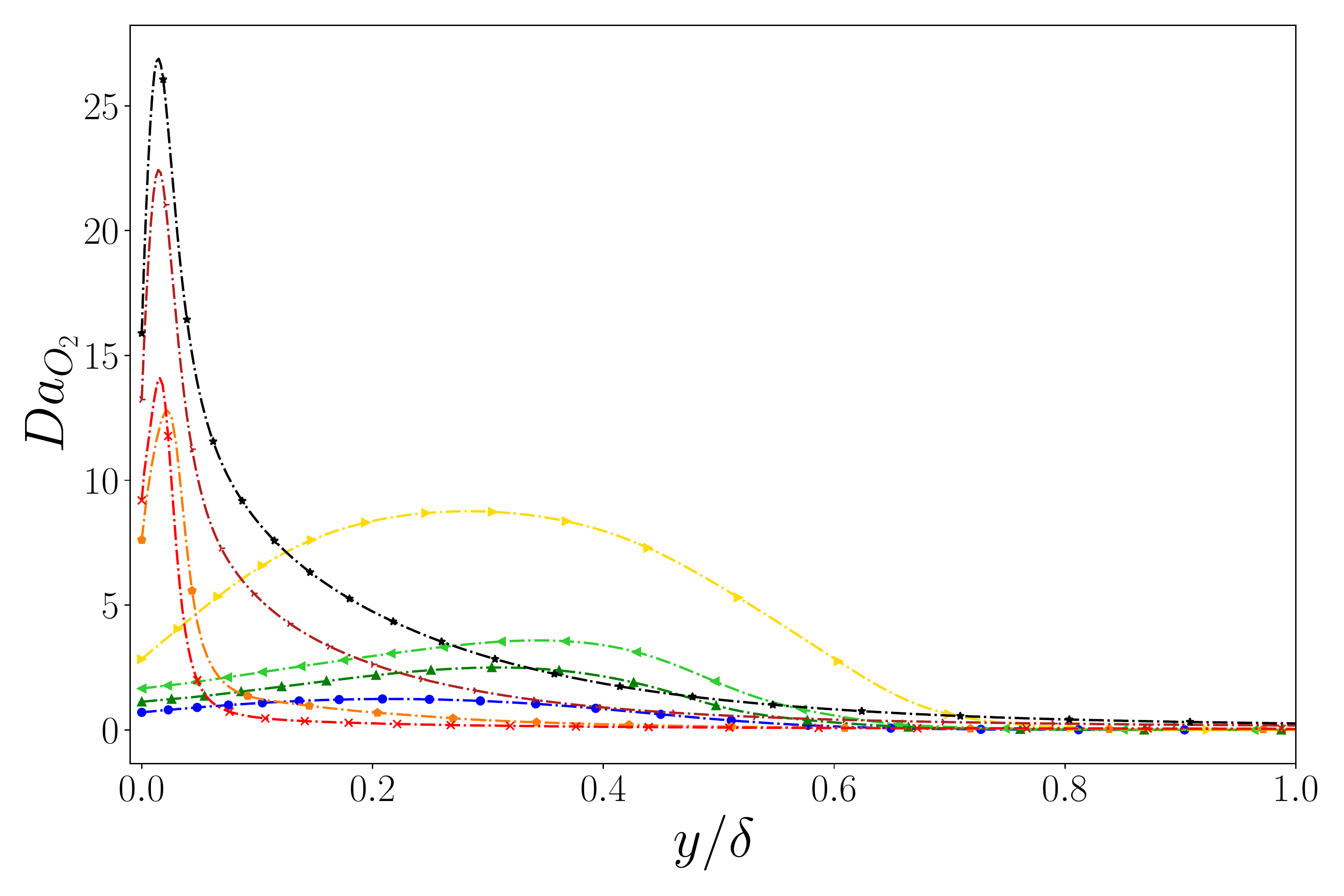}};
   \begin{scope}[x={(a.south east)},y={(a.north west)}]
     \node [align=center] at (0.03,0.95) {(b)};
    \end{scope}
 \end{tikzpicture}

 \begin{tikzpicture}
   \node[anchor=south west,inner sep=0] (a) at (0,0) {\includegraphics[width=0.49\textwidth]{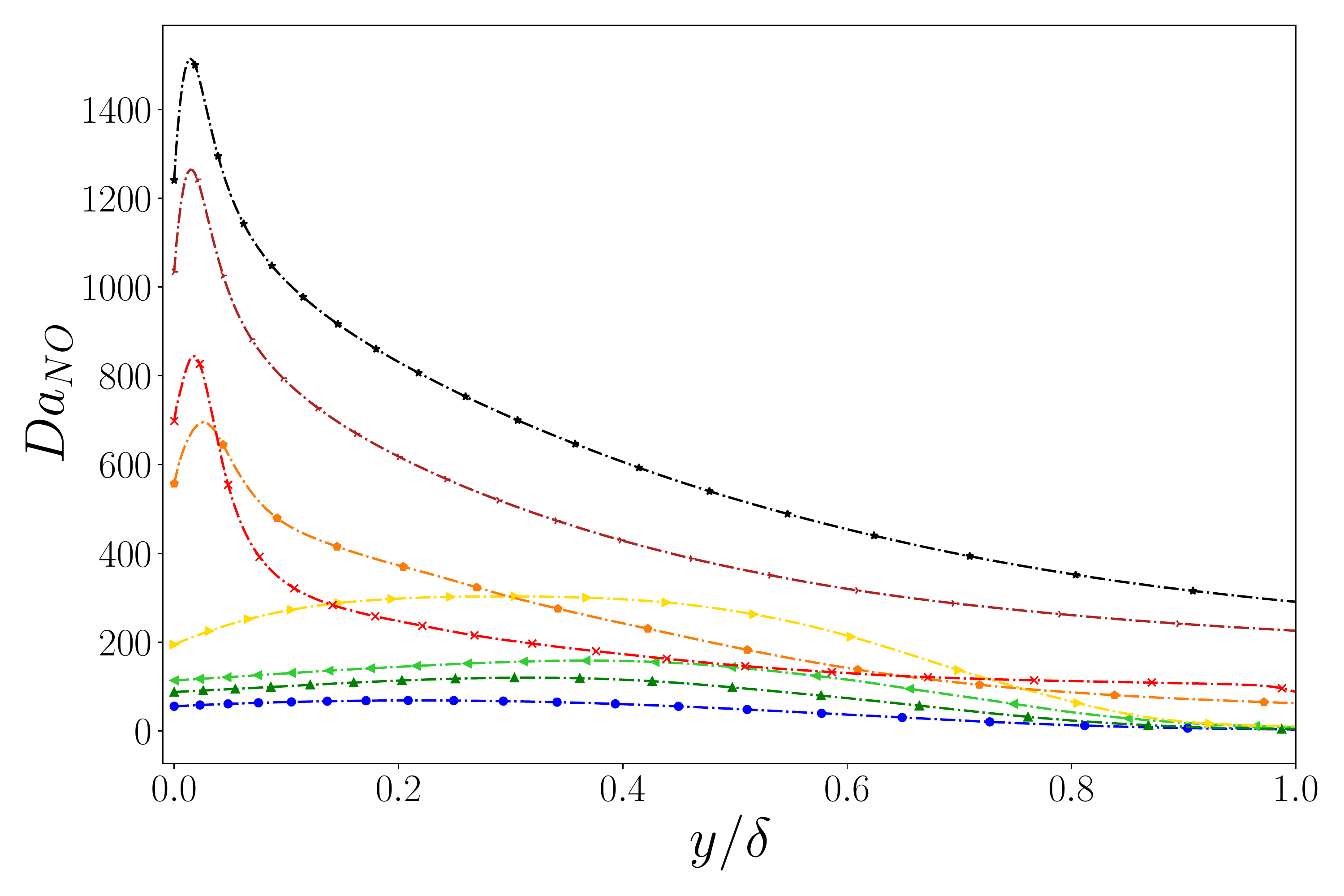}};
   \begin{scope}[x={(a.south east)},y={(a.north west)}]
     \node [align=center] at (0.03,0.95) {(c)};
    \end{scope}
 \end{tikzpicture}
  \centering
 \caption{Wall-normal evolution of vibrational Damk{\"o}hler numbers.}\label{fig:vibr_damkholer}
\end{figure}
The severe thermal non-equilibrium state is confirmed by inspection of the vibrational Damk{\"o}hler numbers computed with respect to the flow residence time, i.e.:
\begin{equation}
Da_m = \frac{x/u_\delta}{t_m}
\end{equation}
and shown in figure~\ref{fig:vibr_damkholer} for the three molecules of the mixture. While vibrational equilibrium is instantly reached for nitric oxide, molecular nitrogen and oxygen are shown to achieve non-equilibrium conditions in different regions. Specifically, $Da_{\text{O}_2}$ displays values of order unity in the outer part of the boundary layer ($0.3 < y/\delta < 0.5$), from the laminar region up to the recirculation bubble. The peaks are shifted towards the wall in the transitional region, and afterwards N$_2$ enters a non-equilibrium state with $Da_{\text{N}_2}$ values approaching unity towards the end of the computational domain.
Figure~\ref{fig:vibr_source_terms} reports the two contributions to the vibrational energy variation, namely, the translational-vibrational energy exchange term $Q_{TV}$ (panel a) and the variation due to chemical production and depletion $\dot{\omega}_m e_{Vm}$ (panels b-d). An order-of-magnitude analysis shows that $e_V$ variations are almost uniquely due to translational-vibrational energy exchanges, which are particularly large in the transitional and fully turbulent regions. A high exchange rate is registered across the whole boundary layer thickness at $\hat{x} = 38$, which corresponds to the skin friction coefficient peak location and to the emergence of  the energetic structures observed in figure~\ref{fig:wall_inst}. Here, the maximum interaction with the chemical activity is also registered, its net effect being however negligible.

\begin{figure}
\centering
 \begin{tikzpicture}
   \node[anchor=south west,inner sep=0] (a) at (0,0) {\includegraphics[width=0.49\textwidth]{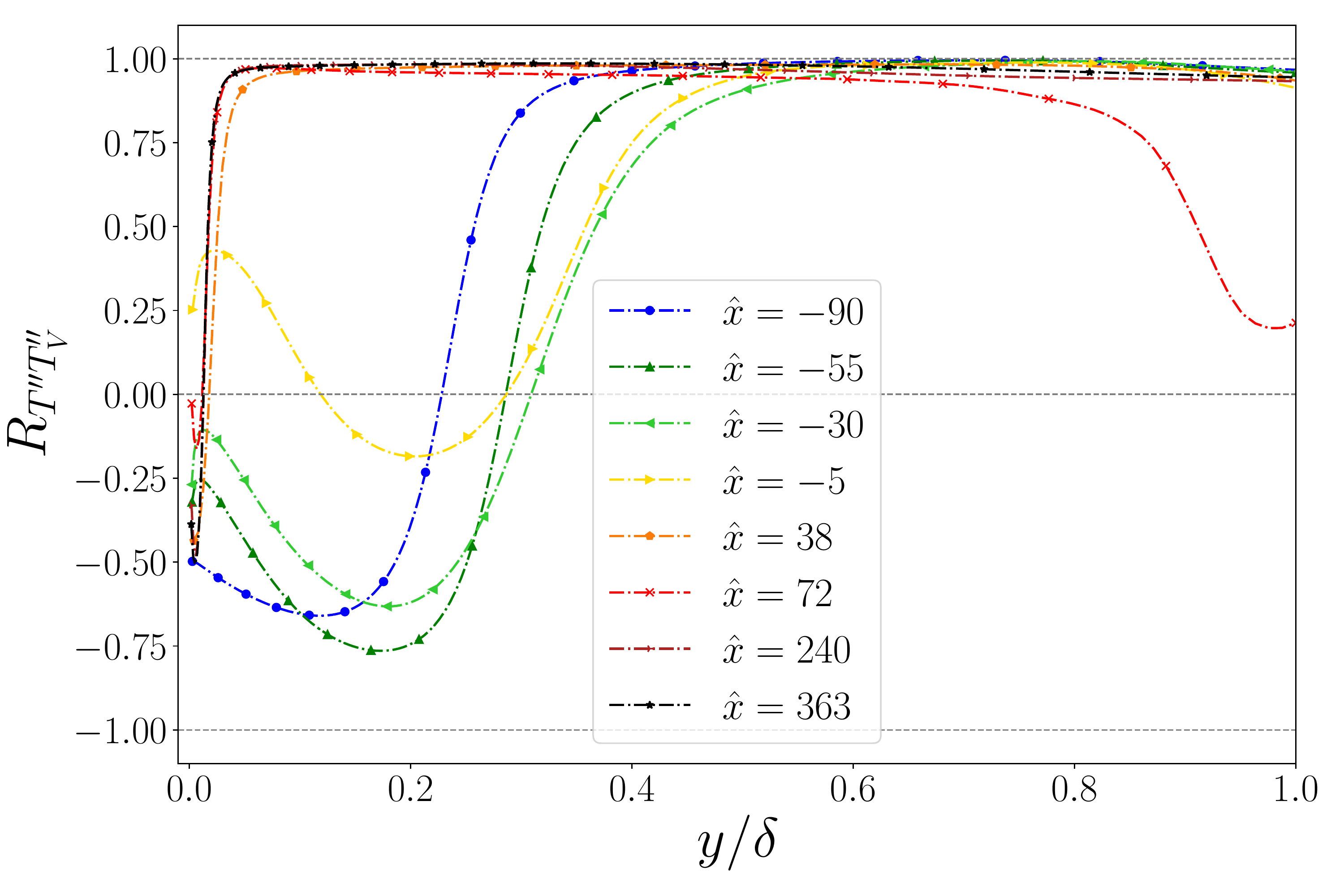}};
   \begin{scope}[x={(a.south east)},y={(a.north west)}]
     \node [align=center] at (0.03,0.95) {(a)};
   \end{scope}
 \end{tikzpicture}

  \centering
   \begin{tikzpicture}
   \node[anchor=south west,inner sep=0] (a) at (0,0) {\includegraphics[width=0.49\textwidth]{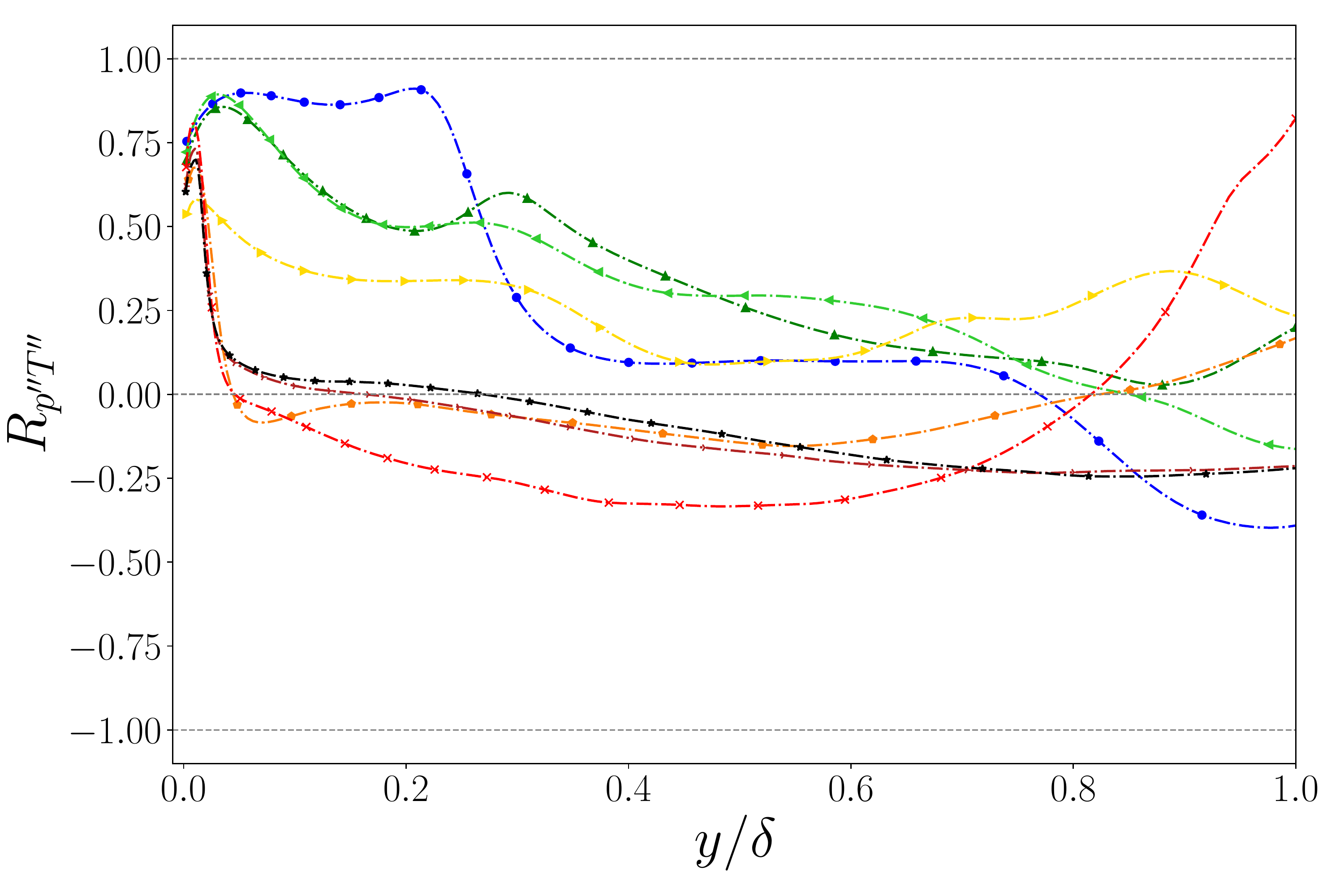}};
   \begin{scope}[x={(a.south east)},y={(a.north west)}]
     \node [align=center] at (0.03,0.95) {(b)};
   \end{scope}
 \end{tikzpicture}
    \begin{tikzpicture}
   \node[anchor=south west,inner sep=0] (a) at (0,0) {\includegraphics[width=0.49\textwidth]{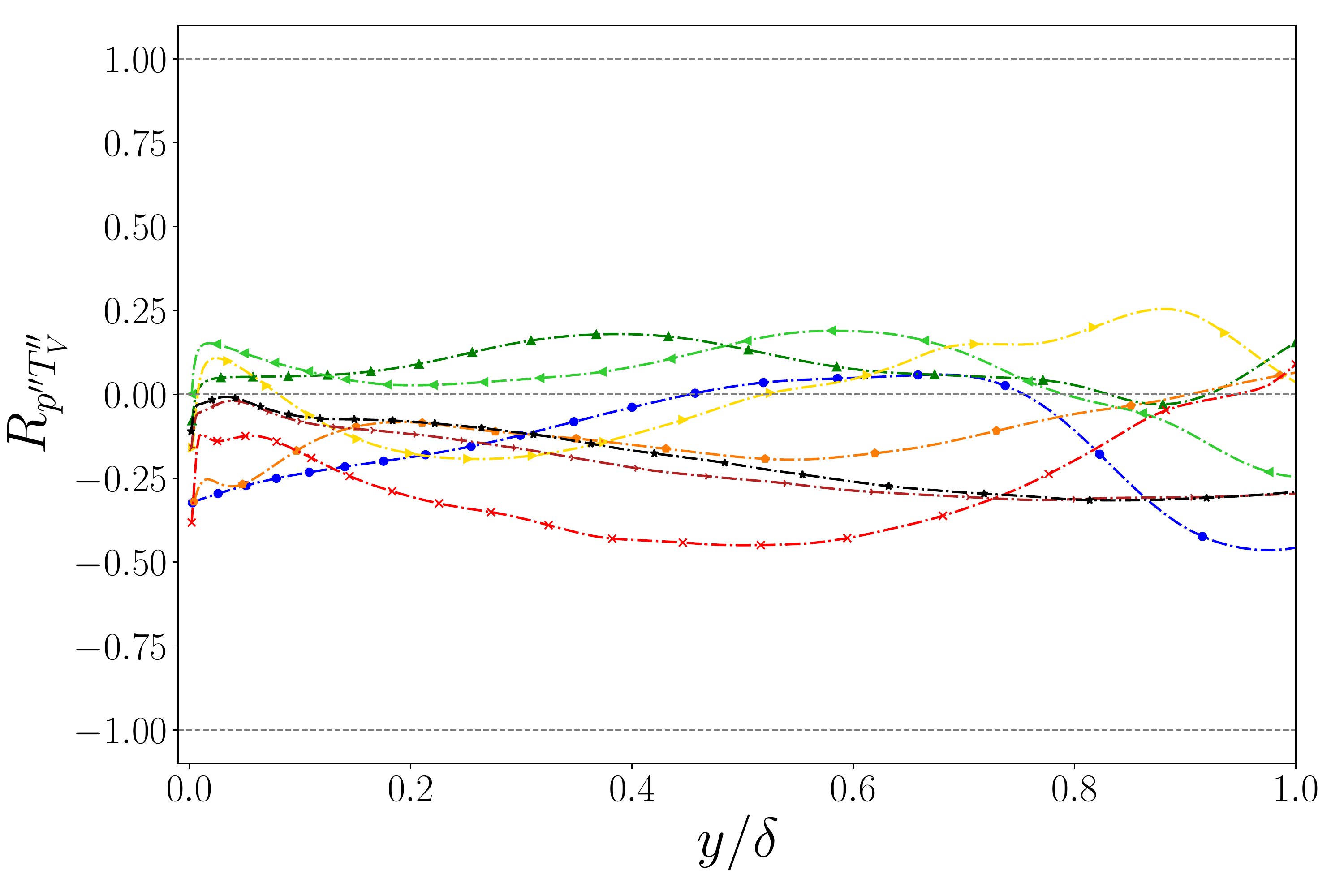}};
   \begin{scope}[x={(a.south east)},y={(a.north west)}]
     \node [align=center] at (0.03,0.95) {(c)};
   \end{scope}
 \end{tikzpicture}

  \centering
  \begin{tikzpicture}
   \node[anchor=south west,inner sep=0] (a) at (0,0) {\includegraphics[width=0.49\textwidth]{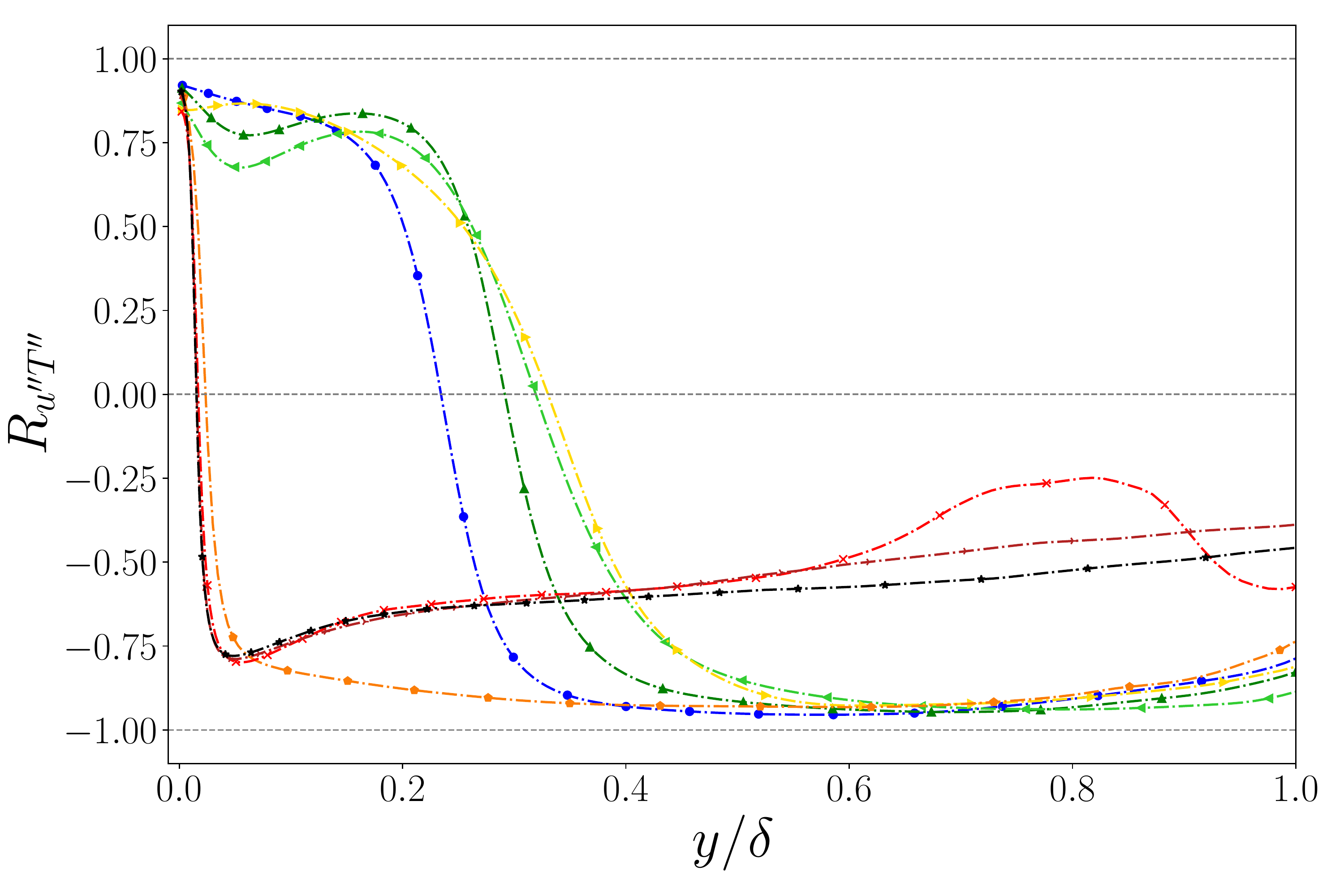}};
   \begin{scope}[x={(a.south east)},y={(a.north west)}]
     \node [align=center] at (0.03,0.95) {(d)};
   \end{scope}
 \end{tikzpicture}
 \begin{tikzpicture}
   \node[anchor=south west,inner sep=0] (a) at (0,0) {\includegraphics[width=0.49\textwidth]{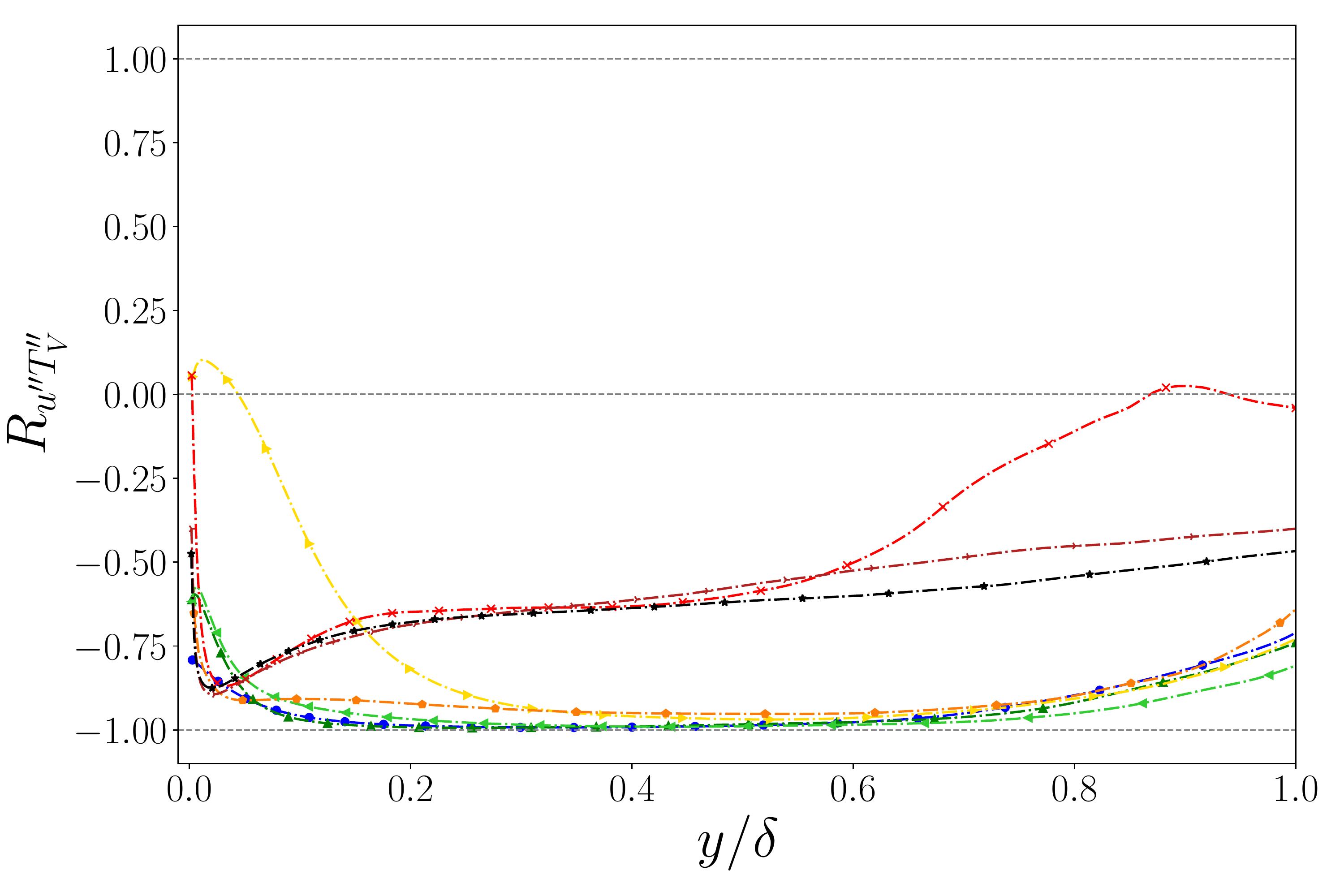}};
   \begin{scope}[x={(a.south east)},y={(a.north west)}]
     \node [align=center] at (0.03,0.95) {(e)};
   \end{scope}
 \end{tikzpicture}

\centering
 \begin{tikzpicture}
   \node[anchor=south west,inner sep=0] (a) at (0,0) {\includegraphics[width=0.49\textwidth]{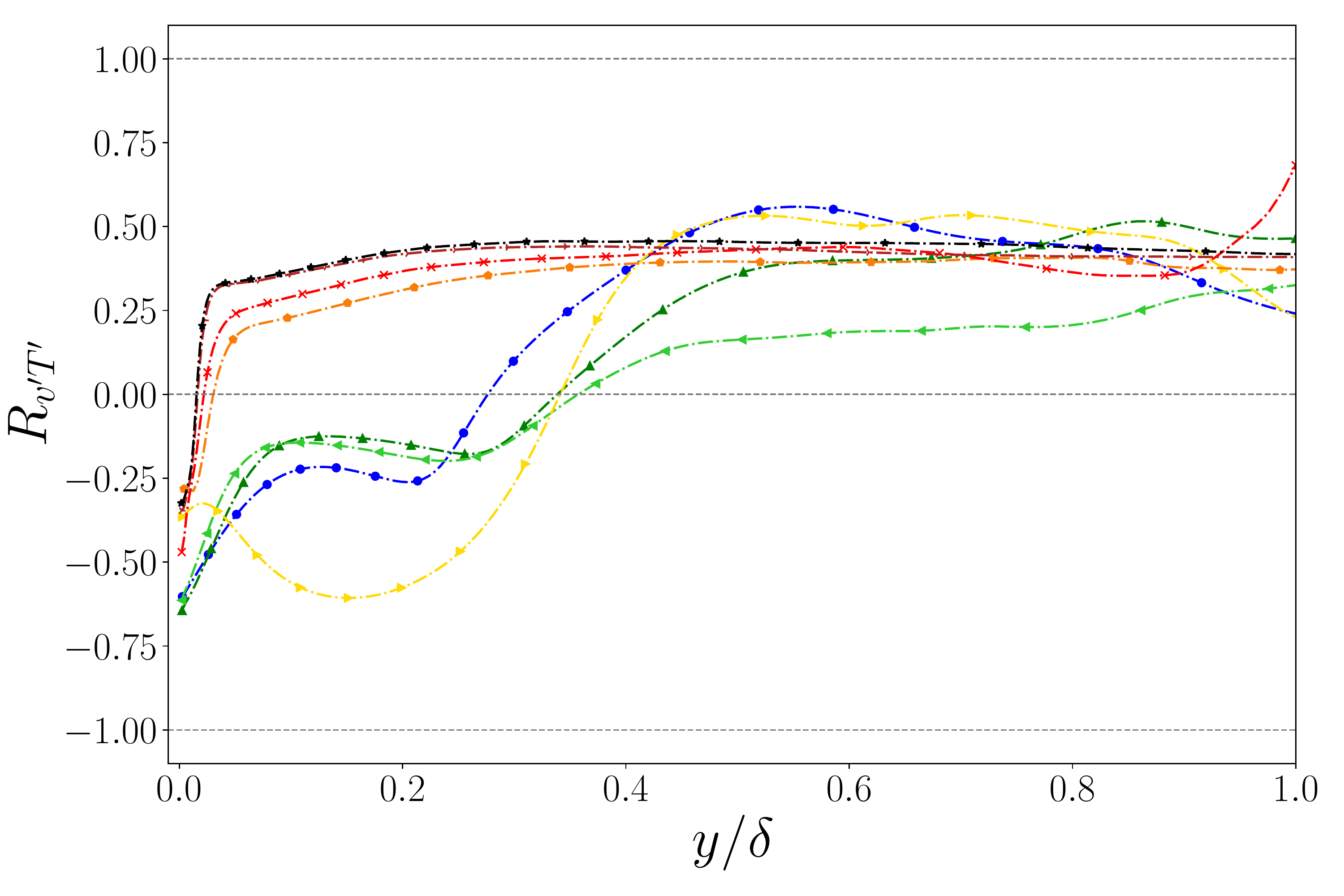}};
   \begin{scope}[x={(a.south east)},y={(a.north west)}]
     \node [align=center] at (0.03,0.95) {(f)};
   \end{scope}
 \end{tikzpicture}
 \begin{tikzpicture}
   \node[anchor=south west,inner sep=0] (a) at (0,0) {\includegraphics[width=0.49\textwidth]{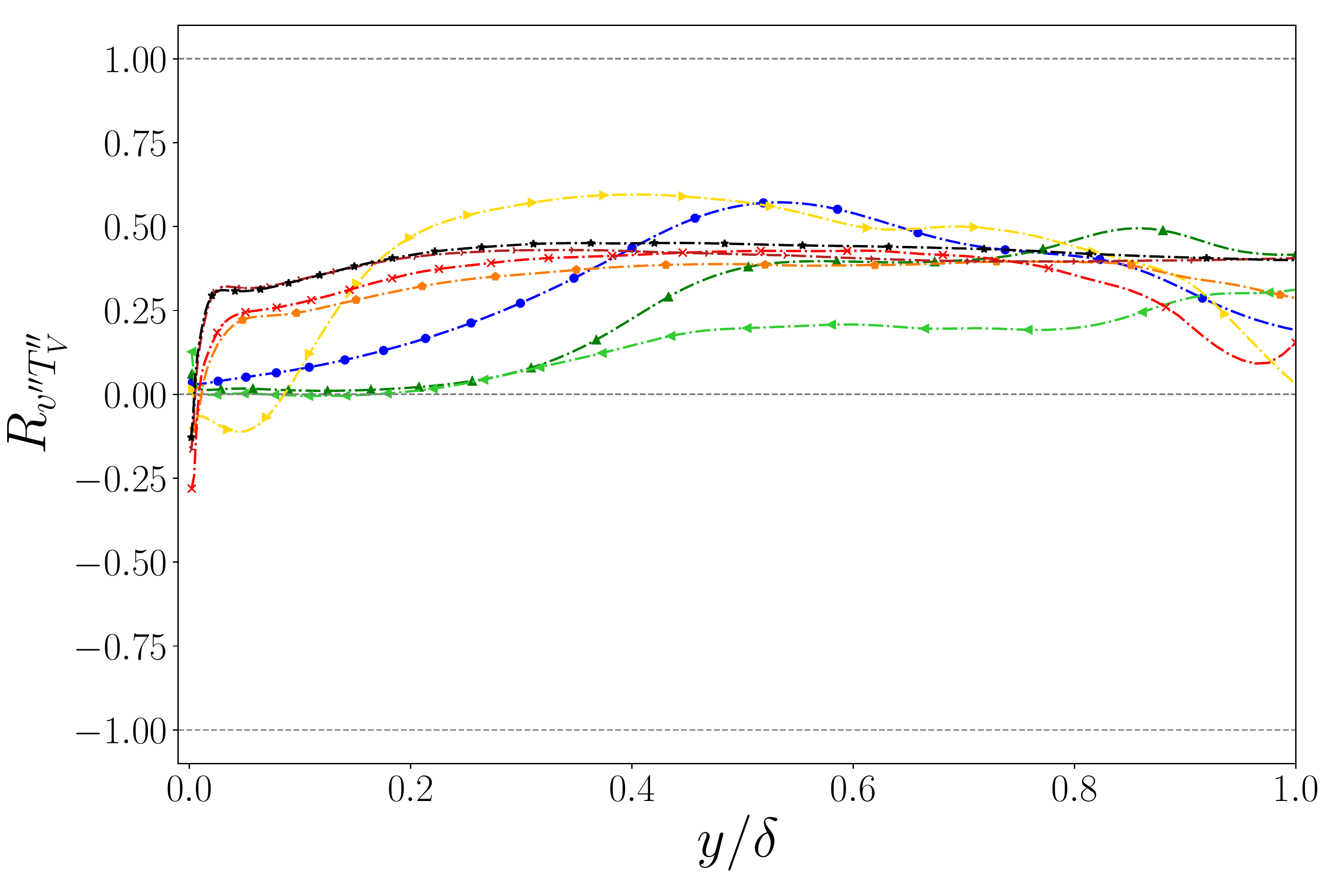}};
   \begin{scope}[x={(a.south east)},y={(a.north west)}]
     \node [align=center] at (0.03,0.95) {(g)};
   \end{scope}
 \end{tikzpicture}
 \caption{Wall-normal profiles of correlation coefficients at the different stations listed in table~\ref{tab:stations}.}\label{fig:correlation}
 \end{figure}

Figure~\ref{fig:correlation} shows correlation coefficients between the Favre-averaged fluctuations of different quantities. First, $T''$ and $T_V''$ are strongly correlated almost everywhere except close to the wall, where a slight anticorrelation is registered notwithstanding the same temperature values applied at the walls. This is again linked to the simultaneous vibrational wall heating and translational wall cooling, as shown by the wall heat fluxes and mean temperature profiles in figures~\ref{fig:wall_heat} and \ref{fig:deltaT}, respectively. The trends of the correlations of $p''$, $u''$ and $v''$ with $T''$ and $T_V''$ are qualitatively similar in the outer region of the boundary layer, while significant variations are observed for the inner zone in particular with $p''$ and $u''$. Specifically, $R_{p''T''}$ is shown to be predominantly positive (albeit in the turbulent region it tends rapidly to zero starting from the buffer layer) whereas $p''$ and $T_V''$ are loosely correlated highlighting the important decoupling between the internal vibrational and dynamic fields. The strong correlation observed for $R_{u''T''}$ in the near-wall region is not observed in the profiles of $R_{u''T_V''}$, which are shown to be strongly anticorrelated everywhere. Globally, $T''$ and $T_V''$ exhibit a drastically different behaviour not only in the turbulent region \cite[as previously shown by][]{passiatore2022thermochemical} but also and above all in the interaction zone, where the amount of thermal nonequilibrium is the largest. Of note, the turbulent Prandtl and vibrational prandtl numbers are $\approx$ 0.85 and 0.9 respectively, consistently with the values obtained under largely different thermodynamic conditions \cite{passiatore2021finite,direnzo2021direct,passiatore2022thermochemical}.

\section{Conclusions}\label{sec:conclusions}
A Mach 9 shock-wave/transitional boundary layer interaction is investigated for the first time by means of direct numerical simulations. The present thermodynamic conditions are such that the boundary layer is thermally and mildly chemically out-of-equilibrium. The impinging shock is generated by a wedge angle of $\vartheta \approx 5 ^\circ$, leading to a shock angle of $\beta = 10^\circ$; the resulting base flow is then perturbed with freestream inflow disturbances that are found to feed the classical Mack mode instability. Rope-like structures and 2D modes are indeed recognized in the pre-shock region, with acoustic waves trapped between the wall and the sonic line even in the recirculation bubble. The dominant frequencies and wave lengths are also found to be in accordance with the second-mode instability. Concurrently, streaky structures are formed in the initial part of the domain, but are then weakened by the shock impingement. The latter creates a much smaller separation region with respect to the unperturbed configuration. The combination of the instability mechanisms and incident shock is such that transition to turbulence is promoted only after the reattachment point. This is clearly shown by the evolution of the skin friction coefficient, which exhibits an anomalous peak due to the foot of the incident shock. The total wall heat flux follows approximately the same trend, albeit the vibrational contribution is one order of magnitude smaller its rototranslational counterpart and mainly of opposite sign. The correlation between $C_f$ and the Stanton number still stands, except in the interaction region. In the fully turbulent portion downstream the impinging shock, turbulent statistics reveal reasonable self-similarity and corroborate the results previously obtained for turbulent boundary layers. 
Thermal non-equilibrium is quantified by means of mean and fluctuating temperature values, vibrational Damk{\"o}hler numbers and contributions of the vibrational source terms. Both molecular oxygen and molecular nitrogen exhibit vibrational excitation, but in different portion of the domain; namely, $Da_{\text{O}_2}$ is of order unity in the laminar region up to the recirculation bubble, whereas $Da_{\text{N}_2}$ attains similar values in the fully turbulent zones. Vibrational modes are found to be almost everywhere in an under-excited state, the largest amount of thermal nonequilibrium being achieved in the recirculating bubble. The correlation coefficients of the two temperatures with respect to $p$, $u$ and $v$ drastically differ and highlight the important decoupling between the internal vibrational and dynamic fields.\\
The current study represents a first step towards the understanding of the influence of high-enthalpy effects on shocked turbulent flow configurations. Future investigations on the subject will mainly focus on three different aspects; namely, i) the characterization of possible low-frequency unsteady motions detected by considering longer integration times, ii) the exploration of different regimes, in particular taking into account higher free-stream total stagnation enthalpies, and iii) the analysis of the interaction with fully-turbulent incoming boundary layers.
\acknowledgements
This work was granted access to the HPC resources of IDRIS and TGCC under
the allocation A0092B10947 made by GENCI (Grand Equipement National de Calcul Intensif). D. Passiatore and G. Pascazio were partially supported by the Italian Ministry of Education, University and Research under the Program Department of Excellence Legge 232/2016 (Grant No. CUP - D94I18000260001).

\revappendix

\section{validation}\label{app:sandham}
\begin{figure}
\centering
 \begin{tikzpicture}
   \node[anchor=south west,inner sep=0] (a) at (0,0) {\includegraphics[width=0.49\textwidth]{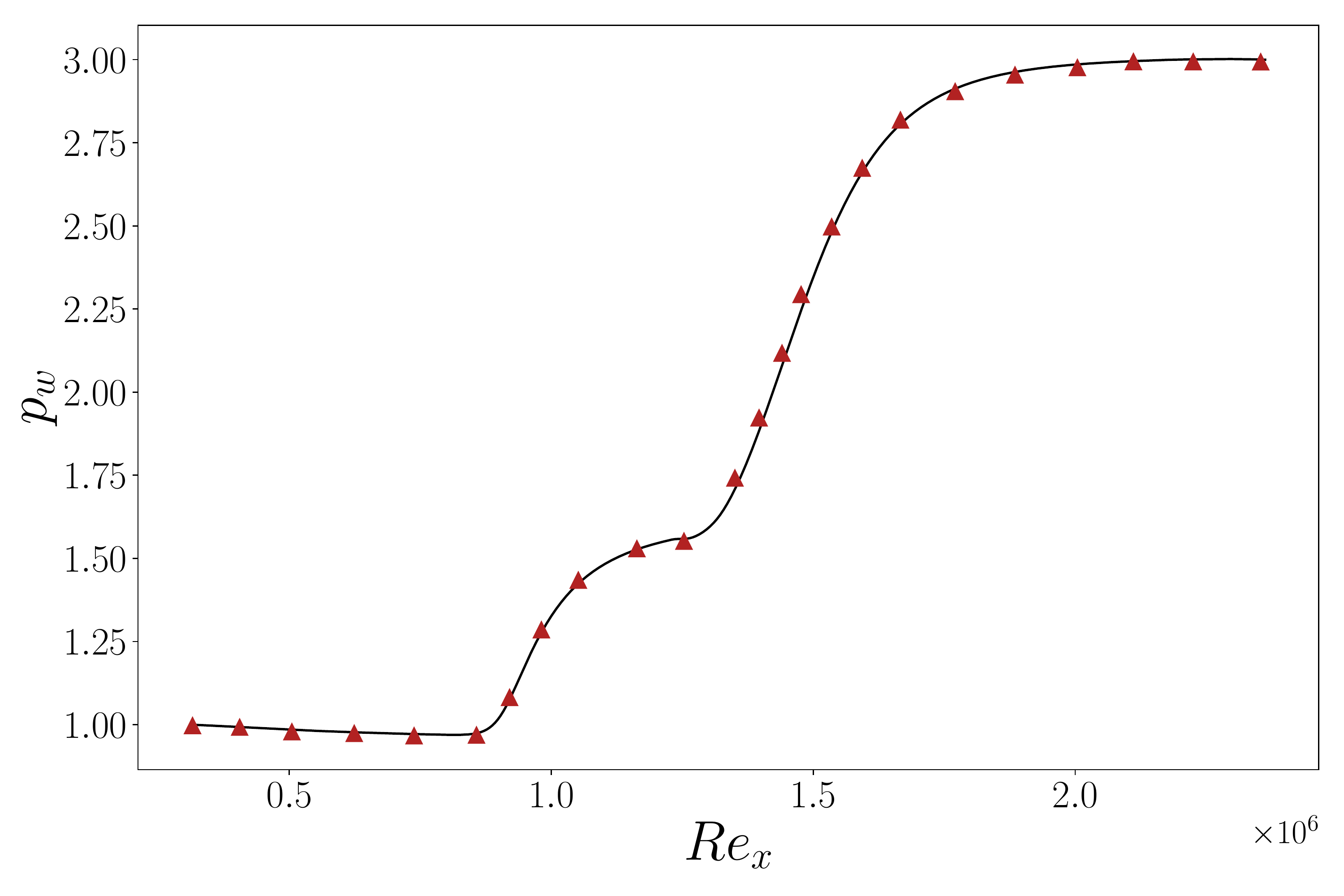}};
   \begin{scope}[x={(a.south east)},y={(a.north west)}]
     \node [align=center] at (0.03,0.95) {(a)};
   \end{scope}
 \end{tikzpicture}
  \begin{tikzpicture}
   \node[anchor=south west,inner sep=0] (a) at (0,0) {\includegraphics[width=0.49\textwidth]{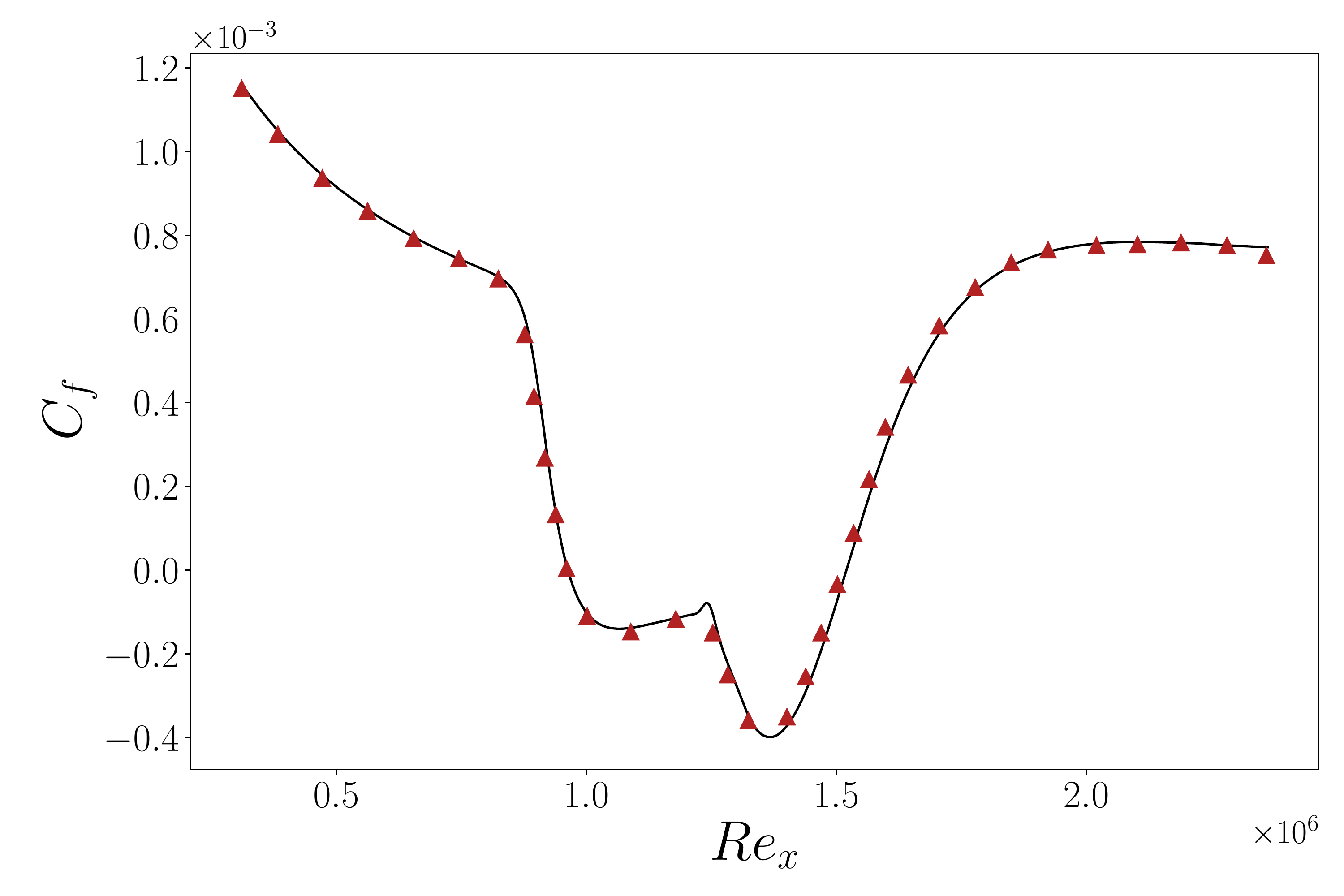}};
   \begin{scope}[x={(a.south east)},y={(a.north west)}]
     \node [align=center] at (0.03,0.95) {(b)};
   \end{scope}
 \end{tikzpicture}
\caption{Verification test cases for the two-dimensional SWBLI configuration. Evolution of normalized wall pressure (a) and skin friction coefficient (b), compared with the reference solution \cite[symbols from][]{sandham2014transitional}.}
\label{fig:swbli_valid_lam}
\end{figure}

For the purpose of validation, we consider the configuration analyzed by Sandham \textit{et al.} \cite{sandham2014transitional} of a shock-wave interacting with a flat-plate boundary layer at Mach 6. The freestream conditions are $T_\infty=\SI{65}{K}$, $p_\infty=\SI{335.24}{Pa}$, $M_\infty = 6$; the wall temperature is fixed equal to $T_w = \SI{292.5}{K}$. In order to minimize the differences with respect to the reference case, Sutherland's law is used to compute the dynamic viscosity with $T_\text{ref}=\SI{65}{K}$ and $\mu_\text{ref}=\SI{4.335 e-6}{Pa\,s}$, along with a constant Prandtl number equal to 0.72 and a specific heat ratio $\gamma = 1.4$. For the considered freestream conditions, a deflection angle of $ \vartheta  = \SI{4}{\degree}$ generates a shock angle of $\beta \approx \SI{12}{\degree}$. The shock impinges on the boundary layer at a distance from the leading edge equal to $\SI{0.344}{m}$. The profiles of the similarity theory are imposed as inflow boundary condition at a Reynolds number $Re_{x_\text{in}}=31.36 \times 10^4$, corresponding to $Re_{\delta^\star_\text{in}}=6830$. The dimensions of the computational domain are $L_x \times L_y = 300 \delta^\star_\text{in} \times 35 \delta^\star_\text{in}$, discretized with 2400 and 300 grid points in the streamwise and wall-normal directions, respectively. A constant grid stretching is applied in the wall-normal direction. Figure~\ref{fig:swbli_valid_lam} depicts the evolution of the normalized wall pressure and skin friction coefficient, respectively. The results are in excellent agreement with the reference data and faithfully reconstruct the pressure jump, the separation region and the reattachment zone.\\
Afterwards, we have compared the results of the three-dimensional perturbed field. We have selected the case referred to as S-H by the authors. The results reported in figure~\ref{fig:swbli_valid_turb} show a quite acceptable agreement and the transition to turbulence is well captured. In the present simulation there is a mild separation bubble with respect to the authors results. Since the separation is extremely weak, this can be attributed to the statistically averaging of the skin friction coefficient. We also show, in figure~\ref{fig:sandham_3D}, the instantaneous flow field colored by of the numerical schlieren in a $xy$-plane and the normalized streamwise velocity in a plane parallel to the wall. The results are in accordance with the imposed perturbation and with the analyses in Sandham \textit{et al.} \cite{sandham2014transitional}.
\begin{figure}
\centering
 \begin{tikzpicture}
   \node[anchor=south west,inner sep=0] (a) at (0,0) {\includegraphics[width=0.49\textwidth]{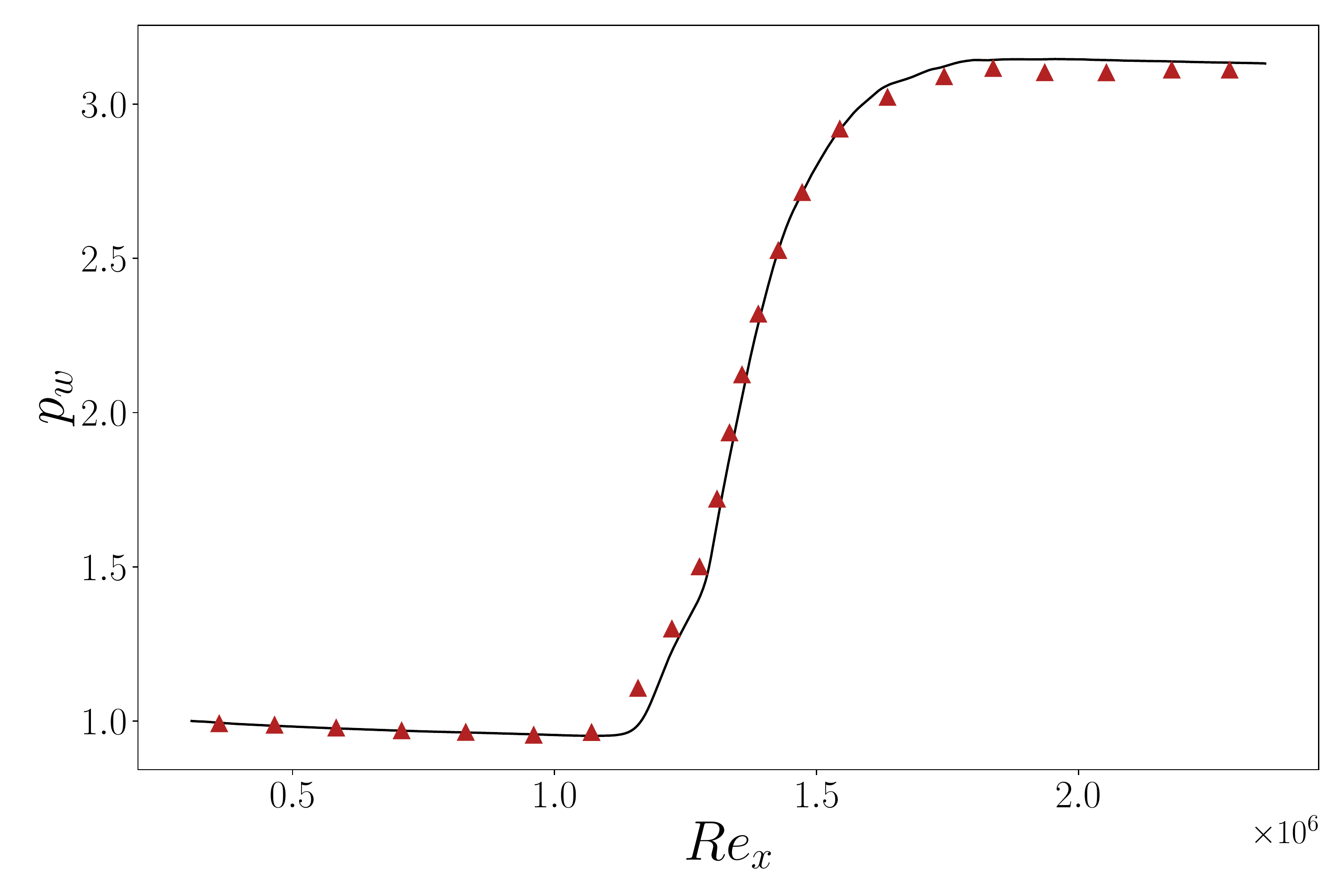}};
   \begin{scope}[x={(a.south east)},y={(a.north west)}]
     \node [align=center] at (0.03,0.95) {(a)};
   \end{scope}
 \end{tikzpicture}
  \begin{tikzpicture}
   \node[anchor=south west,inner sep=0] (a) at (0,0) {\includegraphics[width=0.49\textwidth]{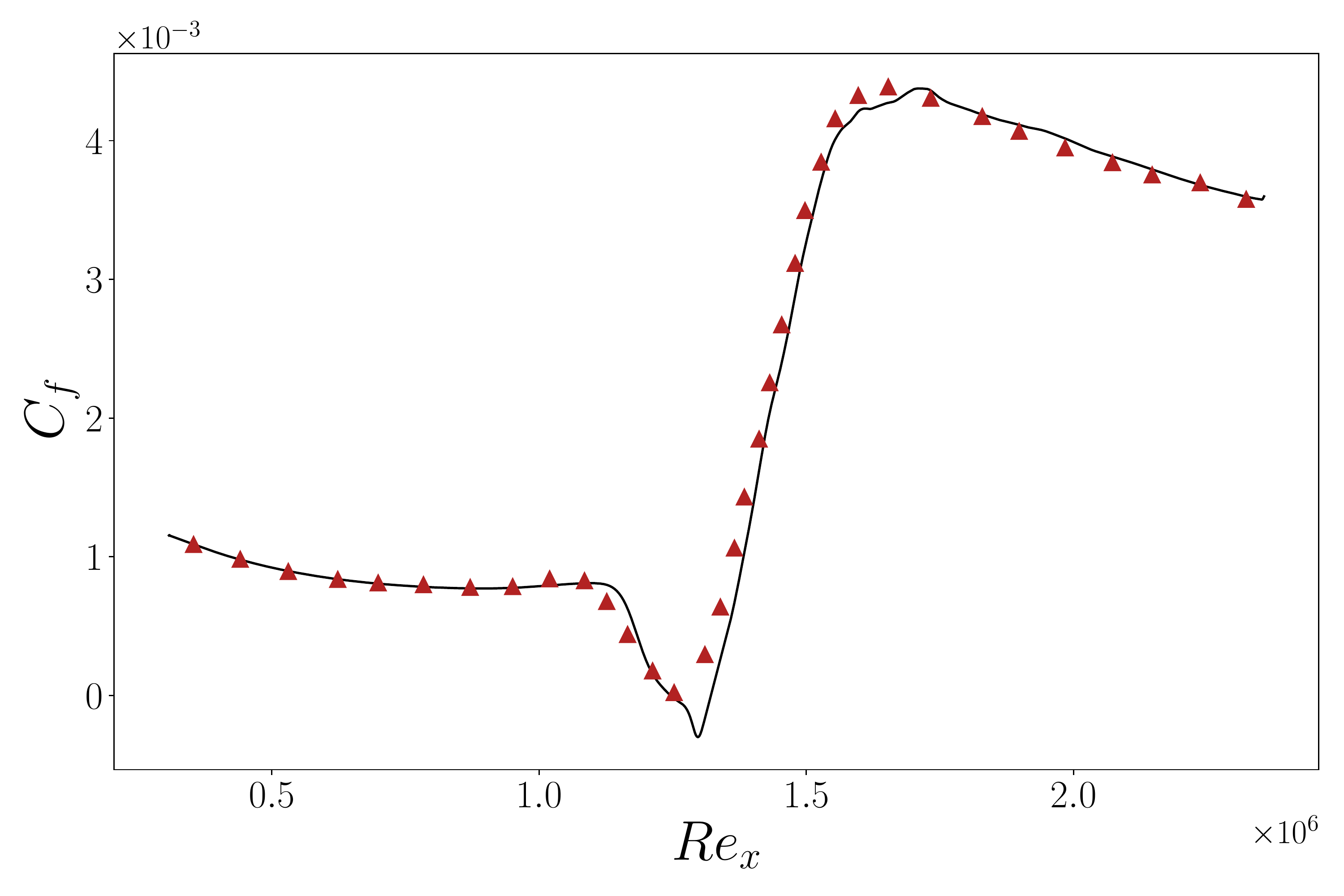}};
   \begin{scope}[x={(a.south east)},y={(a.north west)}]
     \node [align=center] at (0.03,0.95) {(b)};
   \end{scope}
 \end{tikzpicture}
\caption{Verification test cases for the three-dimensional perturbed SWBLI configuration. Evolution of normalized wall pressure (a) and skin friction coefficient (b), compared with the reference solution \cite[symbols from][]{sandham2014transitional}.}
\label{fig:swbli_valid_turb}
\end{figure}
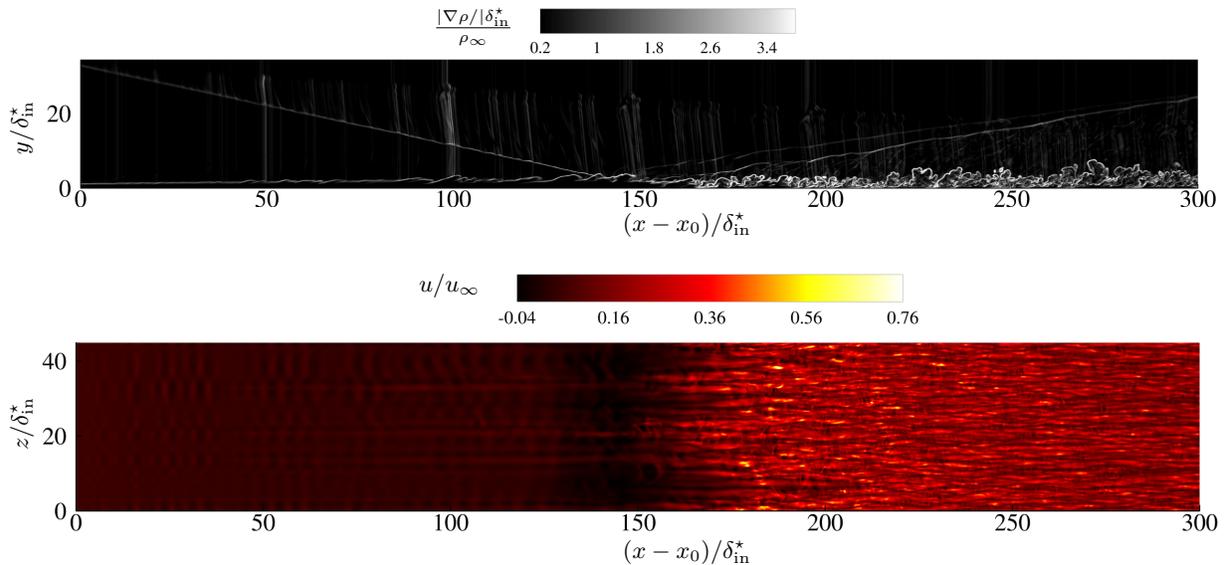
\begin{figure}
 \centering
 \begin{tikzpicture}
   \node[anchor=south west,inner sep=0] (image) at (0,0) {
   \includegraphics[width=0.88\columnwidth,trim={2 3 2 5},clip]{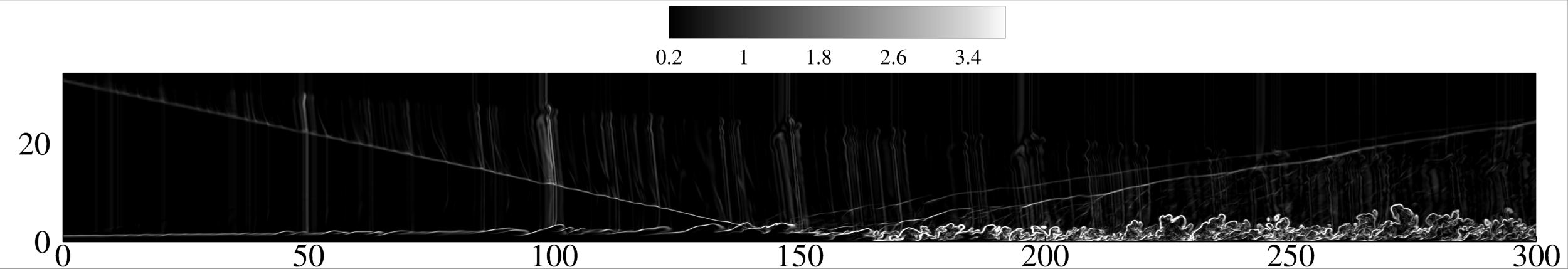}};
   \begin{scope}[x={(image.south east)},y={(image.north west)}]
    \node [align=center] at (0.37,0.9) {\small $\frac{|\nabla \rho/|\delta^\star_\text{in}}{\rho_\infty}$};
    \node [align=center] at (0.55,-0.085) {\small $(x-x_0)/\delta^\star_\text{in}$};
    \node [align=center, rotate =90] at (-0.01,0.4) {$y/\delta^\star_\text{in}$};
   \end{scope}
  \end{tikzpicture}
 \centering
    \begin{tikzpicture}
   \node[anchor=south west,inner sep=0] (image) at (0,0) {
   \includegraphics[width=0.88\columnwidth,trim={2 3 2 5},clip]{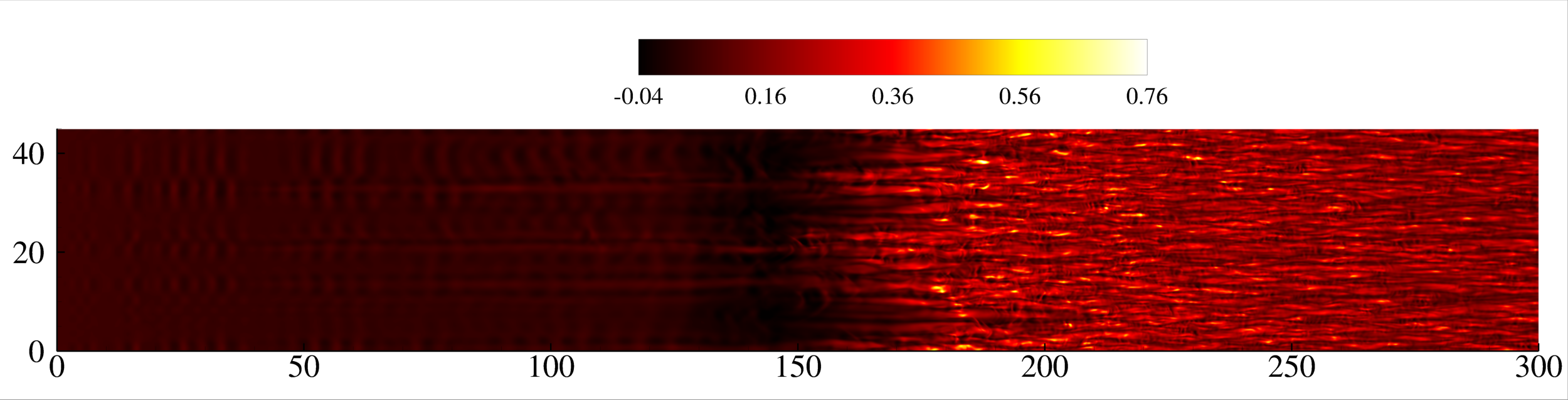}};
   \begin{scope}[x={(image.south east)},y={(image.north west)}]
    \node [align=center] at (0.35,0.87) {\small $u/u_\infty$};
    \node [align=center] at (0.55,-0.02) {\small $(x-x_0)/\delta^\star_\text{in}$};
    \node [align=center, rotate =90] at (-0.01,0.41) {$z/\delta^\star_\text{in}$};
   \end{scope}
  \end{tikzpicture}
 \caption{Flow visualization. Numerical Schlieren (top) and normalized streamwise velocity in a \textit{xz}-plane near the wall (bottom).}
\label{fig:sandham_3D}
\end{figure}
\bibliography{biblio}
\end{document}